\documentclass[aps,pra,superscriptaddress,reprint,floatfix]{revtex4-2}

\usepackage{graphicx}
\usepackage{dcolumn}
\usepackage{bm}
\usepackage{amsmath,amssymb}
\usepackage{float}
\usepackage{booktabs}
\usepackage{array}
\usepackage{relsize}
\usepackage{braket}
\usepackage[colorlinks=true,citecolor=blue,urlcolor=blue]{hyperref}

\newcommand{\cmark}{\textcolor{blue}{\checkmark}}
\newcommand{\xmark}{\textcolor{blue}{\times}}
\newcommand{\Hket}{\ket{H}}
\newcommand{\Vket}{\ket{V}}

\begin{document}

\title{\textbf{Ternary Quantum Eraser Cryptography} 
}

\author{Ahmed Halawani}
\email{ahalawani@kacst.gov.sa}
 \affiliation{Institute of Quantum Technologies and Advanced Computing, KACST, Riyadh, 11442, Saudi Arabia}
 
\author{Yahya Meshalwi Khabrani}
\affiliation{
 Department of Physics, College of Science, Imam Mohammad Ibn Saud Islamic University
(IMSIU), P.O. Box 65892, Riyadh 11566, Saudi Arabia
}

\author{A. Al-Mogeeth}
\affiliation{
 Department of Physics, College of Science, King Khalid University, P.O Box 9004, Abha 61413, Saudi Arabia
}

\author{Zheng-Hong Li}
\email{crefirefox@shu.edu.cn}
\affiliation{
 Institute for Quantum Science and Technology and Department of Physics, Shanghai University, Shanghai, 200444, China
}
\affiliation{
 Shanghai Key Laboratory of High Temperature Superconductors, Shanghai University, Shanghai, 200444, China
}
\author{M. Al-Amri}
\email{mdalamri@kacst.gov.sa}
 \affiliation{Institute of Quantum Technologies and Advanced Computing, KACST, Riyadh, 11442, Saudi Arabia}
 \affiliation{
 NCQOQI, KACST, P.O. Box 6086, Riyadh, 11442, Saudi Arabia
}

\date{\today}

\begin{abstract}
Quantum key distribution protocols based on the quantum eraser phenomenon offer an operational advantage: automatic identification of matching and mismatching encoding choices through interference, eliminating basis reconciliation. However, binary quantum eraser implementations permit an eavesdropper to recover Alice's encoded bit with $85\%$ probability. To overcome this constraint, we introduce a ternary quantum eraser protocol employing three polarization states with $120^\circ$ angular separation, transmitted in three-photon groups with randomized temporal ordering. This extension achieves enhanced security through two complementary mechanisms. First, the reduced distinguishability of symmetrically-arranged quantum states limits single-photon discrimination. Second, the combinatorial complexity of unknown photon ordering constrains multi-photon eavesdropping strategies. Security analysis against individual eavesdropping attacks within the four-dimensional path-polarization Hilbert space establishes that an eavesdropper's maximum success probability is bounded at 54\%, substantially below the binary discrimination bound. The protocol maintains a binary-equivalent efficiency of 0.30 bits per photon, comparable to established binary QKD protocols at the sifted-rate level, while preserving the operational simplicity inherent to quantum eraser cryptography.
\end{abstract}

\maketitle

\section{Introduction}

Quantum key distribution (QKD) enables parties to establish shared secret keys with security guaranteed by quantum mechanics rather than computational assumptions~\cite{BB84, Ekert1991, Gisin2002}. Since the introduction of the BB84 protocol~\cite{BB84}, extensive research has advanced quantum cryptography from theoretical foundations to commercial deployments spanning hundreds of kilometers via fiber and satellite links~\cite{Xu_2020, Pirandola_2020, Tang_2014, Boaron2018, Yin2017, Cao2023}.

The security of QKD protocols fundamentally relies on the inability of an eavesdropper to perfectly discriminate or clone non-orthogonal quantum states~\cite{Wootters1982ASQ, DIEKS1982271}. When information is encoded in such states, any measurement by an eavesdropper necessarily disturbs the quantum channel in a detectable manner. However, the degree of security varies significantly across protocols, depending on the encoding scheme, the number and arrangement of quantum states employed, and resilience to practical imperfections~\cite{Scarani_2009, Tamaki_2014, pereira2019}. Device imperfections, including source flaws, detector inefficiencies, and side channels, can be exploited through sophisticated attacks. Trojan-horse intrusion, photon-number-splitting, and detector blinding can all compromise security guarantees derived from idealized analyses~\cite{Huang_2018, Jain_2014, Sun2021}.

Quantum eraser cryptography \cite{Salih_2016} represents a distinctive approach to secure key distribution that exploits the fundamental complementarity between which-path information and interference visibility~\cite{PhysRevA.25.2208, PhysRevLett.84.1, AharonovZubairy2005}. In the quantum eraser framework, encoding operations that mark the photon path destroy interference at the output beam splitter. Matched operations by sender and receiver restore interference through effective erasure of path information. This mechanism provides an operational advantage: the interference pattern itself distinguishes matched from mismatched encoding choices, eliminating the basis reconciliation step that conventional protocols require. Conventional protocols require public comparison of basis choices and discard mismatched measurements; quantum eraser protocols instead achieve automatic sifting through interference physics, streamlining key distribution, a goal also pursued via non-interference-based schemes~\cite{Gao_2006}.

Despite the operational elegance, binary quantum eraser implementations face a fundamental security limitation. When information is encoded in two
non-orthogonal quantum states, optimal eavesdropping strategies achieve identification probability approaching 85\% ~\cite{Helstrom1969}. Section~\ref{sec:security-binary} derives this value and shows that the four-state and randomized-polarization variants yield the same bound.

These inherent security limitations in binary quantum eraser protocols motivate exploration of higher-dimensional encoding schemes. Theoretical analysis of quantum state discrimination establishes that symmetric arrangements of multiple non-orthogonal states can reduce distinguishability compared to two-state systems with overlap $1/\sqrt{2}$. Ternary protocols employing three symmetrically-arranged states have emerged as promising candidates for enhanced security in both standard QKD and quantum secure direct communication~\cite{BechmannPasquinucci2000, LongLiu2002, Deng2003}. In the continuous-variable regime, security proofs for ternary coherent-state protocols with $120^\circ$ phase separation have demonstrated the viability of three-state encoding for practical implementations~\cite{Bradler2019}. High-dimensional approaches using ternary Hadamard gates have similarly demonstrated security advantages over binary protocols through exploitation of mutually unbiased bases~\cite{Chen2022}. The challenge lies in designing protocols that capture these security benefits while preserving practical efficiency and the operational advantages characteristic of quantum eraser cryptography.

We emphasize that the novelty of this work does not lie in the abstract properties of ternary quantum states, whose minimum-error discrimination bounds are well established~\cite{Helstrom1969, BarnettCroke2009}. Rather, our contribution is to embed ternary encoding within the quantum eraser cryptographic architecture, preserving its defining operational feature, automatic basis reconciliation through interference, while overcoming the security ceiling inherent to binary implementations. Beyond this specific application, the work addresses a broader question relevant to QKD generally: can symmetric multi-state encodings provide security advantages that justify increased implementation complexity~\cite{Bruss1998, ErhardKrennZeilinger2020, FicklerPrabhakar2021}? Our affirmative answer, with quantified bounds, informs protocol design choices across diverse QKD architectures.

In this work, we develop a ternary quantum eraser protocol that reduces eavesdropping success probability while maintaining the automatic basis reconciliation that distinguishes the quantum eraser approach. Our protocol employs three polarization states arranged with $120^\circ$ angular separation, transmitted in groups of three photons with randomized temporal ordering. This design achieves security through two complementary mechanisms: the quantum mechanical indistinguishability of symmetrically-arranged non-orthogonal states limits single-photon discrimination, while the combinatorial complexity introduced by unknown photon ordering constrains multi-photon eavesdropping strategies. Our security analysis establishes that the ternary protocol limits eavesdropping success under individual attacks to 54\% while maintaining efficiency of approximately 0.30 sifted bits per photon, within the range of established binary QKD protocols at the sifted level and comparable to recently proposed encoding-decoding optimizations~\cite{Bebrov2021}.

The remainder of this paper is organized as follows. Section~\ref{sec:fundamentals} establishes the quantum eraser cryptography framework. Section~\ref{sec:security-binary} analyzes binary protocol security across multiple variants. Section~\ref{sec:efficiency-binary} develops the general security-efficiency trade-off theory. Section~\ref{sec:ternary} presents the ternary quantum eraser protocol. Section~\ref{sec:security-ternary} derives the ternary security bound. Section~\ref{sec:conclusion} is the conclusion.

\section{Quantum Eraser Cryptography Fundamentals}
\label{sec:fundamentals}

\subsection{Basic Protocol}

The quantum eraser exploits complementarity between which-path information and interference visibility~\cite{PhysRevA.25.2208, PhysRevLett.84.1, Qureshi_2021, Englert}. Recording which-path information destroys interference; erasing this information restores it~\cite{violaris2025counterfactualsmacroscopicquantumphysics,Wheeler1978MFQT, Wheeler:1984dy}. The phenomenon relies on coherent mapping between path and polarization degrees of freedom.

\begin{figure}[ht]
\centering
\includegraphics[width=1\columnwidth]{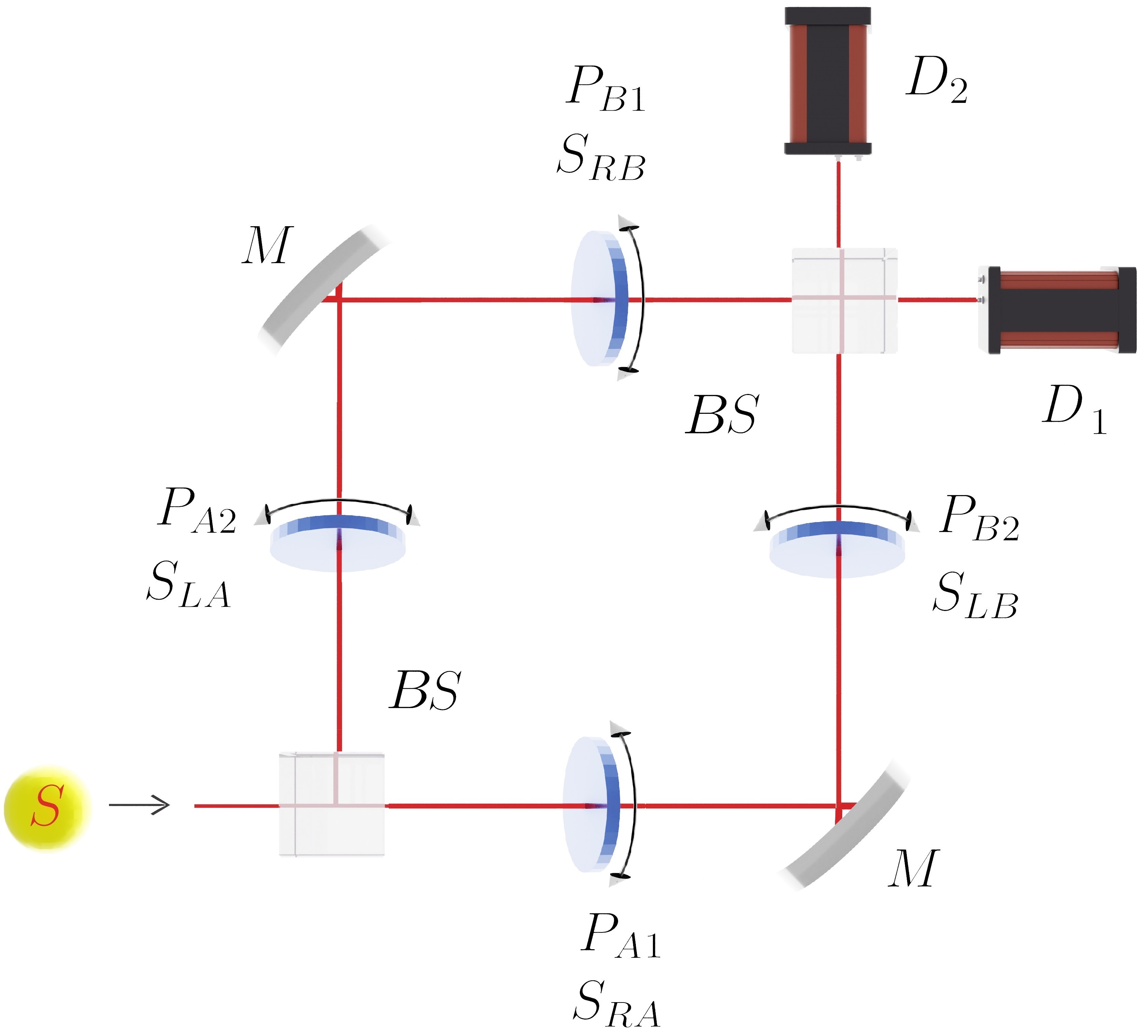}
\caption{Schematic of the binary quantum eraser cryptography setup. A single photon enters a balanced Mach--Zehnder interferometer (MZI) through input port S, where it is separated into the two spatial paths $|U\rangle$ and $|L\rangle$ by the first beam splitter (BS$_1$). Alice's station applies a polarization rotation $S_L$ on the upper path, encoding her bit value by choosing whether the photon acquires which-path information. Bob independently encodes his bit using the rotation $S_R$ on the lower path. When Alice and Bob choose matching settings, the path--polarization tags introduced by $S_L$ and $S_R$ cancel at BS$_2$, restoring full interference between $|U\rangle$ and $|L\rangle$. In this case constructive interference directs the photon deterministically to detector $D_1$, while $D_2$ remains dark. When Alice and Bob use mismatched settings, their operations imprint distinct polarization states on the two paths, preventing the recombination amplitudes at BS$_2$ from interfering. The destruction of interference yields equal intensities at the output ports, hence detector $D_2$ clicks with probability $1/2$. Because only mismatched settings lead to non-interfering amplitudes, a click at $D_2$ uniquely identifies opposite encoding choices. This enables key generation without basis reconciliation.}
\label{fig:setup}
\end{figure}

Figure~\ref{fig:setup} illustrates the basic configuration of the quantum eraser cryptography system. A polarized photon (with input state $|\psi_i\rangle$) from source S enters a Mach-Zehnder interferometer through beam splitter BS1, creating a superposition of upper and lower path states. Alice controls polarization rotators $P_{A1}$ and $P_{A2}$ placed in the upper and lower paths respectively. When activated, these rotators transform the photon polarization such that the two paths carry orthogonal polarization states, destroying the interference pattern. Bob's rotators $P_{B1}$ and $P_{B2}$ can precisely cancel Alice's transformations when both parties apply the same operation. The second beam splitter BS2 recombines the paths, and detectors $D_1$ and $D_2$ register the output.

Alice encodes binary information by controlling her polarization rotators: encoding “1” by activating both rotators (making the path polarizations orthogonal), and encoding “0” by leaving them inactive. Bob independently makes the same choice. The key insight is that detector $D_2$ registers photons only when Alice and Bob make different choices, automatically identifying cases suitable for key generation without requiring basis reconciliation.

\subsection{Mathematical Formulation}

\begin{figure}[ht]
\centering
\includegraphics[width=1\columnwidth]{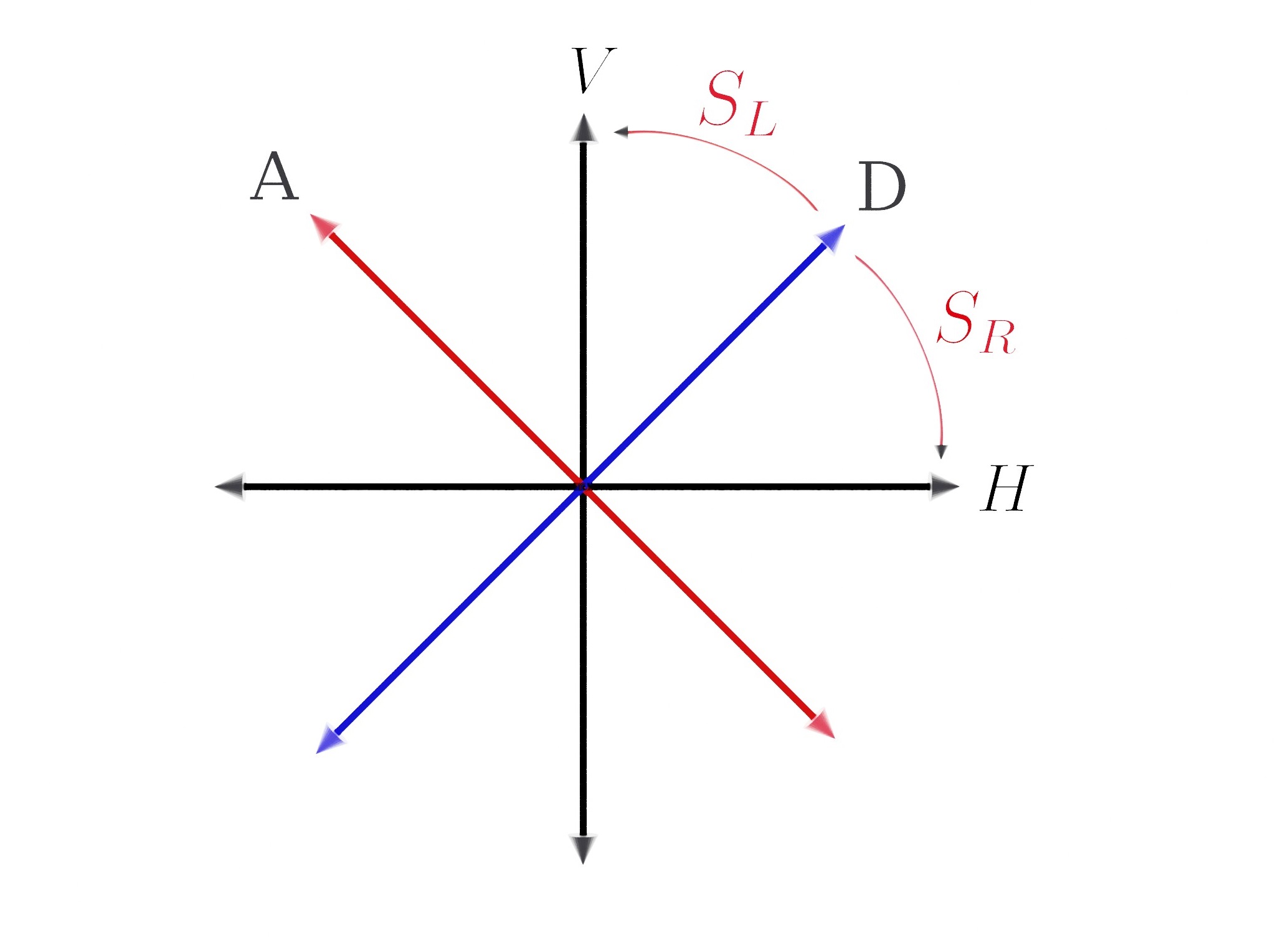}
\caption{Path--polarization channel states generated by Alice's and Bob's encoding operations. The four states shown correspond to the combinations of Alice's operation $S_L$ (applied to the upper path) and Bob's operation $S_R$ (applied to the lower path). Each operation either preserves or rotates the photon's polarization, thereby controlling whether the two paths carry identical or distinguishable polarization tags. When Alice and Bob choose the same operation, the resulting channel state maintains path--polarization symmetry, enabling full interference at BS$_2$ and producing deterministic detection at $D_1$. When they choose different operations, the upper and lower paths acquire orthogonal or partially distinguishable polarization states, preventing interference and yielding equal probabilities at detectors $D_1$ and $D_2$. These four channel states constitute the set $\{|\phi_0^+\rangle, |\phi_0^-\rangle, |\phi_1^+\rangle, |\phi_1^-\rangle\}$ used in the security analysis, and their geometry determines Eve's optimal POVM and the 85\% discrimination bound.}
\label{fig:pol}
\end{figure}

Consider a photon initially prepared in state $|D\rangle = \frac{1}{\sqrt{2}}(|H\rangle + |V\rangle)$ entering the interferometer, where $|H\rangle$ and $|V\rangle$ denote horizontal and vertical polarization states respectively, as shown in Figure~\ref{fig:pol}. 

The beam splitter transformation can be represented as:
\begin{equation}
    \begin{aligned}
        |U\rangle &\rightarrow \cos\theta|U\rangle + \sin\theta|L\rangle ,\\
|L\rangle &\rightarrow \sin\theta|U\rangle - \cos\theta|L\rangle,
    \label{eq:beamsplitter}
    \end{aligned}
\end{equation}
where $|U\rangle$ and $|L\rangle$ denote the upper and lower path states respectively, and $\theta = \pi/4$ for a 50:50 beam splitter.

Suppose the input state from the light-source is $|U\rangle |D\rangle$. After the first beam splitter, the quantum state becomes:
\begin{equation}
|\psi_{\text{initial}}\rangle = \frac{1}{\sqrt{2}}(|U\rangle + |L\rangle) \otimes \frac{1}{\sqrt{2}}(|H\rangle + |V\rangle).
\label{eq:initial_state}
\end{equation}

The polarization manipulation in each arm is performed by polarization rotators, labeled $P_{A1}$ and $P_{A2}$ at Alice's station and $P_{B1}$ and $P_{B2}$ at Bob's station (see Figure~\ref{fig:setup}). These devices implement one of two unitary operations: $S_L$, which rotates the polarization counterclockwise by $45^\circ$, and $S_R$, which rotates clockwise by $45^\circ$. Specifically, Alice's upper-path rotator $P_{A1}$ and Bob's lower-path rotator $P_{B2}$ implement $S_L$, while Alice's lower-path rotator $P_{A2}$ and Bob's upper-path rotator $P_{B1}$ implement $S_R$. The complementary assignment ensures that when both parties activate their rotators, the net transformation on each path is the identity, restoring interference at the output beam splitter. The polarization rotators are characterized by the operators:
\begin{equation}
S_L = \frac{1}{\sqrt{2}}(|H\rangle\langle H| + |V\rangle\langle V| - |H\rangle\langle V| + |V\rangle\langle H|),
\end{equation}
and:
\begin{equation}
S_R = S_L^\dagger = \frac{1}{\sqrt{2}}(|H\rangle\langle H| + |V\rangle\langle V| + |H\rangle\langle V| - |V\rangle\langle H|),
\end{equation}
in the $\{|H\rangle, |V\rangle\}$ basis. These operators satisfy $S_L = S_R^\dagger$, indicating their complementary nature. The action of the rotators on the polarization degree of freedom can be described as:
\begin{equation}
\begin{aligned}
    &S_L: \begin{pmatrix} |H\rangle \\ |V\rangle \end{pmatrix} \to \frac{1}{\sqrt{2}}\begin{pmatrix} |H\rangle + |V\rangle \\ - |H\rangle + |V\rangle \end{pmatrix},\\
    &S_R: \begin{pmatrix} |H\rangle \\ |V\rangle \end{pmatrix} \to \frac{1}{\sqrt{2}}\begin{pmatrix} |H\rangle - |V\rangle \\ |H\rangle + |V\rangle \end{pmatrix}.
\end{aligned}
\label{eq:rotators}
\end{equation}

Alice's encoding operation is implemented by the transformation:
\begin{equation}
T_A =  |U\rangle\langle U| \otimes S_L  + |L\rangle\langle L| \otimes S_R,
\label{eq:alice_operator}
\end{equation}
which applies $S_L$ to the upper path and $S_R$ to the lower path. Similarly, Bob's operation is:
\begin{equation}
T_B = |U\rangle\langle U| \otimes S_R  + |L\rangle\langle L|  \otimes S_L.
\label{eq:bob_operator}
\end{equation}

These operators satisfy the property $T_B T_A = I$, demonstrating that Bob's operation precisely cancels Alice's when both are applied sequentially. This complementarity forms the basis for the cryptographic protocol; Appendix~\ref{app:operators} derives the residual distinguishability when the rotators are non-ideal.

\subsection{Four encoding cases and detection statistics}

We analyze the complete quantum state evolution for all four possible combinations of Alice and Bob's encoding choices. There are four cases for Alice and Bob's different choices, which reveal how the detector statistics automatically identify matching versus mismatching encodings. Let “0” denote \emph{inactive} (no rotations) and “1” denote \emph{active} (apply $T_A$ or $T_B$). After BS1 and the (in)active operations, the state is recombined by BS2 and routed to $D_1$ or $D_2$. For a balanced interferometer:

\paragraph{(0,0) and (1,1): matched choices.}
When Alice and Bob both choose 0 (neither activates their rotators), the polarization is unmarked and the constructive interference at $BS_2$ directs all photons to detector $D_1$,

\begin{equation}
\cos\theta |U\rangle|D\rangle + \sin\theta |L\rangle|D\rangle \xrightarrow{BS_2} |U\rangle|D\rangle.
\label{eq:case_00}
\end{equation}

When both choose 1 (both activate their rotators), $T_B T_A=\mathbb{I}$ restores the unmarked state before BS2. The combined operation leaves the state unchanged:
\begin{equation}
T_B T_A[\cos\theta |U\rangle|D\rangle + \sin\theta |L\rangle|D\rangle] \xrightarrow{BS_2} |U\rangle|D\rangle.
\label{eq:case_11}
\end{equation}

In both cases, all photons go to $D_1$: $P(D_1)=1$, $P(D_2)=0$.

\paragraph{(1,0) and (0,1): mismatched choices.}
If only one party is active, the two paths carry orthogonal polarizations at BS2, thus interference is erased. 

When only Alice's rotators are active, transforming the state to create which-path information:
\begin{equation}
\begin{aligned}
  & T_A[\cos\theta |U\rangle|D\rangle + \sin\theta |L\rangle|D\rangle] \\
 & \qquad \xrightarrow{BS_2} \ \cos ^2 \theta|U\rangle|V\rangle+\sin ^2 \theta|U\rangle|H\rangle \\
  & \qquad\qquad+\cos \theta \sin \theta|L\rangle|V\rangle-\sin \theta \cos \theta|L\rangle|H\rangle \\
& \qquad\qquad=\frac{1}{2}(|U\rangle|V\rangle+|L\rangle|V\rangle-|L\rangle|H\rangle+|U\rangle|H\rangle)\Big|_{\theta=45^{\circ}}\!\!\!.
\label{eq:case_10}
\end{aligned}
\end{equation}

Similarly, when only Bob's rotators are active:
\begin{equation}
\begin{aligned}
& T_B[\cos\theta |U\rangle|D\rangle + \sin\theta |L\rangle|D\rangle] \\
 &\qquad \xrightarrow{BS_2} \ \cos ^2 \theta|U\rangle|H\rangle+\sin ^2 \theta|U\rangle|V\rangle \\
&  \qquad \qquad+\cos \theta \sin \theta|L\rangle|H\rangle-\sin \theta \cos \theta|L\rangle|V\rangle \\
& \qquad\qquad=\left.\frac{1}{2}(|U\rangle|H\rangle+|L\rangle|H\rangle-|L\rangle|V\rangle+|U\rangle|V\rangle)\right|_{\theta=45^{\circ}}\!\!\!.
\label{eq:case_01}
\end{aligned}
\end{equation}

In these two cases, the state contains which-path information encoded in the orthogonal polarizations. The outputs split evenly, $P(D_1)=P(D_2)=\tfrac{1}{2}$.

\subsection{Key sifting rule and consequence}
Since $D_2$ clicks \emph{only} for mismatched choices, Alice and Bob assign raw-key bits exclusively to $D_2$ events and discard $D_1$ events or use them for channel monitoring. Because the detectors are located at Bob's station, he must announce which rounds produced $D_2$ clicks so that Alice can identify the key-generating events. This announcement is the only public communication required; it reveals which rounds are usable but carries no information about Alice's or Bob's encoding choices, since both mismatched configurations produce identical detection statistics. The key bit value is determined by a predetermined convention mapping the mismatched configuration to either ``0'' or ``1'' (for instance, assigning ``1'' when Alice encoded ``1'' and Bob encoded ``0'', and ``0'' for the reverse case).

A crucial security feature is that both mismatched cases produce identical detection statistics: $P(D_1) = P(D_2) = 1/2$. An eavesdropper observing only detection outcomes cannot distinguish which mismatched configuration occurred. Combined with the non-orthogonality of the transmitted quantum states, this indistinguishability provides the foundation for secure key distribution.

The protocol thus achieves key sifting without public basis reconciliation, a significant operational advantage over standard protocols, meaningful only if the protocol resists eavesdropping attacks.

\section{Security Analysis of Binary Protocol}
\label{sec:security-binary}

We establish security through two results: first, that an eavesdropper (Eve) cannot copy the in-flight states without disturbance (no-cloning), and second, that even with optimal measurements her success probability is bounded. We begin with the simplest two-state setting before analyzing multi-state variants.

\subsection{Eve cannot copy the photon states in the public transmission channel}
\label{subsec:no-cloning}

We begin by proving that an eavesdropper cannot perfectly copy the quantum states in the transmission channel without disturbing them. In the transmission channel, the protocol produces four pure states corresponding to Alice's bit and the sign of the initial polarization,
\begin{equation}
\begin{aligned}
|\phi_0^+\rangle &= \cos\theta|U\rangle|D\rangle + \sin\theta|L\rangle|D\rangle ,\\
|\phi_0^-\rangle &= \cos\theta|U\rangle|A\rangle + \sin\theta|L\rangle|A\rangle, \\
|\phi_1^+\rangle &= \cos\theta|U\rangle|V\rangle + \sin\theta|L\rangle|H\rangle, \\
|\phi_1^-\rangle &= \cos\theta|U\rangle|H\rangle - \sin\theta|L\rangle|V\rangle,
\label{states}
\end{aligned}
\end{equation}
where $\theta = \pi/4$. The subscript indicates Alice's encoding of “0” and “1” and superscript indicates the initial polarization.

For these states, pairs $\{|\phi_0^+\rangle,|\phi_1^+\rangle\}$ and $\{|\phi_0^-\rangle,|\phi_1^-\rangle\}$ are non-orthogonal:

\begin{equation}
\begin{aligned}
& \left\langle\phi_0^- | \phi_0^+\right\rangle=\left\langle\phi_1^- |\phi_1^+\right\rangle=0, \\
& \left\langle\phi_1^+ |\phi_0^+\right\rangle=\left\langle\phi_1^- |\phi_0^-\right\rangle=\frac{1}{\sqrt{2}}, \\
& \left\langle\phi_1^- |\phi_0^+\right\rangle=\frac{1}{\sqrt{2}}\left(\cos ^2 \theta-\sin ^2 \theta\right)=\left.0\right|_{\theta=45^{\circ}}, \\
& \left\langle\phi_1^+ | \phi_0^-\right\rangle=\frac{1}{\sqrt{2}}\left(-\cos ^2 \theta+\sin ^2 \theta\right)=\left.0\right|_{\theta=45^{\circ}},
\end{aligned}
\end{equation}
demonstrating their non-orthogonality. According to the no-cloning theorem, non-orthogonal quantum states cannot be perfectly copied. If Eve attempts to clone these states via $U_E(|\phi_i\rangle \otimes |E\rangle) = |\phi_i\rangle \otimes |\phi_i\rangle$, where $|E\rangle$ is her initial ancilla states, the unitarity requirement leads to:
\begin{equation} 
\begin{aligned} 
    \langle\phi_j|\langle E|U_E^\dagger U_E|\phi_i\rangle|E\rangle &=\langle\phi_j|\phi_i\rangle\langle\phi_j|\phi_i\rangle, \\           
    |\langle\phi_i|\phi_j\rangle| &= |\langle\phi_i|\phi_j\rangle|^2.
\end{aligned}                 \end{equation}                     

This equation is satisfied only when $|\langle\phi_i|\phi_j\rangle| \in \{0, 1\}$, i.e., only orthogonal or     
identical states can be cloned. Since our states have inner product $|\langle\phi_i|\phi_j\rangle| = 1/\sqrt{2}$, perfect cloning is impossible. This proves that Eve cannot perfectly clone the quantum states without disturbing them~\cite{Wootters1982ASQ, DIEKS1982271}, indicating that any attempt to learn the state necessarily introduces disturbance. Appendix~\ref{app:operators} presents the explicit no-cloning argument and the transition to the Helstrom bound.     

\subsection{The additional error caused by Eve}
\label{subsec:eve-error}

We next quantify Eve’s information gain under optimal measurement strategies.

We evaluate Eve's information gain in three binary-eraser scenarios and compare with BB84:
\begin{itemize}
    \item \textbf{Two non-orthogonal states} (Section~\ref{subsubsec:two-state}): The minimal case where Alice transmits only $|D\rangle$-polarized photons, producing two possible channel states.
    \item \textbf{Four non-orthogonal states} (Section~\ref{subsubsec:four-state}): Alice randomly transmits both $|D\rangle$ and $|A\rangle$ polarizations, doubling the number of channel states.
    \item \textbf{Randomized initial polarization} (Section~\ref{subsubsec:random-pol}): Alice selects the initial polarization angle randomly, potentially obscuring which-path information from Eve.
    \item \textbf{Comparison with BB84} (Section~\ref{subsubsec:bb84}): Contextualizing our results against the canonical quantum key distribution protocol.
\end{itemize}

All three eraser scenarios and the BB84 comparison yield $\approx 85\%$. In the basic two-state case this is the optimal identification probability for states with overlap $1/\sqrt{2}$, namely $(1+1/\sqrt{2})/2$; the other cases give the same value via the case-specific analyses that follow.

To analyze Eve's optimal measurement strategy, we express the four channel states from Eq.~(\ref{states}) in an orthonormal basis $\{|\varphi_1^+\rangle,|\varphi_1^-\rangle,|\varphi_2^+\rangle,|\varphi_2^-\rangle\}$:

\begin{equation}
\begin{aligned}
& \left|\varphi_1^+\right\rangle=\cos \theta|U\rangle|V\rangle+\sin \theta|L\rangle|H\rangle, \\
& \left|\varphi_1^-\right\rangle=\cos \theta|U\rangle|H\rangle-\sin \theta|L\rangle|V\rangle, \\
& \left|\varphi_2^-\right\rangle=\sin \theta|U\rangle|V\rangle-\cos \theta|L\rangle|H\rangle, \\
& \left|\varphi_2^+\right\rangle=\sin \theta|U\rangle|H\rangle+\cos \theta|L\rangle|V\rangle.
\end{aligned}
\end{equation}

Then, $\left|\varphi_0^+\right\rangle$ and $\left|\varphi_0^-\right\rangle$ can be written as:
\begin{equation}
\begin{aligned}
\left|\varphi_0^+\right\rangle&=\frac{1}{\sqrt{2}}\left(\cos 2 \theta\left|\varphi_1^-\right\rangle+\sin 2 \theta\left|\varphi_2^+\right\rangle+\left|\varphi_1^+\right\rangle\right)\\
&=\left.\frac{1}{\sqrt{2}}\left(\left|\varphi_2^+\right\rangle+\left|\varphi_1^+\right\rangle\right)\right|_{\theta=45^{\circ}} ,\\
 \left|\varphi_0^-\right\rangle &=\frac{1}{\sqrt{2}}\left(\left|\varphi_1^-\right\rangle-\cos 2 \theta\left|\varphi_1^+\right\rangle-\sin 2 \theta\left|\varphi_2^-\right\rangle\right)\\
&=\left.\frac{1}{\sqrt{2}}\left(\left|\varphi_1^-\right\rangle-\left|\varphi_2^-\right\rangle\right)\right|_{\theta=45^{\circ}}.
\end{aligned}
\end{equation}

Note that $|\varphi_1^+\rangle$ coincides with the channel state $|\phi_1^+\rangle$, as one of the transmitted states is itself a basis vector. Suppose Eve measures in a different orthonormal basis $|\alpha_j\rangle$, using four detectors corresponding to four orthogonal states. The most general detection bases can be expressed as $
\left|\alpha_j\right\rangle=\sum_i C_i^{(j)}\left|\varphi_i\right\rangle,$
where $j=1,2,3,4 ; \; i=1+, 1-, 2+, 2-$ and $\sum_i C_i^{(j)^*} C_i^{(j)}=1$.

\subsubsection{Alice transporting two non-orthogonal states to Bob}
\label{subsubsec:two-state}

To analyze Eve's optimal strategy, we focus on the simplified case where Alice sends only the $|D\rangle$ polarized photons. Then, the original bases can be reduced as $\{|\varphi_0^+\rangle, |\phi_1^+\rangle\}$. The two possible states in the public channel are:

\begin{equation}
\left\{\begin{array}{l}
\left|\varphi_0^+\right\rangle=\frac{1}{\sqrt{2}}\left(\left|\varphi_2^+\right\rangle+\left|\varphi_1^+\right\rangle\right) , \\
\left|\varphi_1^+\right\rangle.
\end{array}\right.
\label{eq:binary_states}
\end{equation}

The most general measurement Eve can perform is described by a positive operator-valued measure (POVM). Eve's measurement could be described as $A=\sum_{i=1,2}\left|\alpha_i\right\rangle\left\langle\alpha_i\right|$. For distinguishing between two non-orthogonal states $|\varphi_0^+\rangle$ and $|\varphi_1^+\rangle$, the optimal measurement can be parameterized as:

\begin{equation}
\begin{aligned}
& \left|\alpha_1\right\rangle=\cos \kappa\left|\varphi_1^+\right\rangle+\sin \kappa\left|\varphi_2^+\right\rangle, \\
& \left|\alpha_2\right\rangle=\sin \kappa\left|\varphi_1^+\right\rangle-\cos \kappa\left|\varphi_2^+\right\rangle,
\end{aligned}
\label{eq:eve_measurement_binary}
\end{equation}
where $|\varphi_1^+\rangle$ and $|\varphi_2^+\rangle$ form an orthonormal basis, and $\kappa$ is the measurement parameter that Eve optimizes to maximize her information gain. The probability that Eve correctly identifies Alice's state, assuming equal probability of sending $|\varphi_0^+\rangle$ or $|\varphi_1^+\rangle$, is:

\begin{equation}
P_{\text{correct}} = \frac{1}{2}|\langle \alpha_1|\varphi_0^+\rangle|^2 + \frac{1}{2}|\langle \alpha_2|\varphi_1^+\rangle|^2.
\end{equation}

After Eve's measurement, the states transform to:
\begin{align}
|\alpha_1\rangle\langle\alpha_1|\varphi_0^+\rangle &= \frac{1}{\sqrt{2}}[(\cos\kappa + \sin\kappa)]|\alpha_1\rangle, \\
|\alpha_2\rangle\langle\alpha_2|\varphi_1^+\rangle &=  \sin\kappa|\alpha_2\rangle.
\end{align}

Supposing $|\alpha_1\rangle$ represents Eve's detector for $|\varphi_0^+\rangle$ and $|\alpha_2\rangle$ for $|\varphi_1^+\rangle$, the probability that Eve correctly identifies the state (assuming equal prior probabilities) is:

\begin{equation}
P_{\text{correct}} = \frac{1}{4}(\cos\kappa + \sin\kappa)^2 + \frac{1}{2}\sin^2\kappa.
\label{eq:P_correct_binary}
\end{equation}

Setting $dP_{\text{correct}}/d\kappa = 0$ gives $(\sin 2\kappa + \cos 2\kappa)/2 = 0$, hence $\tan 2\kappa = -1$ and $\kappa = 67.5^\circ$. Substituting:

\begin{equation}                   P_{\text{max}} = P_{\text{correct}}(\kappa = 67.5^\circ) = \frac{1+\tfrac{1}{\sqrt{2}}}{2} \approx 0.85.
\label{eq:binary_85bound}
\end{equation}

We designate this value the \textit{binary discrimination bound} and reference it as such throughout the remainder of this work. The full closed-form derivation and optimization over $\kappa$ are given in Appendix~\ref{app:kappa}.

\begin{figure}[ht]
\centering
\includegraphics[width=1\columnwidth]{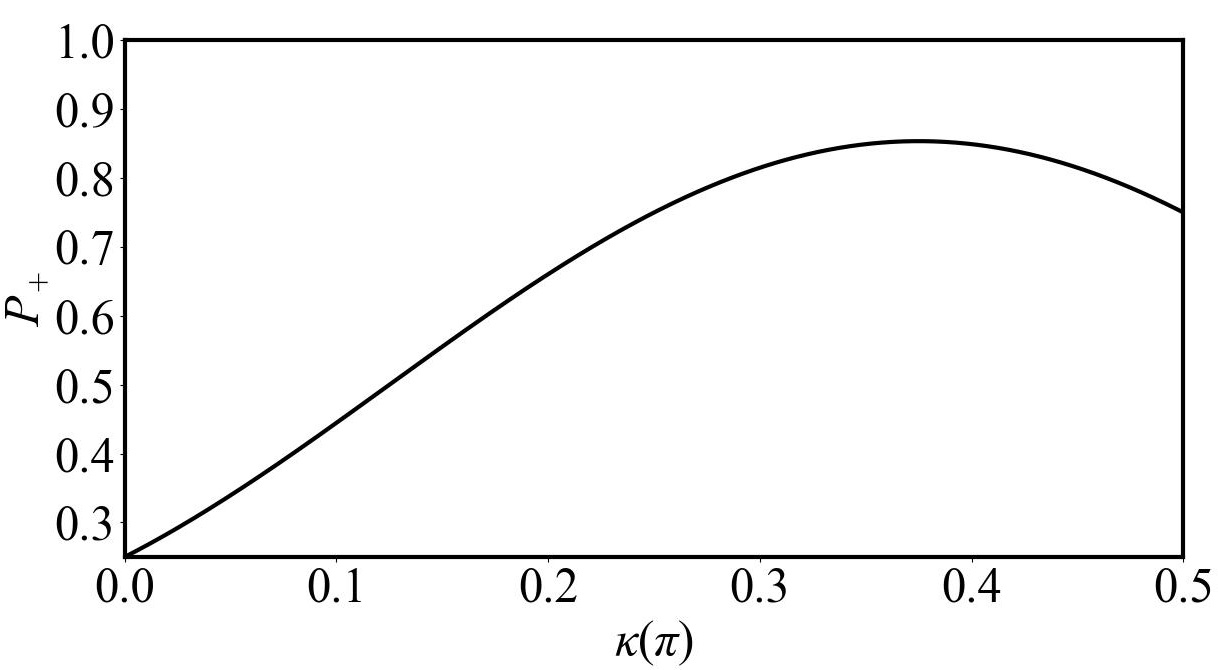} \caption{Probability of correct identification as a function of measurement angle $\kappa$. The maximum occurs at $\kappa = 67.5^\circ$, yielding $P_{\text{max}} = 0.85$.}
\label{fig:eve_probability}
\end{figure}

An eavesdropper employing optimal measurements correctly identifies Alice's state with probability 85\%, as shown in Figure~\ref{fig:eve_probability}. Furthermore, with this choice of measurement angle, Eve's two detectors click with equal probability:

\begin{equation}
    P(\alpha_i) = \frac{1}{2}|\langle \alpha_i|\varphi_0^+\rangle|^2 + \frac{1}{2}|\langle \alpha_i|\varphi_1^+\rangle|^2 = 0.5.
\end{equation}
  
She can therefore forward ``0'' and ``1'' with equal probabilities to Bob without introducing detectable statistical anomalies that would reveal her presence.

\subsubsection{Alice transporting four non-orthogonal states to Bob}
\label{subsubsec:four-state}

We now extend the analysis to the case where Alice randomly sends both $|D\rangle$ and $|A\rangle$ polarized photons to Bob. This introduces four non-orthogonal states in the transmission channel, potentially increasing the protocol's security. The four states are:
\begin{equation}
\begin{aligned}
|\varphi_0^+\rangle &= \frac{1}{\sqrt{2}}(|\varphi_1^+\rangle + |\varphi_2^+\rangle)\\ 
&= \cos\theta|U\rangle|D\rangle + \sin\theta|L\rangle|D\rangle, \\
|\varphi_1^+\rangle &= \cos\theta|U\rangle|V\rangle + \sin\theta|L\rangle|H\rangle, \\
|\varphi_0^-\rangle &= \frac{1}{\sqrt{2}}(|\varphi_1^+\rangle - |\varphi_2^+\rangle) \\
&= \cos\theta|U\rangle|A\rangle + \sin\theta|L\rangle|A\rangle, \\
|\varphi_1^-\rangle &=  \cos\theta|U\rangle|H\rangle - \sin\theta|L\rangle|V\rangle.
\end{aligned}
\end{equation}

Eve’s goal is to learn Alice’s bit ($0$ vs.\ $1$). While a complete optimization over all possible measurement strategies is complex, we can leverage the result from the two-state case to simplify the analysis. Since we found $\kappa_1 = 3\pi/8$ to be optimal for distinguishing between $|\phi_0^+\rangle$ and $|\phi_1^+\rangle$, we consider Eve's measurement operators:
\begin{equation}
    \begin{aligned}
|\alpha_1\rangle &= \cos\kappa_1|\varphi_1^+\rangle + \sin\kappa_1|\varphi_2^+\rangle, \\
|\alpha_2\rangle &= -\sin\kappa_1|\varphi_1^+\rangle + \cos\kappa_1|\varphi_2^+\rangle, \\
|\alpha_3\rangle &= \cos\kappa_2|\varphi_1^-\rangle + \sin\kappa_2|\varphi_2^-\rangle, \\
|\alpha_4\rangle &= \sin\kappa_2|\varphi_1^-\rangle - \cos\kappa_2|\varphi_2^-\rangle,
\end{aligned}
\end{equation}
where $\{|\varphi_1^-\rangle, |\varphi_2^-\rangle\}$ form an orthonormal basis for the $|A\rangle$ subspace, with:
\begin{equation}
|\varphi_i^-\rangle = \cos\theta|U\rangle|A\rangle \pm \sin\theta|L\rangle|A\rangle,
\end{equation}
where $i=1$ is associated with $(+)$ and $i=2$ is associated with $(-)$. With $\kappa_1 = 3\pi/8$ already determined, we only need to find the optimal $\kappa_2$. Now, Eve's measurement is:  $A = \sum_{j=1}^{4}|\alpha_j\rangle\langle\alpha_j|$. Projecting each input onto the $\left\{\left|\alpha_j\right\rangle\right\}$ basis gives the (non-normalized) post-measurement states:
\begin{equation}
    \begin{aligned}
        A\left|\varphi_0^+\right\rangle &=\frac{1}{\sqrt{2}}\left(\cos \kappa_1+\sin \kappa_1\right)\left|\alpha_1\right\rangle
        \\& +\frac{1}{\sqrt{2}}\left(\sin \kappa_1-\cos \kappa_1\right)\left|\alpha_2\right\rangle, \\
A\left|\varphi_1^+\right\rangle & =\cos \kappa_1\left|\alpha_1\right\rangle+\sin \kappa_1\left|\alpha_2\right\rangle, \\
A\left|\varphi_0^-\right\rangle &=\frac{1}{\sqrt{2}}\left(\cos \kappa_2-\sin \kappa_2\right)\left|\alpha_3\right\rangle\\ &+\frac{1}{\sqrt{2}}\left(\sin \kappa_2+\cos \kappa_2\right)\left|\alpha_4\right\rangle, \\
A\left|\varphi_1^-\right\rangle &=\cos \kappa_2\left|\alpha_3\right\rangle+\sin \kappa_2\left|\alpha_4\right\rangle.
    \end{aligned}
\end{equation}

Supposing $|\alpha_1\rangle$ represents the ``correct click'' of Eve's detector for $|\varphi_0^+\rangle$, $|\alpha_2\rangle$ for $|\varphi_1^+\rangle$, $|\alpha_3\rangle$ for $|\varphi_0^-\rangle$, and $|\alpha_4\rangle$ for $|\varphi_1^-\rangle$, the four single-shot correctness probabilities are:
\begin{equation}
\begin{aligned}
& \operatorname{Pr}\left(\operatorname{correct} \mid \varphi_0^+\right)=\frac{1}{2}\left(\cos \kappa_1+\sin \kappa_1\right)^2,\\
& \operatorname{Pr}\left(\operatorname{correct} \mid \varphi_1^+\right)=\sin ^2 \kappa_1, \\
&\operatorname{Pr}\left(\operatorname{correct} \mid \varphi_0^-\right)=\frac{1}{2}\left(\cos \kappa_2-\sin \kappa_2\right)^2, \\
& \operatorname{Pr}\left(\operatorname{correct} \mid \varphi_1^-\right)=\sin ^2 \kappa_2,
\end{aligned}
\end{equation}
where the factor 
$1/2$ in the first and third terms comes from the $1/\sqrt{2}$. Averaging over the four equiprobable inputs (each with prior 
$1/4$) gives the probability that Eve correctly identifies the state is:
\begin{equation}
\begin{aligned}P_{\text{correct}} &= \frac{1}{4}\left[\frac{1}{2}(\cos\kappa_1 + \sin\kappa_1)^2 \right. \\
    &\left. + \sin^2\kappa_1 + \frac{1}{2}(\cos\kappa_2 - \sin\kappa_2)^2 + \sin^2\kappa_2\right].
\end{aligned}
\end{equation}

Differentiating with respect to $\kappa_2$ gives $\partial P_{\text{correct}}/\partial \kappa_2 = (\sin 2\kappa_2-\cos 2\kappa_2)/4= 0$, yielding $\kappa_2 = -67.5^\circ$ (equal in magnitude to $\kappa_1$ but opposite in sign). Therefore:
\begin{equation}
P_{\text{MAX}} = \frac{1}{4}\left[2 \times \left(\frac{1 + \sqrt{2}}{2}\right) + 2 \times \frac{1}{2}\right] = \tfrac{1+ \frac{1}{\sqrt{2}}}{2}  \approx 0.85.
\label{eq:85-four}
\end{equation}

Extending to four channel states leaves the binary discrimination bound unchanged:

\begin{equation}
\begin{aligned}
& P_{\alpha_1}=\frac{1}{4}\left[\frac{1}{2}\left(\cos \kappa_1+\sin \kappa_1\right)^2+\cos ^2 \kappa_1\right]=0.25, \\
& P_{\alpha_2}=\frac{1}{4}\left[\frac{1}{2}\left(\cos \kappa_1-\sin \kappa_1\right)^2+\sin ^2 \kappa_1\right]=0.25, \\
& P_{\alpha_3}=\frac{1}{4}\left[\frac{1}{2}\left(\cos \kappa_2-\sin \kappa_2\right)^2+\cos ^2 \kappa_2\right]=0.25, \\
& P_{\alpha_4}=\frac{1}{4}\left[\frac{1}{2}\left(\cos \kappa_2+\sin \kappa_2\right)^2+\sin ^2 \kappa_2\right]=0.25.
\end{aligned}
\end{equation}
    
This means Eve can forward her measurement results to Bob with the correct statistical distribution, making her presence undetectable through simple statistical analysis. The addition of two more states does not improve the security of the binary quantum eraser protocol.

\subsubsection{Alice randomly selects her photon's polarization when she prepares the photon}
\label{subsubsec:random-pol}

A natural question arises: can Alice enhance security by randomizing the initial photon polarization? In this scenario, Alice randomly selects her photon's polarization angle $\phi_0$ when preparing it (for example, by placing an additional polarization rotator before BS1), potentially making it more difficult for Eve to extract information without knowing the initial polarization.

The photon states transported in the transmission channel become:
\begin{equation}
\begin{aligned}
    |\varphi_0^{+(\theta_0)}\rangle &= \cos\theta|U\rangle \tfrac{\sqrt{2}}{2}(|\phi_0 +90^\circ\rangle + |\phi_0 +0^\circ\rangle) \\
    &+ \sin\theta|L\rangle\tfrac{\sqrt{2}}{2}(|\phi_0 +90^\circ\rangle + |\phi_0 +0^\circ\rangle), \\
    |\varphi_1^{+(\theta_0)}\rangle &= \cos\theta|U\rangle|\phi_0 +90^\circ\rangle + \sin\theta|L\rangle|\phi_0+0^\circ\rangle,
\end{aligned}
\end{equation}
where $\phi_0$ is a random polarization angle selected by Alice, $|\phi_0+90^\circ\rangle$ means the polarization is rotated counterclockwise by $90^\circ$ from $\phi_0$, and $|\phi_0+0^\circ\rangle$ means the polarization remains at angle $\phi_0$.

Since Bob's measurement is independent of the photon polarization in the ideal case, this randomization does not affect the legitimate communication. However, to distinguish Alice's different photon states, Eve seemingly needs to know the original polarization.

Nevertheless, we now demonstrate that Eve can still extract information with high probability without knowing $\phi_0$. Eve employs a measurement device similar to Bob's but with rotation angle $\omega$. The polarization rotation operator is:

\begin{equation}
\begin{aligned}
     S_L^{(\omega)} &= \cos\omega(|\phi_0+0^\circ\rangle\langle\phi_0+0^\circ| + |\phi_0+90^\circ\rangle\langle\phi_0+90^\circ|) \\
     &+ \, \sin\omega(|\phi_0+90^\circ\rangle\langle\phi_0+0^\circ| - |\phi_0+0^\circ\rangle\langle\phi_0+90^\circ|).
\end{aligned}
\end{equation}

After Eve's polarization rotators (applying $S_L(\omega)$ to the upper path and $S_R(\omega)$ to the lower path), the states transform to:
\begin{equation}
\begin{aligned}
|\varphi_0^{+{(\theta_0)}}\rangle &\rightarrow \tfrac{1}{\sqrt{2}}\cos\theta|U\rangle[(\cos\omega - \sin\omega)|\phi_0+0^\circ\rangle\\
&\quad+ (\cos\omega + \sin\omega)|\phi_0+90^\circ\rangle] \\
&\quad + \tfrac{1}{\sqrt{2}}\sin\theta|L\rangle[(\cos\omega + \sin\omega)|\phi_0+0^\circ\rangle\\
&\quad+ (\cos\omega - \sin\omega)|\phi_0+90^\circ\rangle], \\
|\varphi_1^{+{(\theta_0)}}\rangle &\rightarrow \cos\theta|U\rangle(\cos\omega|\phi_0+90^\circ\rangle \\
& \quad- \sin\omega|\phi_0+0^\circ\rangle) \\
&\quad + \sin\theta|L\rangle(\cos\omega|\phi_0+0^\circ\rangle \\
& \quad+ \sin\omega|\phi_0+90^\circ\rangle).
\end{aligned}
\end{equation}

After Eve's second beam splitter (BS2):
\begin{equation}
\begin{aligned}
|\varphi_0^{+{(\theta_0)}}\rangle & \xrightarrow{B S_2} \frac{1}{\sqrt{2}}\left[\cos ^2 \theta(\cos \omega+\sin \omega)\right. \\
& \left. +\sin ^2 \theta(\cos \omega-\sin \omega)\right]\left|\phi_0+90\right\rangle|U\rangle \\
& +\frac{1}{\sqrt{2}}\left[\cos ^2 \theta(\cos \omega-\sin \omega)\right.\\
& \left. +\sin ^2 \theta(\cos \omega+\sin \omega)\right]\left|\phi_0+0\right\rangle|U\rangle  \\
& +\sqrt{2} \cos \theta \sin \theta \sin \omega\left|\phi_0+90\right\rangle|L\rangle\\
&-\sqrt{2} \cos \theta \sin \theta \sin \omega\left|\phi_0+0\right\rangle|L\rangle, \\
|\varphi_1^{+{(\theta_0)}}\rangle & \rightarrow\left(\cos ^2 \theta \cos \omega-\sin ^2 \theta \sin \omega\right)\left|\phi_0+90\right\rangle|U\rangle \\
& +\left(\sin ^2 \theta \cos \omega-\cos ^2 \theta \sin \omega\right)\left|\phi_0+0\right\rangle|U\rangle \\
& +\cos \theta \sin \theta(\cos \omega+\sin \omega)\left|\phi_0+90\right\rangle|L\rangle \\
& -\sin \theta \cos \theta(\cos \omega+\sin \omega)\left|\phi_0+0\right\rangle|L\rangle.
\end{aligned}
\end{equation}

If Eve associates detector $U$ clicking with $|\varphi_0^{+{(\theta_0)}}\rangle$ and detector $L$ clicking with $|\varphi_1^{+{(\theta_0)}}\rangle$, the probability that Eve correctly identifies the state is (setting $\theta = 45^\circ$):
\begin{equation}
P_{\text{correct}} = \frac{1}{2}\left[\cos^2\omega + \frac{1}{2}(\cos\omega + \sin\omega)^2\right].
\end{equation}

\begin{figure}[htp]
\centering
\includegraphics[width=1\columnwidth]{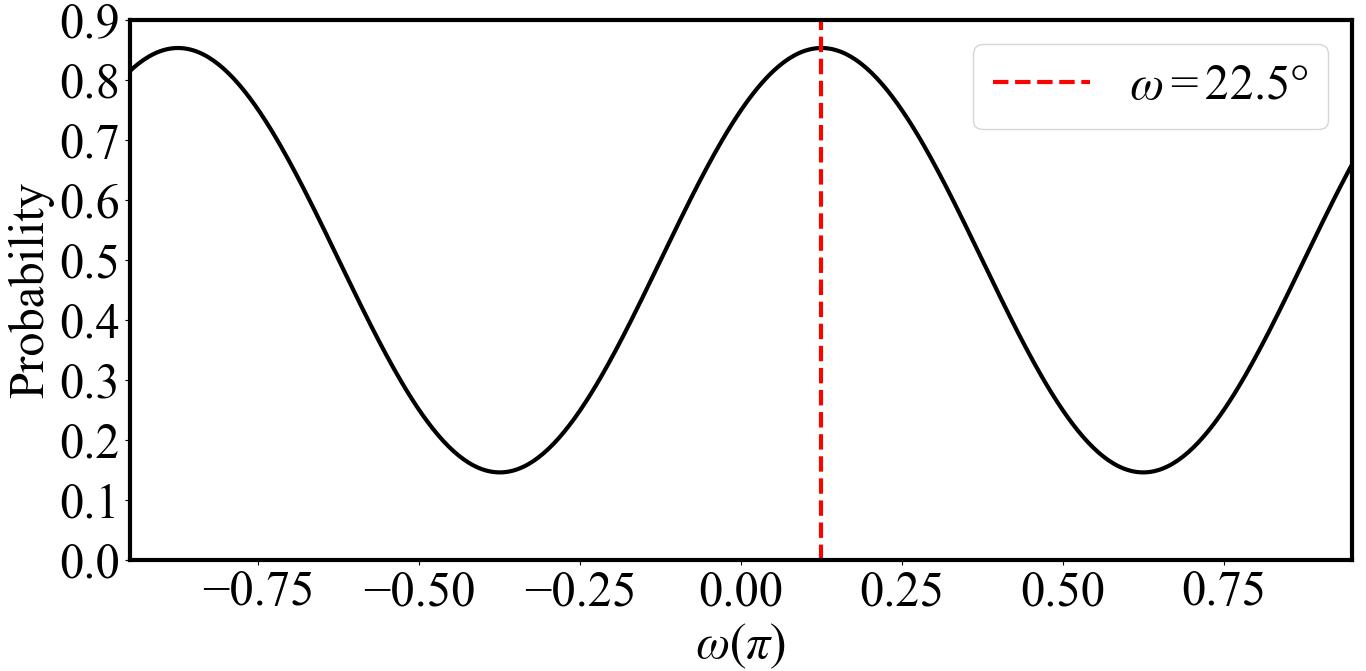} \caption{Eve's correct identification probability $P_{\text{correct}}$ as a function of measurement angle $\omega$
for randomized initial polarization. The maximum of $0.85$ occurs at $\omega = 22.5^\circ$.}
\label{fig:eve_random_probability}
\end{figure}

The coefficient $1/2$ is due to the fact Alice sends ``0'' and ``1'' equally. Differentiating and setting to zero gives $-\tfrac{1}{2}\sin 2\omega + \cos 2\omega = 0$, hence $\tan 2\omega = 2$ and $\omega = 22.5^\circ$ (Figure~\ref{fig:eve_random_probability}). With this choice:
\begin{equation}
\begin{aligned}
    P_{\text{MAX}} &= \frac{1}{2}\left[\cos^2(22.5^\circ) + \frac{1}{2}(\cos 22.5^\circ + \sin 22.5^\circ)^2\right]\\
    &\approx 0.85.
\end{aligned}
\end{equation}

One can verify that at $\omega = 22.5^\circ$ Eve's two detectors click with equal probabilities: $\cos^2\omega -\tfrac{1}{2}(\cos\omega + \sin\omega)^2\big|_{\omega=22.5^\circ}=0$. Thus polarization randomization still lets Eve reach the binary discrimination bound, and her balanced detector statistics remain undetectable.

\subsubsection{BB84 case}
\label{subsubsec:bb84}

For comparison, we analyze the security of the BB84 protocol, which uses four states in two mutually unbiased bases. The four states are:
\begin{equation}
\begin{aligned}
|\phi_1\rangle &= |H\rangle \quad \text{encoding “0”} ,\\
|\phi_2\rangle &= |V\rangle \quad \text{encoding “1”} ,\\
|\phi_3\rangle &= \frac{1}{\sqrt{2}}(|H\rangle + |V\rangle) \quad \text{encoding “0”} ,\\
|\phi_4\rangle &= \frac{1}{\sqrt{2}}(|H\rangle - |V\rangle) \quad \text{encoding “1”}.
\end{aligned}
\end{equation}

Eve cannot distinguish all four states with only two detectors, as $|\phi_1\rangle$ and $|\phi_2\rangle$ are orthogonal, as are $|\phi_3\rangle$ and $|\phi_4\rangle$, but states from different bases are non-orthogonal. We analyze Eve's capability to extract bit values (“0” or “1”) rather than exact state identification.

Suppose Eve's measurement operators are:
\begin{equation}
    \begin{aligned}
        |\alpha_1'\rangle &= \cos\kappa'|H\rangle + \sin\kappa'|V\rangle, \\
|\alpha_2'\rangle &= \sin\kappa'|H\rangle - \cos\kappa'|V\rangle.
    \end{aligned}
\end{equation}

Eve's measurement can be written as:
\begin{equation}
\begin{aligned}
    A' &= |\alpha_1'\rangle(\cos\kappa'\langle H| + \sin\kappa'\langle V|) \\
    & + |\alpha_2'\rangle(\sin\kappa'\langle H| - \cos\kappa'\langle V|).
\end{aligned}
\end{equation}

After Eve's measurement, the states transform to:
\begin{equation}
    \begin{aligned}
        A'|\phi_1\rangle &= \cos\kappa'|\alpha_1'\rangle + \sin\kappa'|\alpha_2'\rangle, \\
A'|\phi_2\rangle &= \sin\kappa'|\alpha_1'\rangle - \cos\kappa'|\alpha_2'\rangle, \\
A'|\phi_3\rangle &= \frac{1}{\sqrt{2}}[(\cos\kappa' + \sin\kappa')|\alpha_1'\rangle + (\sin\kappa' - \cos\kappa')|\alpha_2'\rangle], \\
A'|\phi_4\rangle &= \frac{1}{\sqrt{2}}[(\cos\kappa' - \sin\kappa')|\alpha_1'\rangle + (\sin\kappa' + \cos\kappa')|\alpha_2'\rangle].
    \end{aligned}
\end{equation}

Suppose $|\alpha_1'\rangle$ represents detection of “0” and $|\alpha_2'\rangle$ represents detection of “1”. We can verify that Eve's detectors click with equal probability:
\begin{equation}
\begin{aligned}
P_{\alpha_1'} &= \tfrac{1}{4}\bigl[\cos^2\kappa' + \sin^2\kappa' \\
    &+ \tfrac{1}{2}(\cos\kappa' + \sin\kappa')^2 + \tfrac{1}{2}(\cos\kappa' - \sin\kappa')^2\bigr] \\
    &= 0.5,\\
P_{\alpha_2'} &= \tfrac{1}{4}\bigl[\cos^2\kappa' + \sin^2\kappa' \\
    &+ \tfrac{1}{2}(\cos\kappa' - \sin\kappa')^2 + \tfrac{1}{2}(\cos\kappa' + \sin\kappa')^2\bigr] \\
    &= 0.5.
\end{aligned}
\end{equation}

The probability that Eve correctly identifies the bit value is:
\begin{equation}
P_+' = \frac{1}{2}(\cos\kappa' + \sin\kappa')\cos\kappa' + \frac{1}{4}.
\end{equation}

Differentiating gives $dP_+'/d\kappa' = \tfrac{1}{2}\cos 2\kappa' = 0$, yielding $\kappa' = 22.5^\circ$ and:
\begin{equation}                   
    P_{\text{MAX}}' = \tfrac{1}{2}(\cos 22.5^\circ + \sin 22.5^\circ)\cos 22.5^\circ + \tfrac{1}{4} \approx 0.85.   
\end{equation}

Thus, Eve can determine the bit value with a probability that's identical to the quantum eraser protocol, consistent with the individual-attack bounds derived in~\cite{FuchsGisin1997}. However, a crucial difference emerges when considering Eve's ability to reproduce the exact quantum state. Since Eve cannot distinguish between $|\phi_1\rangle$ and $|\phi_3\rangle$ (both encoding “0”), or between $|\phi_2\rangle$ and $|\phi_4\rangle$ (both encoding “1”), her maximum success rate for exact state reproduction is only 42\%. Furthermore, in BB84, even random guessing yields 50\% correct bits due to the binary nature of the key.

\subsection{Summary of binary protocol security}
\label{subsec:security-summary}

The binary discrimination bound, $(1+1/\sqrt{2})/2 \approx 85\%$, appears in all three binary-eraser variants. For the two-state case it is Helstrom's optimal identification probability at overlap $1/\sqrt{2}$; the four-state and randomized-polarization variants yield the same value. Reducing this bound requires a larger encoding alphabet, motivating the ternary extension developed in Section~\ref{sec:ternary}.
The BB84 comparison clarifies the geometric origin. BB84 permits the same $85\%$ bit-value extraction because its signal-state pairs also have overlap $1/\sqrt{2}$; the match is a shared overlap, not a consequence of non-orthogonality in general. BB84's use of mutually unbiased bases further restricts exact state reproduction to $42\%$, whereas the eraser offers no analogous restriction beyond the bit-value bound.
The quantum eraser framework compensates through operational efficiency: interference-based sifting eliminates basis reconciliation, converting every mismatched detection into a potential key bit without the classical overhead of basis comparison. This tension between security vulnerability and operational advantage motivates the ternary extension developed in Section~\ref{sec:ternary}.

\section{General analysis for the efficiency and security of transporting two non-orthogonal states}
\label{sec:efficiency-binary}

This section parameterizes the security and efficiency of a general two-state encoding by the overlap, placing the binary quantum eraser at a specific point on the resulting trade-off curve.

\subsection{Optimal Measurement Strategy}

Consider two arbitrary states in a two-dimensional Hilbert space spanned by orthogonal states $|X\rangle$ and $|Y\rangle$:
\begin{equation}
    \begin{aligned}
        |\psi_1\rangle &= |X\rangle, \\
|\psi_2\rangle &= \cos\gamma_A|X\rangle + \sin\gamma_A|Y\rangle,
    \end{aligned}
\end{equation}
where $\gamma_A$ parameterizes the overlap between the states. An eavesdropper employs a projective measurement $M = |F_1\rangle\langle F_1| + |F_2\rangle\langle F_2|$ with:
\begin{equation}                   \begin{aligned}
    |F_1\rangle &= \cos\alpha|X\rangle + \sin\alpha|Y\rangle, \\                   |F_2\rangle &= \cos\alpha|Y\rangle - \sin\alpha|X\rangle,
\end{aligned}                      \end{equation}
where the angle $\alpha$ is optimized to maximize the probability of correct identification.  

Applying the measurement to the two states yields:
\begin{equation}
\begin{aligned}
M|X\rangle &= \cos\alpha|F_1\rangle - \sin\alpha|F_2\rangle,\\
M(\cos\gamma_A|X\rangle + \sin\gamma_A|Y\rangle) &= (\cos\alpha\cos\gamma_A + \sin\alpha\sin\gamma_A)|F_1\rangle \\
&+ (\cos\alpha\sin\gamma_A - \sin\alpha\cos\gamma_A)|F_2\rangle.
\end{aligned}
\end{equation}

If $|F_1\rangle$ represents detection of $|X\rangle$ and $|F_2\rangle$ represents detection of $\cos\gamma_A|X\rangle + \sin\gamma_A|Y\rangle$, the probability of correct identification is:
\begin{equation}
B = \frac{1}{2}[\sin^2(\alpha + \gamma_A) + \cos^2\alpha].
\end{equation}

\subsection{Optimization of Measurement Angle}

To find the optimal measurement angle, we differentiate with respect to $\alpha$:
\begin{equation}
\begin{aligned}
    \frac{dB}{d\alpha} &= \sin(\alpha + \gamma_A)\cos(\alpha + \gamma_A) - \cos\alpha\sin\alpha \\
    &= \tfrac{1}{2}[\sin 2(\alpha + \gamma_A) - \sin 2\alpha].
\end{aligned}
\end{equation}

Setting the derivative to zero yields several cases:

\textbf{Case 1:} If $\gamma_A = k\pi$ (orthogonal states), then $\alpha$ is arbitrary, and $B = 1$ (perfect discrimination).

\textbf{Case 2:} If $\gamma_A = k\pi + \pi/2$, then $\alpha = k''\pi/2$.

\textbf{Case 3:} For general $\gamma_A \neq k\pi/2$, we have,
$\tan 2\alpha = \tan\left(\frac{\pi}{2} - \gamma_A\right).$

\begin{figure}[H]
\centering
\includegraphics[width=1\columnwidth]{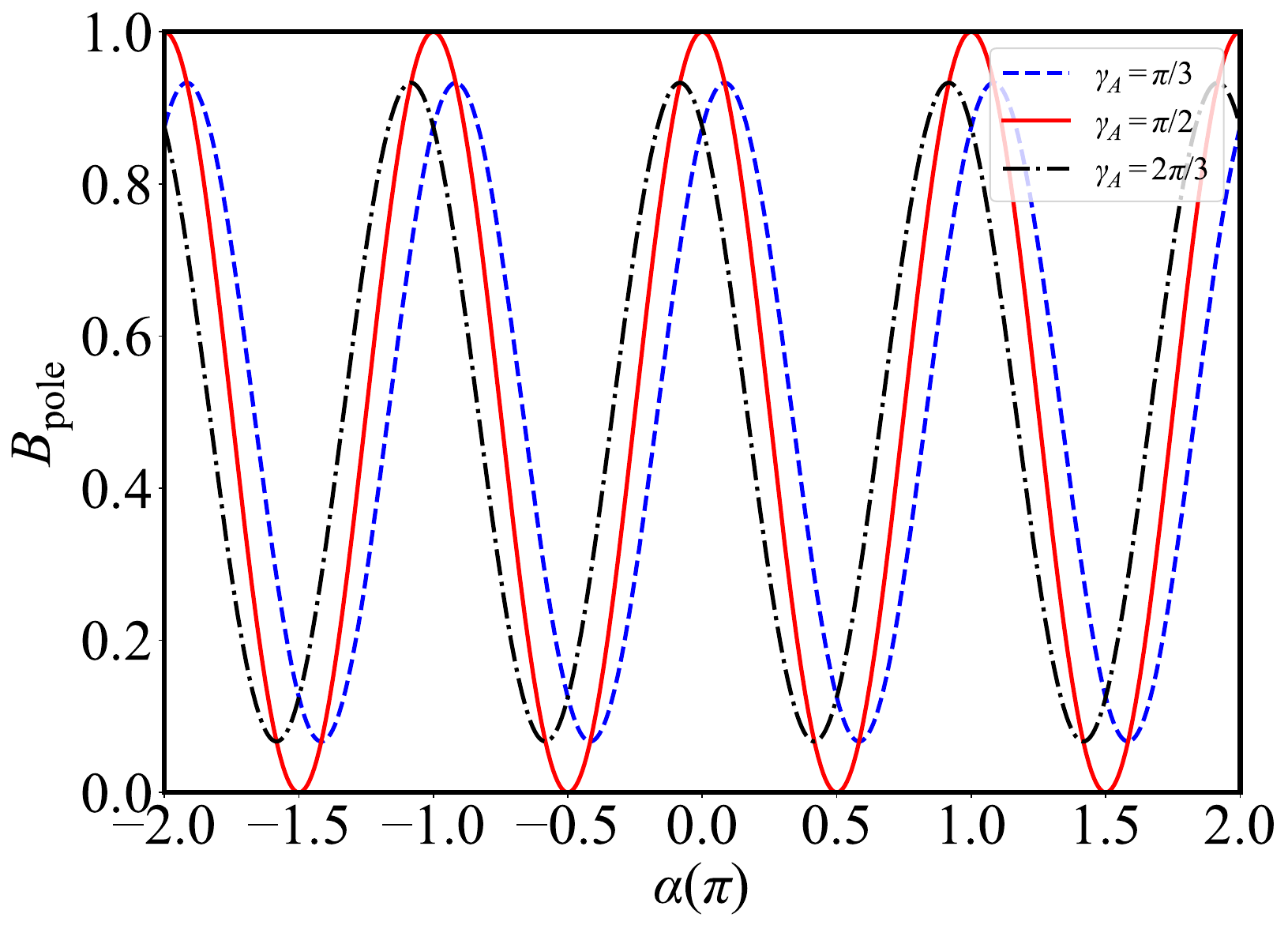}
\caption{Probability of correct identification $B_{\text{pole}}$ as a function of the measurement angle for different fixed values of $\gamma_A$: $\gamma_A = \pi/3$ (dashed blue), $\gamma_A = \pi/2$ (solid red), and $\gamma_A = 2\pi/3$ (dashed-dotted black).}
\label{fig:bpole}
\end{figure}

This gives $\alpha = \frac{k\pi}{2} + \frac{\pi}{4} - \frac{\gamma_A}{2}.$ The extremal values are simulated in Figure~\ref{fig:bpole}, and are given by:
\begin{equation}
B_{\text{pole}} = \tfrac{1}{2}\Bigl[\sin^2\tfrac{k\pi + \frac{\pi}{2} + \gamma_A}{2} + \cos^2\tfrac{k\pi + \frac{\pi}{2} - \gamma_A}{2}\Bigr].
\end{equation}

\subsection{Security-Efficiency Trade-off}

For the quantum eraser protocol, the two states in the transmission channel are:
\begin{align}
|X\rangle &= \cos\theta|U\rangle|D\rangle + \sin\theta|L\rangle|D\rangle, \\
|\psi_2\rangle &= \cos\gamma_A|X\rangle + \sin\gamma_A|Y\rangle,
\end{align}
where $|Y\rangle = -\cos\theta|U\rangle|A\rangle + \sin\theta|L\rangle|A\rangle$ is orthogonal to $|X\rangle$ in the polarization degree of freedom.

The efficiency depends on Bob's measurement outcomes, we have two situations. When Alice and Bob encode the same message, Bob's operation cancels Alice's transformation (since $T_B T_A = \mathbb{I}$), and the state at his detector is simply $|U\rangle|D\rangle$, independent of $\theta$. All photons are directed to detector $D_1$, producing no key bits. When they encode different messages, the state after the second beam splitter becomes:
\begin{equation}
\cos\gamma_A|U\rangle|D\rangle + \sin\gamma_A\cos 2\theta|U\rangle|A\rangle + \sin\gamma_A\sin 2\theta|L\rangle|A\rangle.
\end{equation}

The probability that detector $D_1$ clicks (corresponding to the $|U\rangle$ path) is obtained by summing the squared amplitudes of all $|U\rangle$ components. Since $|D\rangle$ and $|A\rangle$ are orthogonal polarization states, we have:
\begin{equation}
\begin{split}
    P(D_1|\text{mismatch}) &= |\cos\gamma_A|^2 + |\sin\gamma_A\cos 2\theta|^2\\
    &= \cos^2\gamma_A + \sin^2\gamma_A\cos^2 2\theta.
\end{split}
\end{equation}

Key bits are generated only when $D_2$ clicks during mismatched encoding. Since the two encoding cases (matched and mismatched) occur with equal probability, the protocol efficiency is:
\begin{equation}
\begin{split}
    E_{ff} &= \tfrac{1}{2} \times 0 + \tfrac{1}{2} \times P(D_2|\text{mismatch})\\
    &= \tfrac{1}{2}\left[1 - P(D_1|\text{mismatch})\right].
\end{split}
\end{equation}

For the standard quantum eraser configuration with $\theta = \pi/4$, we have $\cos 2\theta = 0$, which simplifies the efficiency to:
\begin{equation}
E_{ff} = \tfrac{1}{2}(1 - \cos^2\gamma_A) = \tfrac{1}{2}\sin^2\gamma_A.
\label{eq:efficiency}
\end{equation}

Alternatively, considering the full parameter space and both parties' rotation angles, the general expression is:
\begin{equation}
E_{ff} = 1 - \left(\cos\gamma_A\cos^2\theta + \cos\gamma_B\sin^2\theta\right)^2,
\end{equation}
which reduces to $E_{ff} = 1 - \cos^2\gamma_A$ when $\gamma_A = \gamma_B$ and $\theta = \pi/4$.

This reveals a fundamental trade-off. Larger $\gamma_A$ increases state distinguishability, improving efficiency. The increasing distinguishability, however, also increases Eve's success probability $B_{\text{pole}}$, reducing security.

\begin{figure}[H]
\centering
\includegraphics[width=1\columnwidth]{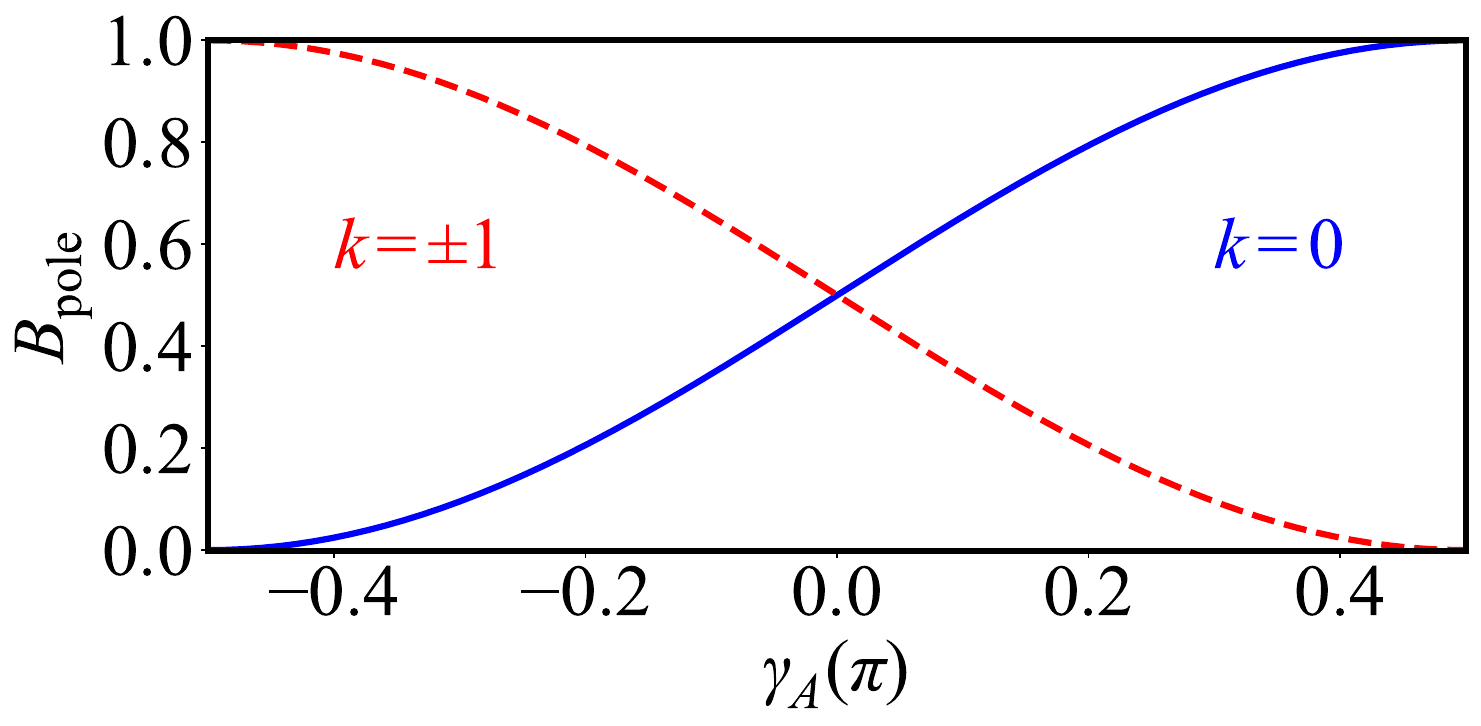}
\caption{Probability of correct identification $B_{\text{pole}}$ as a function of $\gamma_A(\pi)$ for $k=0$ (blue) and $k=\pm 1$ (dashed red). For any $\gamma_A$, Eve can select whichever strategy yields higher success; the upper envelope of the two curves therefore never falls below $1/2$, establishing a geometric lower bound on her discriminability.}
\label{fig:bpole_k}
\end{figure}

\subsection{Implications for Protocol Design}

Regardless of $\gamma_A$, Eve achieves at least 50\% discrimination success (Figure~\ref{fig:bpole_k}), a geometric constraint inherent to two-state systems.
\begin{figure}[H]
\centering
\includegraphics[width=1\columnwidth]{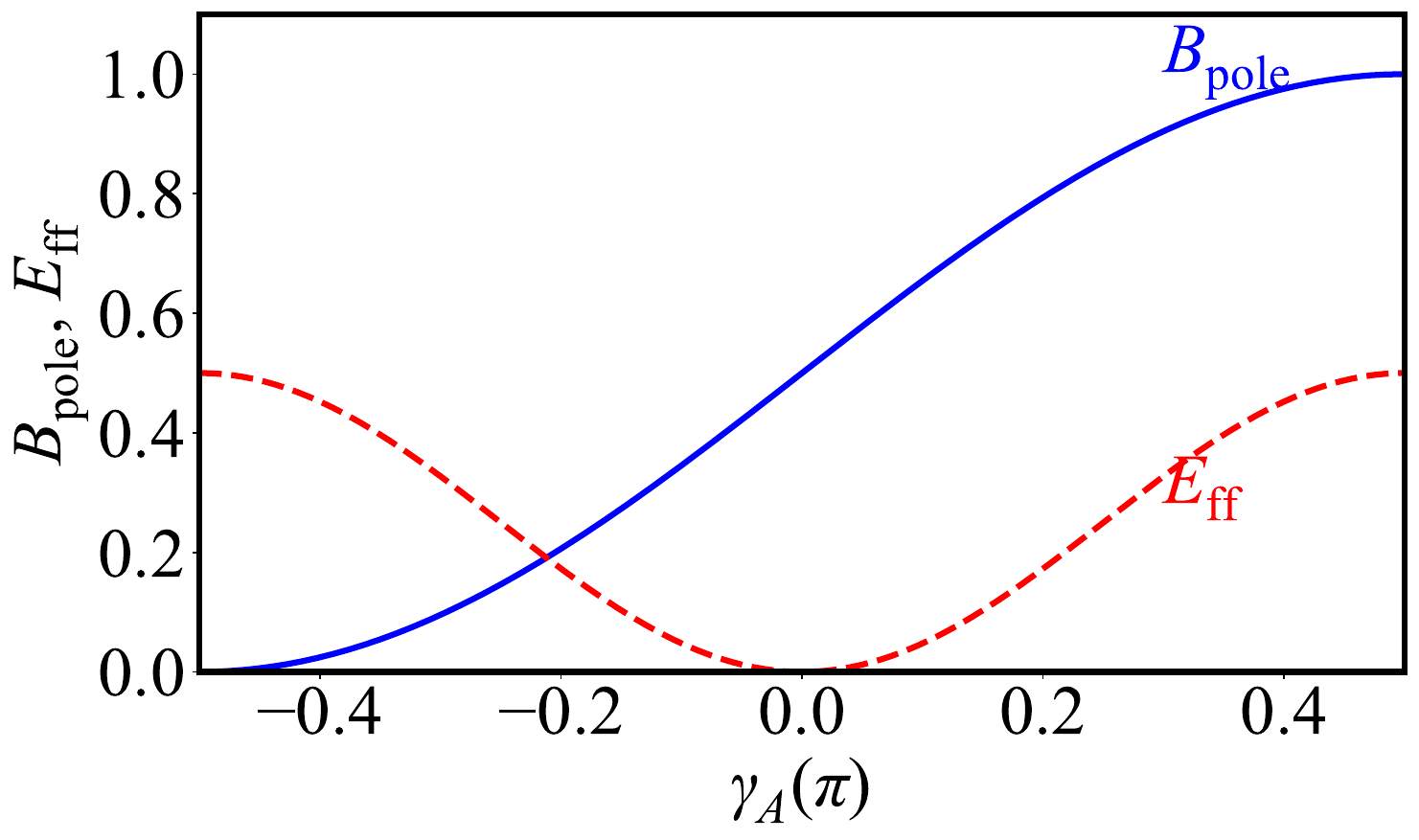}
\caption{Security parameter $B_{\text{pole}}$ (blue, shown for $k=0$) and efficiency $E_{ff}$ (dashed red) as functions of $\gamma_A \in [-\pi/2, \pi/2]$. For $\gamma_A \geq 0$, increasing efficiency is accompanied by higher Eve success probability, illustrating the security--efficiency trade-off.}
\label{fig:combined}
\end{figure}

Figure~\ref{fig:combined} plots $B_{\text{pole}}(\gamma_A)$ and $E_{ff}(\gamma_A)$ for the binary eraser, with $\gamma_A$ the angle between the two transmitted states (overlap $\cos\gamma_A$). The standard configuration sits at $\gamma_A = \pi/4$, where $B_{\text{pole}}$ reaches the binary discrimination bound ($0.85$) and $E_{ff} = 0.25$. $B_{\text{pole}}$ follows from the overlap of the two transmitted states (a general two-state discrimination result); $E_{ff}$ reflects the eraser's sifting mechanism, in which matched encodings yield deterministic $D_1$ clicks (no key bit) and mismatched encodings yield $D_2$ clicks with probability depending on $\gamma_A$.

Achieving substantially enhanced security within the binary quantum eraser framework is therefore constrained by this bound. Overcoming it requires extending the encoding alphabet beyond two states, which motivates the ternary protocol developed in the following section.

\section{Ternary Quantum Eraser Protocol}
\label{sec:ternary}

The preceding analysis motivates expanding the encoding alphabet beyond two states. The ternary protocol employs three polarization states with $120^\circ$
angular separation, transmitted in groups with randomized ordering. Security enhancement arises from two synergistic mechanisms: reduced quantum distinguishability among symmetrically-arranged states and combinatorial complexity from unknown photon ordering, thereby mitigating, without eliminating, the structural vulnerability of binary quantum eraser protocols.

\subsection{Three-State Polarization System}

Consider three polarization states created by passing horizontally polarized photons through different polarization rotators:

\begin{equation}
    \begin{aligned}
A_1 &: |H\rangle \quad {\color{red}\longrightarrow} \quad \text{(Horizontal polarization)},\\
A_2 &: \cos\left(\frac{2\pi}{3}\right)|H\rangle + \sin\left(\frac{2\pi}{3}\right)|V\rangle \\
&= -\frac{1}{2}|H\rangle + \frac{\sqrt{3}}{2}|V\rangle \quad {\color{red}\longrightarrow} \quad \text{($+120^\circ$ rotation)}, \\
A_3 &: \cos\left(-\frac{2\pi}{3}\right)|H\rangle + \sin\left(-\frac{2\pi}{3}\right)|V\rangle \\
&= -\frac{1}{2}|H\rangle - \frac{\sqrt{3}}{2}|V\rangle \quad {\color{red}\longrightarrow} \quad \text{($-120^\circ$ rotation)}.
\end{aligned}
\end{equation}

\begin{figure}[ht]
\centering
\includegraphics[width=0.65\columnwidth]{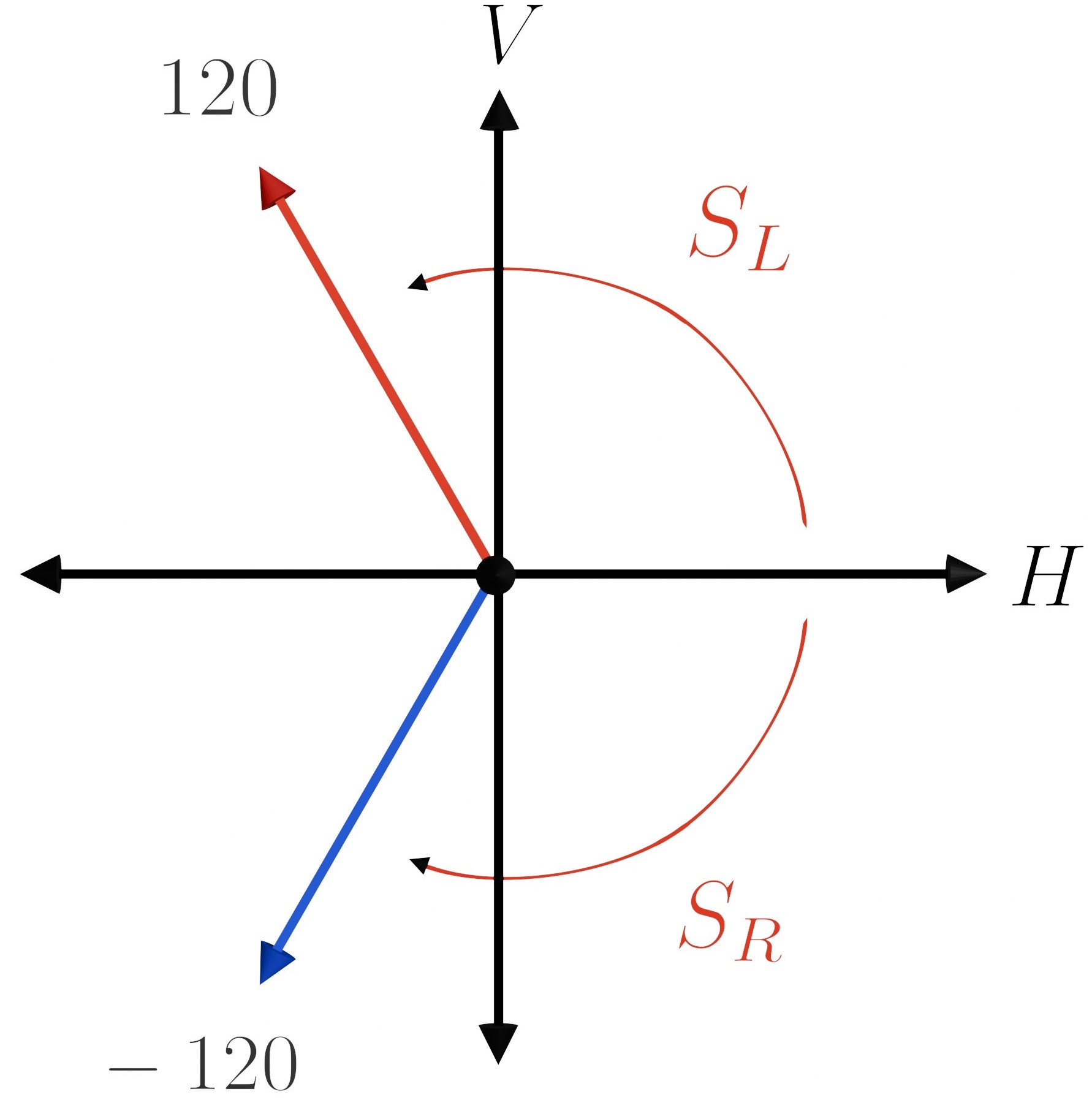}
\caption{Polarization state geometric representation for the ternary protocol. The three states are separated by $120^\circ$ angles, with $S_L$ and $S_R$ representing the rotation operators for $\pm120^\circ$ transformations, exhibiting three-fold rotational symmetry.}
\label{fig:ternary_polarization}
\end{figure}

These states correspond to polarization angles of $0^\circ$, $120^\circ$, and $-120^\circ$ respectively, maintaining $120^\circ$ angular separation between adjacent states, shown geometrically in Figure~\ref{fig:ternary_polarization}. 

The three states will appear in the transmission channel and will be attacked by Eve. The symmetric arrangement ensures that the overlap between any two states is identical:
\begin{equation}
\langle A_i | A_j \rangle = \begin{cases}
1 & \text{if } i = j,\\
-\frac{1}{2} & \text{if } i \neq j.
\end{cases}
\label{eq:trine_overlap_main}
\end{equation}

This uniform overlap contrasts with the binary protocol where states are either orthogonal or have varying overlaps, providing a more balanced security foundation.

\subsection{Protocol Operation Without Interference}

We first analyze the protocol operation considering only the polarization degree of freedom, temporarily setting aside the interference effects inherent in the Mach-Zehnder interferometer. This simplified analysis provides insight into the security improvements offered by the ternary encoding.

Alice encodes information by selecting one of three polarization states for each transmitted photon. Bob independently chooses one of three measurement configurations, implemented through polarization rotators:
\begin{equation}
\begin{aligned}
B_1 &: \text{Identity operation (no rotation)}, \\
B_2 &: S_L \text{ operator (rotate anticlockwise by }+120^\circ), \\
B_3 &: S_R \text{ operator (rotate clockwise by }-120^\circ).
\end{aligned}
\end{equation}

The rotation operators are defined as:
\begin{equation}
\begin{aligned}
 S_L(\phi)&=\cos \phi(|H\rangle\langle H|+|V\rangle\langle V|)\\
 &+ \,\sin \phi(|V\rangle\langle H|-|H\rangle\langle V|) ,\\
 S_R(\phi)&=\cos \phi(|H\rangle\langle H|+|V\rangle\langle V|)\\
 &+ \,\sin \phi(|H\rangle\langle V|-|V\rangle\langle H|),
\end{aligned}
\end{equation}
where we set the angle $\phi = \pm 120^\circ$.

\subsection{Detection Statistics and Key Generation}

The interaction between Alice's encoding and Bob's measurement creates nine possible configurations. When Bob measures in the $\{|H\rangle, |V\rangle\}$ basis after applying his chosen rotation, the detection probabilities reveal the correlation between their choices. Table~\ref{tab:detection_outcomes} shows the result for each combination.

\begin{table*}[t]
\centering
\caption{Detection outcomes for different state combinations.}
\begin{tabular}{c|ccc}
\hline
States & $B_1$ & $B_2$ & $B_3$ \\
\hline
$A_1$ & $|H\rangle$ & $\cos(-\frac{2\pi}{3})|H\rangle + \sin(-\frac{2\pi}{3})|V\rangle$ & $\cos(\frac{2\pi}{3})|H\rangle + \sin(\frac{2\pi}{3})|V\rangle$ \\
$A_2$ & $\cos(\frac{2\pi}{3})|H\rangle + \sin(\frac{2\pi}{3})|V\rangle$ & $|H\rangle$ & $\cos(-\frac{2\pi}{3})|H\rangle + \sin(-\frac{2\pi}{3})|V\rangle$ \\
$A_3$ & $\cos(-\frac{2\pi}{3})|H\rangle + \sin(-\frac{2\pi}{3})|V\rangle$ & $\cos(\frac{2\pi}{3})|H\rangle + \sin(\frac{2\pi}{3})|V\rangle$ & $|H\rangle$ \\
\hline
\end{tabular}
\label{tab:detection_outcomes}
\end{table*}

For matched configurations where Alice and Bob select corresponding operations ($(A_1, B_1)$, $(A_2, B_2)$, or $(A_3, B_3)$), the photon is always detected in the $|H\rangle$ state: $P(H|\text{matched}) = 1$, $P(V|\text{matched}) = 0$.

This deterministic outcome occurs because Bob's rotation precisely compensates for Alice's encoding when they choose corresponding operations. For instance, when Alice encodes with $A_2$ (rotating by $+120^\circ$) and Bob applies $B_2$ (rotating by $-120^\circ$), the net transformation is the identity, preserving the initial horizontal polarization.

For mismatched configurations, the detection probabilities become:
\begin{equation}
\begin{aligned}
   &P(H|\text{mismatched}) = \cos^2\left(\frac{2\pi}{3}\right) = \frac{1}{4},\\
    &P(V|\text{mismatched}) = \sin^2\left(\frac{2\pi}{3}\right) = \frac{3}{4}.
\end{aligned}
\end{equation}

The asymmetric detection probabilities for the mismatched cases enable key generation without basis reconciliation. A $|V\rangle$ detection can only arise from a mismatched configuration (matched configurations yield $|H\rangle$ deterministically), so every $|V\rangle$ event guarantees that Alice and Bob selected different operations. By contrast, an $|H\rangle$ detection is ambiguous: it occurs with certainty for matched configurations but also with probability $1/4$ for mismatched ones. The sifting criterion therefore retains only unambiguous events. Since each configuration pair has probability $1/9$ and six of nine pairs are mismatched, the probability of detecting $|V\rangle$ when Alice and Bob make random independent choices is $P(V) = \frac{6}{9} \times \frac{3}{4} = \frac{1}{2}$.

Alice and Bob establish their key by assigning values only to detection events where Bob measures $|V\rangle$. The key extraction rule maps the six mismatched configuration pairs to binary values, with Alice's choice determining the key bit according to a predetermined convention. This approach automatically filters out matched configurations without requiring classical communication about basis choices.

However, a single $|V\rangle$ detection event does not uniquely determine which mismatched configuration occurred; multiple Alice-Bob combinations produce indistinguishable detection outcomes. To establish unambiguous shared keys, Alice must transmit multiple photons as a coordinated group, enabling Bob's detection pattern to reveal sufficient information for key extraction. The design of these photon groups presents a trade-off between protocol complexity, efficiency, and security.

We now present two methods for implementing the ternary key distribution protocol. Method I employs pairs of photons, each prepared in two of the three available polarization states, providing a straightforward extension of binary concepts. Method II transmits all three polarization states in groups of three photons with randomized ordering, achieving superior efficiency and security through combinatorial complexity. While Method I serves to illustrate the principles, Method II represents our primary practical contribution and forms the basis for the security analysis in Section~\ref{sec:security-ternary}.

\subsection{Efficiency}
\label{subsec:efficiency-metrics}

Comparing the efficiency of ternary and binary quantum key distribution protocols requires care, since a ternary symbol carries more information than a binary one. We therefore introduce two complementary efficiency metrics that facilitate meaningful comparison.

The raw efficiency quantifies the probability that a transmitted photon contributes to key generation:
\begin{equation}
\eta_{\text{raw}} = \frac{P_{\text{sift}}}{m},
\label{eq:raw_efficiency}
\end{equation}
where $P_{\text{sift}}$ is the probability that a group of $m$ photons passes the sifting criteria and yields a valid key symbol. This metric captures the photon utilization rate but does not account for the information content of each symbol.

The binary-equivalent efficiency expresses the key generation rate in bits per photon, enabling direct comparison with binary protocols:
\begin{equation}
\eta_{\text{bin}} = \frac{\log_2 3}{m} P_{\text{sift}}.
\label{eq:bin_efficiency}
\end{equation}

The factor $\log_2 3 \approx 1.585$ converts ternary symbols to their binary information content. This metric represents the effective number of secret key bits generated per transmitted photon, placing ternary and binary protocols on equal footing.

For reference, the ideal BB84 protocol achieves $\eta_{\text{bin}} = 0.5$ bits per photon (since half the photons are discarded during basis reconciliation). The binary quantum eraser at the standard configuration ($\gamma_A = \theta = \pi/4$) yields $\eta_{\text{bin}} = E_{ff} = 0.25$ bits per photon (Sec.~\ref{sec:efficiency-binary}), half the BB84 rate due to automatic sifting on $D_2$ clicks.

\subsection{Key distribution Method I}

Method I employs photon pairs as the fundamental signal unit, with each pair containing two of the three available polarization states. This approach provides a natural first step from binary to ternary encoding while maintaining conceptual simplicity.

In this method, Alice sends pairs of photons to Bob as one signal unit. For each pair, Alice randomly selects two different polarization states from the three available options $(A_1,A_2,A_3)$. This pairing strategy ensures that the two photons within each signal unit always carry distinguishable polarization information. Alice's encoding scheme for the ternary signals uses six possible photon pairs:

\begin{equation}
\text { Alice }\left(S_L\right):\left\{\begin{array}{cc}
\text { signal } & \text { polarization } \\
0: & (0,120), \\
0^{\prime}: & (120,0), \\
1: & (-120,0), \\
1^{\prime}: & (0,-120), \\
2: & (-120,120), \\
2^{\prime}: & (120,-120),
\end{array}\right.
\end{equation}
where the pairs $(x, y)$ represent the polarization angles of the first and second photons.

Bob applies one of three operations uniformly to both received photons:

\begin{equation}
\text{Bob}\left(S_L\right):\left\{\begin{array}{cl}
\text{signal} & \text{polarization} \\
0: & \text{rotate by } \phantom{-}120^\circ, \\
1: & \text{rotate by } -120^\circ, \\
2: & \text{no rotation}. \\
\end{array}\right.
\end{equation}

After Bob's rotation and measurement in the $\{|H\rangle, |V\rangle\}$ basis, the detection outcomes for all nine combinations are shown in Table \ref{tab:method1_detection}.

\begin{table*}[t]
\centering
\caption{Detection outcomes for Method I. Each cell shows the resulting polarization states of the two photons after Bob's operation. The primed and unprimed states denote the first and second photons respectively.}
\label{tab:method1_detection}
\small
\begin{tabular}{c|ccc}
\hline
Alice $\backslash$ Bob & 0 & 1 & 2 \\
\hline
0 & $\begin{pmatrix} -\frac{1}{2}|H\rangle + \frac{\sqrt{3}}{2}|V\rangle \\ -\frac{1}{2}|H'\rangle - \frac{\sqrt{3}}{2}|V'\rangle \end{pmatrix}$ & 
$\begin{pmatrix} -\frac{1}{2}|H\rangle - \frac{\sqrt{3}}{2}|V\rangle \\ |H'\rangle \end{pmatrix}$ & 
$\begin{pmatrix} |H\rangle \\ -\frac{1}{2}|H'\rangle + \frac{\sqrt{3}}{2}|V'\rangle \end{pmatrix}$ \\
$0^{\prime}$ & $\begin{pmatrix} -\frac{1}{2}|H\rangle - \frac{\sqrt{3}}{2}|V\rangle \\ -\frac{1}{2}|H'\rangle + \frac{\sqrt{3}}{2}|V'\rangle \end{pmatrix}$ & 
$\begin{pmatrix} |H\rangle \\ -\frac{1}{2}|H'\rangle - \frac{\sqrt{3}}{2}|V'\rangle \end{pmatrix}$ & 
$\begin{pmatrix} -\frac{1}{2}|H\rangle + \frac{\sqrt{3}}{2}|V\rangle \\ |H'\rangle \end{pmatrix}$ \\
1 & $\begin{pmatrix} |H\rangle \\ -\frac{1}{2}|H'\rangle + \frac{\sqrt{3}}{2}|V'\rangle \end{pmatrix}$ & 
$\begin{pmatrix} -\frac{1}{2}|H\rangle + \frac{\sqrt{3}}{2}|V\rangle \\ -\frac{1}{2}|H'\rangle - \frac{\sqrt{3}}{2}|V'\rangle \end{pmatrix}$ & 
$\begin{pmatrix} -\frac{1}{2}|H\rangle - \frac{\sqrt{3}}{2}|V\rangle \\ |H'\rangle \end{pmatrix}$ \\
$1^{\prime}$ & $\begin{pmatrix} -\frac{1}{2}|H\rangle + \frac{\sqrt{3}}{2}|V\rangle \\ |H'\rangle \end{pmatrix}$ & 
$\begin{pmatrix} -\frac{1}{2}|H\rangle - \frac{\sqrt{3}}{2}|V\rangle \\ -\frac{1}{2}|H'\rangle + \frac{\sqrt{3}}{2}|V'\rangle \end{pmatrix}$ & 
$\begin{pmatrix} |H\rangle \\ -\frac{1}{2}|H'\rangle - \frac{\sqrt{3}}{2}|V'\rangle \end{pmatrix}$ \\
2 & $\begin{pmatrix}  |H\rangle\\ -\frac{1}{2}|H'\rangle - \frac{\sqrt{3}}{2}|V'\rangle \end{pmatrix}$ & 
$\begin{pmatrix}  -\frac{1}{2}|H\rangle + \frac{\sqrt{3}}{2}|V\rangle \\ |H'\rangle\end{pmatrix}$ & 
$\begin{pmatrix} -\frac{1}{2}|H\rangle - \frac{\sqrt{3}}{2}|V\rangle \\ -\frac{1}{2}|H'\rangle + \frac{\sqrt{3}}{2}|V'\rangle \end{pmatrix}$ \\
$2^{\prime}$ & $\begin{pmatrix} -\frac{1}{2}|H\rangle - \frac{\sqrt{3}}{2}|V\rangle  \\|H'\rangle \end{pmatrix}$ & 
$\begin{pmatrix} |H\rangle \\ -\frac{1}{2}|H'\rangle + \frac{\sqrt{3}}{2}|V'\rangle \end{pmatrix}$ & 
$\begin{pmatrix} -\frac{1}{2}|H\rangle + \frac{\sqrt{3}}{2}|V\rangle \\ -\frac{1}{2}|H'\rangle - \frac{\sqrt{3}}{2}|V'\rangle \end{pmatrix}$ \\
\hline
\end{tabular}
\end{table*}

When Alice and Bob choose matching signals (diagonal entries), both photons yield non-orthogonal states with the same detection probabilities $P(H) = 1/4$ and $P(V) = 3/4$. For mismatched signals (off-diagonal entries), at least one photon produces a pure $|H\rangle$ state with unit probability, while the other yields the characteristic $1/4$ and $3/4$ probabilities. If $|V\rangle$ (for the first photon) and $|V'\rangle$(for the second photon) both click for two measurements, Alice and Bob must know each others’ messages.

\subsubsection{Efficiency Analysis for Method I}

In Method I, Alice encodes each key symbol using a group of $m = 2$ photons prepared in two of the three available polarization states. A valid key symbol is established when Bob's measurement yields $|V\rangle$ detections for both photons, which occurs only when Alice and Bob selected different encoding operations.

The sifting probability factorizes into two independent contributions. First, for any fixed Alice signal, exactly one of Bob's three operations (the ``diagonal'' entry in Table~\ref{tab:method1_detection}) puts both photons in the superposition states $-\tfrac{1}{2}|H\rangle \pm \tfrac{\sqrt{3}}{2}|V\rangle$. For Bob's other two operations, at least one photon is deterministically projected to $|H\rangle$, therefore joint $|V\rangle|V'\rangle$ detection is impossible. Since Bob chooses uniformly at random, only $1/3$ of (Alice, Bob) pairs can in principle yield the double-$|V\rangle$ outcome. Second, conditioned on such a pair, each of the two mixed-state photons independently registers $|V\rangle$ with probability $3/4$, contributing a factor $(3/4)^2 = 9/16$. The combined sifting probability is therefore:

\begin{equation}
    P_{\text{sift}}^{(1)} = \frac{1}{3} \times \frac{9}{16} = \frac{3}{16} \approx 0.188.
\end{equation}

Applying Eq.~(\ref{eq:raw_efficiency}) with $m = 2$:
\begin{equation}
\eta_{\text{raw}}^{(1)} = \frac{3/16}{2} = \frac{3}{32} \approx 0.094.
\label{eq:method1_raw}
\end{equation}

To compare with conventional binary protocols, we convert to binary-equivalent efficiency using Eq.~(\ref{eq:bin_efficiency}):
\begin{equation}
\eta_{\text{bin}}^{(1)} = \frac{\log_2 3}{2}  \frac{3}{16} \approx 0.149 \text{ bits per photon}.
\label{eq:method1_bin}
\end{equation}

For Method I, the binary-equivalent key yield is $\eta_{\text{bin}}^{(1)} \approx 0.15$ bits per photon, or equivalently $\approx 0.30$ bits per 2-photon signal. This modest value motivates the extension developed next, which transmits all three trine states with randomized ordering.

\subsection{Key distribution Method II}

Method II addresses this limitation by transmitting all three polarization states $(A_1, A_2, A_3)$ in every signal group, with Alice randomly permuting their temporal order before transmission. This design achieves both superior efficiency and enhanced security. Using three-photon groups increases information content per transmission. More importantly, including all three states with unknown ordering creates combinatorial uncertainty that significantly constrains Eve's eavesdropping strategies.The non-orthogonality of the trine states limits Eve's single-photon identification probability, while the unknown permutation $\sigma$ introduces additional combinatorial ambiguity that cannot be resolved  through quantum measurements on the individual photon states.

Alice sends groups of three photons, where each photon has one of the three different polarization states $A_1$, $A_2$, and $A_3$-- corresponding to $(0, 2\pi/3, -2\pi/3)$ rotations. The critical feature is that Alice randomly selects a permutation $\sigma \in S_3$ and transmits the photons in the sequence $(|A_{\sigma(1)}\rangle, |A_{\sigma(2)}\rangle, |A_{\sigma(3)}\rangle)$. Bob, on the receiving end, randomly selects one of his three measurement configurations and applies it uniformly to all three photons in the group. Table~\ref{tab:protocol2} shows the measurement outcome for different combinations.

\begin{table}[h]
\caption{Method II protocol operation. Alice sends three photons with states in a permutation $\sigma$ shown in column 1. Bob applies one of three operations uniformly to all photons.  ($\times$) denotes V detection and $(\checkmark)$ denotes H detection. When exactly two V detections occur (highlighted cases), Alice can uniquely determine Bob's operation from the detection pattern and her knowledge of the photon ordering $\sigma$, establishing a shared key bit. }
\hspace{-0.5cm}
\centering
\scalebox{0.95}{%
\renewcommand{\arraystretch}{1.3}
\setlength{\tabcolsep}{4pt}
\setlength{\arrayrulewidth}{0pt}
\begin{tabular}{|c|c|c|c|c|c|}
\hline
 & Alice & $S_{LB}(\theta)$ & Bob & Post-measurement & Published \\
\hline
(1) & 
$\begin{pmatrix} \rightarrow \\ \nwarrow \\ \textcolor{red}{\swarrow} \end{pmatrix}$ &
$120^\circ$ &
$\begin{pmatrix} \nwarrow \\ \swarrow \\ \rightarrow \end{pmatrix}$ &
$\begin{pmatrix} -\frac{1}{2}\Hket + \frac{\sqrt{3}}{2}\textcolor{blue}{\Vket} \\ -\frac{1}{2}\Hket - \frac{\sqrt{3}}{2}\textcolor{blue}{\Vket} \\ \textcolor{blue}{\Hket} \end{pmatrix}$ &
$\begin{pmatrix} \cmark \\ \cmark \\ \xmark \end{pmatrix}$ \\
\hline
(2) & 
$\begin{pmatrix} \rightarrow \\ \textcolor{red}{\nwarrow} \\ \swarrow \end{pmatrix}$ &
$-120^\circ$ &
$\begin{pmatrix} \swarrow \\ \rightarrow \\ \nwarrow \end{pmatrix}$ &
$\begin{pmatrix} -\frac{1}{2}\Hket - \frac{\sqrt{3}}{2}\textcolor{blue}{\Vket} \\ \textcolor{blue}{\Hket} \\ -\frac{1}{2}\Hket + \frac{\sqrt{3}}{2}\textcolor{blue}{\Vket} \end{pmatrix}$ &
$\begin{pmatrix} \cmark \\ \xmark \\ \cmark \end{pmatrix}$ \\
\hline
(3) & 
$\begin{pmatrix} \textcolor{red}{\rightarrow} \\ \nwarrow \\ \swarrow \end{pmatrix}$ &
$0^\circ$ &
$\begin{pmatrix} \rightarrow \\ \nwarrow \\ \swarrow \end{pmatrix}$ &
$\begin{pmatrix} \textcolor{blue}{\Hket} \\ -\frac{1}{2}\Hket + \frac{\sqrt{3}}{2}\textcolor{blue}{\Vket} \\ -\frac{1}{2}\Hket - \frac{\sqrt{3}}{2}\textcolor{blue}{\Vket} \end{pmatrix}$ &
$\begin{pmatrix} \xmark \\ \cmark \\ \cmark \end{pmatrix}$ \\
\hline
(4) & 
$\begin{pmatrix} \nwarrow \\ \rightarrow \\ \textcolor{red}{\swarrow} \end{pmatrix}$ &
$120^\circ$ &
$\begin{pmatrix} \swarrow \\ \nwarrow \\ \rightarrow \end{pmatrix}$ &
$\begin{pmatrix} -\frac{1}{2}\Hket - \frac{\sqrt{3}}{2}\textcolor{blue}{\Vket} \\ -\frac{1}{2}\Hket + \frac{\sqrt{3}}{2}\textcolor{blue}{\Vket} \\ \textcolor{blue}{\Hket} \end{pmatrix}$ &
$\begin{pmatrix} \cmark \\ \cmark \\ \xmark \end{pmatrix}$ \\
\hline
(5) & 
$\begin{pmatrix} \rightarrow \\ \textcolor{red}{\swarrow} \\ \nwarrow \end{pmatrix}$ &
$120^\circ$ &
$\begin{pmatrix} \nwarrow \\ \rightarrow \\ \swarrow \end{pmatrix}$ &
$\begin{pmatrix} -\frac{1}{2}\Hket + \frac{\sqrt{3}}{2}\textcolor{blue}{\Vket}\\ \textcolor{blue}{\Hket} \\ -\frac{1}{2}\Hket - \frac{\sqrt{3}}{2}\textcolor{blue}{\Vket} \end{pmatrix}$ &
$\begin{pmatrix} \cmark \\ \xmark \\ \cmark \end{pmatrix}$ \\
\hline
(6) & 
$\begin{pmatrix} \textcolor{red}{\swarrow} \\ \nwarrow \\ \rightarrow \end{pmatrix}$ &
$120^\circ$ &
$\begin{pmatrix} \rightarrow \\ \swarrow \\ \nwarrow \end{pmatrix}$ &
$\begin{pmatrix} \textcolor{blue}{\Hket} \\ -\frac{1}{2}\Hket - \frac{\sqrt{3}}{2}\textcolor{blue}{\Vket} \\ -\frac{1}{2}\Hket + \frac{\sqrt{3}}{2}\textcolor{blue}{\Vket} \end{pmatrix}$ &
$\begin{pmatrix} \xmark \\ \cmark \\ \cmark \end{pmatrix}$ \\
\hline
\end{tabular}%
}
\label{tab:protocol2}
\end{table}

Note that regardless of Bob's choice, exactly one photon will always result in $H$ detection with certainty, while the other two have probability $3/4$ for $V$ detection. When exactly two $V$ detections occur, Alice can determine Bob's operation from the detection pattern.

The key distribution protocol for Method II operates as follows:

\textbf{Step 1: State Preparation and Transmission}
Alice prepares three photons, one in each of the states $|A_1\rangle$, $|A_2\rangle$, and $|A_3\rangle$, and transmits them in a randomly chosen temporal order. Formally, she selects a permutation $\sigma \in S_3$, where $S_3$ denotes the symmetric group of order $3! = 6$, and transmits the photons in the sequence $(|A_{\sigma(1)}\rangle, |A_{\sigma(2)}\rangle, |A_{\sigma(3)}\rangle)$. The six possible orderings correspond to the six elements of $S_3$:
\begin{equation}
\begin{aligned}
\sigma_1 &: (A_1, A_2, A_3), \quad & \sigma_2 &: (A_1, A_3, A_2), \\
\sigma_3 &: (A_2, A_1, A_3), \quad & \sigma_4 &: (A_2, A_3, A_1), \\
\sigma_5 &: (A_3, A_1, A_2), \quad & \sigma_6 &: (A_3, A_2, A_1).
\end{aligned}
\end{equation}

The permutation $\sigma$ is kept secret; together with the non-orthogonality of the trine states, it prevents Eve from determining Bob's operation from intercepted photons alone.

\textbf{Step 2: Bob's Measurement}
Bob randomly selects one of his three operations $(B_1, B_2, B_3)$ and applies it to all three photons. He then measures each photon in the $\{|H\rangle, |V\rangle\}$ basis.

\textbf{Step 3: Detection Pattern Analysis}
For any combination of Alice's photon group and Bob's operation, the detection pattern follows a specific structure. We denote $\checkmark$ for $V$ detection and $\times$ for $H$ detection.

For example, if Alice sends photons in order $(A_1, A_2, A_3)$ and Bob applies operation $B_1$:
{\relsize{-1}
\begin{equation}
    \begin{aligned}
        &A_1 \xrightarrow{B_1} |H\rangle \Rightarrow \times \text{ (H detection)}, \\
&A_2 \xrightarrow{B_1} -\frac{1}{2}|H\rangle + \frac{\sqrt{3}}{2}|V\rangle \Rightarrow \checkmark \text{ (V detection with prob. 3/4)}, \\
&A_3 \xrightarrow{B_1} -\frac{1}{2}|H\rangle - \frac{\sqrt{3}}{2}|V\rangle \Rightarrow \checkmark \text{ (V detection with prob. 3/4)}.
    \end{aligned}
\end{equation}
}

When Bob applies his measurement configuration $B_j$ (where $j \in \{1, 2, 3\}$) to the received photons and measures in the $\{|H\rangle, |V\rangle\}$ basis, a crucial pattern emerges: the photon whose original state $A_i$ matches Bob's choice (i.e., when $i = j$) will deterministically yield $|H\rangle$, while the other two photons produce probabilistic outcomes with $P(H) = 1/4$ and $P(V) = 3/4$.

\begin{table}[ht]
\centering
\caption{Detection patterns for different permutations $\sigma \in S_3$ and Bob's operations. Each row corresponds to a distinct temporal ordering of the three polarization states.}
\begin{tabular}{c|ccc}
\hline
Alice's Ordering $\sigma$ & $B_1$ & $B_2$ & $B_3$ \\
\hline
$(A_1, A_2, A_3)$ & $(\times, \checkmark, \checkmark)$ & $(\checkmark, \times, \checkmark)$ & $(\checkmark, \checkmark, \times)$ \\
$(A_1, A_3, A_2)$ & $(\times, \checkmark, \checkmark)$ & $(\checkmark, \checkmark, \times)$ & $(\checkmark, \times, \checkmark)$ \\
$(A_2, A_1, A_3)$ & $(\checkmark, \times, \checkmark)$ & $(\times, \checkmark, \checkmark)$ & $(\checkmark, \checkmark, \times)$ \\
$(A_2, A_3, A_1)$ & $(\checkmark, \checkmark, \times)$ & $(\times, \checkmark, \checkmark)$ & $(\checkmark, \times, \checkmark)$ \\
$(A_3, A_1, A_2)$ & $(\checkmark, \times, \checkmark)$ & $(\checkmark, \checkmark, \times)$ & $(\times, \checkmark, \checkmark)$ \\
$(A_3, A_2, A_1)$ & $(\checkmark, \checkmark, \times)$ & $(\checkmark, \times, \checkmark)$ & $(\times, \checkmark, \checkmark)$ \\
\hline
\end{tabular}
\label{tab:detection_patterns}
\end{table}

\textbf{Step 4: Public Announcement}

Bob publicly announces his detection pattern (the sequence of $H$ and $V$ detections). Only groups with exactly two $V$ detections are used for key generation. Crucially, this announcement reveals no information about $\sigma$ or about Bob's operation. To see why, consider the announcement $(\times, \checkmark, \checkmark)$, meaning the matched state occupied temporal slot~1. Inspecting Table~\ref{tab:detection_patterns}, this pattern arises for orderings $(A_1, A_2, A_3)$ and $(A_1, A_3, A_2)$ when Bob chose $B_1$, for $(A_2, A_1, A_3)$ and $(A_2, A_3, A_1)$ when Bob chose $B_2$, and for $(A_3, A_1, A_2)$ and $(A_3, A_2, A_1)$ when Bob chose $B_3$. Since Alice selects $\sigma$ uniformly, each of Bob's three operations is equally likely given the announcement. The same holds, by symmetry, for announcements $(\checkmark, \times, \checkmark)$ and $(\checkmark, \checkmark, \times)$. Therefore the detection pattern constrains the temporal position of the matched state but reveals neither its identity nor $\sigma$.    

\textbf{Step 5: Key Extraction}
When exactly two $V$ detections occur, Alice can uniquely determine Bob's operation from the detection pattern and her knowledge of the photon ordering. For instance:
\begin{itemize}
\item If Alice sent $(A_1, A_2, A_3)$ and Bob announces $(\times, \checkmark, \checkmark)$, Alice knows Bob used $B_1$.
\item If Alice sent $(A_2, A_3, A_1)$ and Bob announces $(\checkmark, \checkmark, \times)$, Alice knows Bob used $B_1$.
\end{itemize}

They establish the shared key bit according to Bob's operation: 0 for $B_1$, 1 for $B_2$, and 2 for $B_3$.

It is worth comparing the role of the public channel in the binary and ternary protocols. In the binary quantum eraser protocol, Bob announces only which rounds produced $D_2$ clicks, a single bit per photon that reveals nothing about either party's encoding choice. The ternary protocol extends this to a richer detection pattern (the sequence of $H$ and $V$ outcomes across three photon slots), but the nature of the communication is the same: detection outcomes, not operational choices. Neither party reveals their encoding choice ($\sigma$ or $B_j$), and as shown above, the announced pattern is equally consistent with all three of Bob's operations. Both protocols therefore achieve key sifting without basis reconciliation, distinguishing them from protocols such as BB84 where basis choices must be publicly compared and mismatched rounds discarded.

\subsubsection{Efficiency Analysis for Method II}

Method II achieves higher efficiency by transmitting all three polarization states in each group of $m = 3$ photons. The sifting criterion requires exactly two $|V\rangle$ detections, which uniquely identifies Bob's measurement choice to Alice.

For any valid key generation event, exactly one photon in the group has polarization matching Bob's measurement basis, yielding $|H\rangle$ with unit probability. The remaining two photons each produce $|V\rangle$ with probability $3/4$. The sifting probability is thus:
\begin{equation}
P_{\text{sift}}^{(2)} = \left(\frac{3}{4}\right)^2 = \frac{9}{16} \approx 0.563.
\end{equation}

Note that unlike Method I, no factor of $1/3$ appears because every photon group contains all three states; Bob's choice determines which photon gives $|H\rangle$, but does not affect whether the group is usable.

The efficiency metrics follow from Eqs.~(\ref{eq:raw_efficiency}) and (\ref{eq:bin_efficiency}):
\begin{equation}
\eta_{\text{raw}}^{(2)} = \frac{9/16}{3} = \frac{3}{16} \approx 0.188.
\label{eq:method2_raw}
\end{equation}

This represents the efficiency in ternary symbols. To compare with binary protocols, we convert to binary-equivalent efficiency. Since each ternary symbol carries $\log_2 3$ bits of information:
\begin{equation}
\eta_{\text{bin}}^{(2)} = \frac{\log_2 3}{3}  \frac{9}{16} \approx 0.297 \text{ bits per photon}.
\label{eq:method2_bin}
\end{equation}

Method II achieves $\eta_{\text{bin}} \approx 0.30$ bits per photon, competitive with deployed QKD systems. This metric accounts for both the ternary alphabet (via the factor $\log_2 3$) and the finite sifting probability, enabling direct comparison with binary protocols.

\subsubsection{Security Analysis for Method II}

The security of Method~II depends on both the quantum indistinguishability of the trine states and the secrecy of $\sigma$. Even if Eve could perfectly identify each photon's polarization state, she would still face uncertainty about Alice's chosen ordering.

To illustrate why this ordering information is crucial, consider what Eve observes when intercepting a photon group. She can determine that the three photons are in states $|H\rangle$, $-\frac{1}{2}|H\rangle + \frac{\sqrt{3}}{2}|V\rangle$, and $-\frac{1}{2}|H\rangle - \frac{\sqrt{3}}{2}|V\rangle$, but she cannot determine their temporal sequence. When Bob announces the detection pattern, say $(\times, \checkmark, \checkmark)$, Eve knows that one photon gave $H$ detection and two gave $V$ detection, but multiple values of $\sigma$
can produce this same detection pattern.

We analyze Eve's maximum success probability through two specific attack strategies:

\paragraph{Strategy (a): Eve focuses on one Alice's photon state}

Suppose Eve optimizes her measurement to reliably identify one specific state, say $|H\rangle$. She designs her detector to click with certainty when this state appears. For simplicity, assume Eve's detectors are $M_1 = |H\rangle$ and $M_2 = |V\rangle$.

The measurement outcomes for Alice's three states are summarized as follows:
\begin{center}
\begin{tabular}{lcc}
\hline
State & $P(M_1)$ & $P(M_2)$ \\
\hline
$|H\rangle$ & $1$ & $0$ \\
$-\tfrac{1}{2}|H\rangle + \tfrac{\sqrt{3}}{2}|V\rangle$ & $\tfrac{1}{4}$ & $\tfrac{3}{4}$ \\
$-\tfrac{1}{2}|H\rangle - \tfrac{\sqrt{3}}{2}|V\rangle$ & $\tfrac{1}{4}$ & $\tfrac{3}{4}$ \\
\hline
\end{tabular}
\end{center}

When measuring three photons, different detection patterns occur:
\begin{itemize}
\item Pattern $(M_1, M_2, M_2)$: probability $1 \times \frac{3}{4} \times \frac{3}{4} = \frac{9}{16}$.
\item Pattern $(M_1, M_1, M_2)$ or $(M_1, M_2, M_1)$: probability $1 \times \frac{1}{4} \times \frac{3}{4} \times 2 = \frac{6}{16}$.
\item Pattern $(M_1, M_1, M_1)$: probability $1 \times \frac{1}{4} \times \frac{1}{4} = \frac{1}{16}$.
\end{itemize}

For pattern $(M_1, M_2, M_2)$, Eve knows the $|H\rangle$ photon position but cannot distinguish between the two $\pm\frac{2\pi}{3}$ states, giving her only 50\% success in determining the complete ordering. For pattern $(M_1, M_1, M_2)$, Eve cannot determine which $M_1$ detection came from the pure $|H\rangle$ state, limiting her success to 25\%. For pattern $(M_1, M_1, M_1)$, Eve can only guess randomly among six orderings, achieving $\frac{1}{6}$ success rate.

The overall success probability is:
\begin{equation}
P_{\text{success}} = \frac{9}{16} \times 0.5 + \frac{6}{16} \times 0.25 + \frac{1}{16} \times \frac{1}{6} = 0.38.
\end{equation}

\paragraph{Strategy (b): Eve attempts to distinguish two states}

Alternatively, Eve might optimize her measurement to distinguish between two of Alice's states, say $|H\rangle$ and $-\frac{1}{2}|H\rangle + \frac{\sqrt{3}}{2}|V\rangle$. The optimal measurement for distinguishing these states uses:
\begin{equation}
\begin{aligned}
M_1 &= \cos 15^\circ|H\rangle + \sin 15^\circ|V\rangle, \\
M_2 &= -\sin 15^\circ|H\rangle + \cos 15^\circ|V\rangle.
\end{aligned}
\end{equation}

With $\cos 15^\circ \approx 0.97$ and $\cos^2 15^\circ \approx 0.93$, Eve can nearly perfectly distinguish these two states. However, she still cannot distinguish either from the third state $-\frac{1}{2}|H\rangle - \frac{\sqrt{3}}{2}|V\rangle$.

In the best case scenario where Eve perfectly distinguishes two states, half of her three-photon measurements will have ambiguous results where she cannot determine the complete ordering. Her maximum success probability remains limited to 50\%.

These examples demonstrate that regardless of Eve's measurement strategy, the random ordering of photons provides an information-theoretic security barrier that cannot be overcome through quantum measurements alone. The complete security analysis in the following section will rigorously prove that Eve's maximum success probability, considering all possible measurement strategies, is limited to 54\%. We designate this the \textit{ternary discrimination bound}. 

The method thus achieves a favorable balance: the ternary encoding with random permutation provides bounded information leakage while maintaining practical efficiency suitable for real-world applications. To our knowledge, this is the first quantum eraser protocol to overcome the binary discrimination bound while preserving interference-based sifting, the defining feature that eliminates classical basis reconciliation. We now derive this bound rigorously.

\section{Security Analysis of Ternary Protocol}
\label{sec:security-ternary}

Before proceeding, we clarify the scope of our security analysis. The results presented in this section quantify an eavesdropper's optimal information gain through minimum-error quantum state discrimination, deriving optimal POVMs acting on the transmitted states. While these bounds rigorously characterize the physical-layer information accessible to an eavesdropper, we do not claim a full composable security proof as defined in modern QKD security frameworks. Establishing composable security under general collective or coherent attacks (including privacy amplification and finite-key effects) remains beyond the scope of this work. We emphasize that the following analysis quantifies Eve’s optimal state-discrimination capability and does not constitute a full composable security proof.

\subsection{Optimal Measurement Strategies}

To determine Eve's maximum information extraction capability, we must analyze her optimal measurement strategies when intercepting groups of three randomly ordered photons. The three states created by Alice exist in a four-dimensional Hilbert space spanned by the basis $\{|U\rangle|H\rangle, |U\rangle|V\rangle, |L\rangle|H\rangle, |L\rangle|V\rangle\}$, where $|U\rangle$ and $|L\rangle$ represent the upper and lower paths in the interferometer, while $|H\rangle$ and $|V\rangle$ denote horizontal and vertical polarization states respectively.

The three quantum states in the transmission channel are:
\begin{equation}
    \begin{aligned}
        |\psi_0\rangle &= \frac{1}{\sqrt{2}}(|U\rangle|H\rangle + |L\rangle|H\rangle), \\
 |\psi_+\rangle &= \frac{1}{\sqrt{2}}\left[|U\rangle\left(-\frac{1}{2}|H\rangle + \frac{\sqrt{3}}{2}|V\rangle\right) \right. \\
 & \left. \ + |L\rangle\left(-\frac{1}{2}|H\rangle - \frac{\sqrt{3}}{2}|V\rangle\right)\right], \\
|\psi_-\rangle &= \frac{1}{\sqrt{2}}\left[|U\rangle\left(-\frac{1}{2}|H\rangle - \frac{\sqrt{3}}{2}|V\rangle\right) \right. \\
& \left. \ + |L\rangle\left(-\frac{1}{2}|H\rangle + \frac{\sqrt{3}}{2}|V\rangle\right)\right].
    \end{aligned}
\end{equation}

These states correspond to no rotation, $+120^\circ$ rotation, and $-120^\circ$ rotation of the polarization respectively. The coefficients $-1/2$ and $\pm\sqrt{3}/2$ arise from the trigonometric values at $120^\circ$, specifically $\cos(120^\circ) = -1/2$ and $\sin(120^\circ) = \sqrt{3}/2$. 
To facilitate the analysis, we construct an orthonormal basis for the four-dimensional space. We define four orthogonal states:
\begin{equation}
    \begin{aligned}
        |\phi_0\rangle &= \frac{1}{\sqrt{2}}(|U\rangle|H\rangle + |L\rangle|H\rangle) ,\\
|\phi_1\rangle &= \frac{1}{\sqrt{2}}(|U\rangle|V\rangle + |L\rangle|V\rangle), \\
|\phi_2\rangle &= \frac{1}{\sqrt{2}}(|U\rangle|V\rangle - |L\rangle|V\rangle) ,\\
|\phi_3\rangle &= \frac{1}{\sqrt{2}}(|U\rangle|H\rangle - |L\rangle|H\rangle).
    \end{aligned}
\end{equation}

The physical interpretation of these basis states is instructive: $|\phi_0\rangle$ and $|\phi_1\rangle$ represent symmetric superpositions where both paths carry the same polarization, while $|\phi_2\rangle$ and $|\phi_3\rangle$ represent antisymmetric superpositions where the paths have opposite phases. These states satisfy the orthonormality condition $\langle\phi_i|\phi_j\rangle = \delta_{ij}$.

Expressing Alice's three protocol states in this orthonormal basis reveals a property relevant to the analysis:
\begin{equation}
    \begin{aligned}
        |\psi_0\rangle &= |\phi_0\rangle ,\\
|\psi_+\rangle &= -\frac{1}{2}|\phi_0\rangle + \frac{\sqrt{3}}{2}|\phi_2\rangle ,\\
|\psi_-\rangle &= -\frac{1}{2}|\phi_0\rangle - \frac{\sqrt{3}}{2}|\phi_2\rangle.
    \end{aligned}
\end{equation}

This representation immediately shows that all three states lie in a two-dimensional subspace of the four-dimensional Hilbert space, specifically the subspace spanned by $\{|\phi_0\rangle, |\phi_2\rangle\}$. Notably, none of Alice's states have components along $|\phi_1\rangle$ or $|\phi_3\rangle$. This dimensional reduction simplifies the analysis of Eve's measurement capabilities, as it constrains the effective dimensionality of the problem.

The geometric structure of these states exhibits a three-fold rotational symmetry. In the two-dimensional subspace, the three states are separated by angles of $120^\circ$, forming an equilateral triangle when visualized on the Bloch sphere representation of the subspace. This symmetry can be expressed through a rotation operator $R$ that cyclically permutes the states:
\begin{equation}
R: |\psi_0\rangle \rightarrow |\psi_+\rangle \rightarrow |\psi_-\rangle \rightarrow |\psi_0\rangle.
\label{eq:symmetry}
\end{equation}

Each of Eve's measurement states can be expressed in the orthonormal basis as:
\begin{equation}
|\alpha_j\rangle = \sum_{i=0}^{3} C_i^{(j)} |\phi_i\rangle,
\end{equation}
where $C_i^{(j)}$ are complex coefficients satisfying the normalization condition $\sum_i |C_i^{(j)}|^2 = 1$.

However, since Alice's states lie entirely in the two-dimensional subspace spanned by $\{|\phi_0\rangle, |\phi_2\rangle\}$, the components of Eve's measurement operators along $|\phi_1\rangle$ and $|\phi_3\rangle$ do not contribute to the measurement outcomes. This effectively reduces Eve's measurement problem to optimizing within a two-dimensional subspace, though she can still utilize the full four-dimensional space for her measurement apparatus.

We parameterize Eve's measurement operators in the basis $ \{|\phi_0\rangle, |\phi_1\rangle, |\phi_2\rangle, |\phi_3\rangle\}$ as:

\begin{equation}
\begin{aligned}
|\alpha_1\rangle &= \lambda_{10}|\phi_0\rangle + \lambda_{11}|\phi_1\rangle + \lambda_{12}|\phi_2\rangle + \lambda_{13}|\phi_3\rangle, \\[6pt]
|\alpha_2\rangle &= \lambda_{20}|\phi_0\rangle + \lambda_{21}|\phi_1\rangle + \lambda_{22}|\phi_2\rangle + \lambda_{23}|\phi_3\rangle, \\[6pt]
|\alpha_3\rangle &= \lambda_{30}|\phi_0\rangle + \lambda_{31}|\phi_1\rangle + \lambda_{32}|\phi_2\rangle + \lambda_{33}|\phi_3\rangle, \\[6pt]
|\alpha_4\rangle &= \lambda_{40}|\phi_0\rangle + \lambda_{41}|\phi_1\rangle + \lambda_{42}|\phi_2\rangle + \lambda_{43}|\phi_3\rangle,
\end{aligned}
\label{eq:measurement_operators}
\end{equation}
where the $\lambda_{ij}$ are coefficients to be determined by the optimization conditions, and,
\begin{equation}
\lambda_{i3} = \sqrt{1 - |\lambda_{i0}|^2 - |\lambda_{i1}|^2 - |\lambda_{i2}|^2}, \quad i = 1,2,3,4.
\end{equation} 

The fourth measurement operator $|\alpha_4\rangle$ is determined by the orthogonality constraints with the first three operators. This means parameters $\lambda_{40}$, $\lambda_{42}$, $\lambda_{41}$ can be described by those parameters in $|\alpha_1\rangle$, $|\alpha_2\rangle$, $|\alpha_3\rangle$. As we will show later, under the symmetry approximation, $|\alpha_4\rangle$ lies entirely in the subspace orthogonal to Alice's states and does not contribute to the measurement outcomes, justifying our focus on only three measurement operators.

The orthogonality conditions between these measurement operators impose constraints on the coefficients:
\begin{equation}
\begin{aligned}
\langle\alpha_i|\alpha_j\rangle &= \sum_{k=0}^{3} \lambda_{ik}^* \lambda_{jk} = \delta_{ij}, \quad i,j \in \{1,2,3,4\}.
\end{aligned}
\end{equation}

Expanding for the first three operators in the relevant subspace:
\begin{equation}
\begin{aligned}
\langle\alpha_2|\alpha_1\rangle &= \lambda_{20}^*\lambda_{10} + \lambda_{21}^*\lambda_{11} + \lambda_{22}^*\lambda_{12} + \lambda_{23}^*\lambda_{13} = 0, \\[4pt]
\langle\alpha_3|\alpha_2\rangle &= \lambda_{30}^*\lambda_{20} + \lambda_{31}^*\lambda_{21} + \lambda_{32}^*\lambda_{22} + \lambda_{33}^*\lambda_{23} = 0, \\[4pt]
\langle\alpha_1|\alpha_3\rangle &= \lambda_{10}^*\lambda_{30} + \lambda_{11}^*\lambda_{31} + \lambda_{12}^*\lambda_{32} + \lambda_{13}^*\lambda_{33} = 0.
\end{aligned}
\end{equation}

Since Alice's three states lie in the same plane (subspace $\{|\phi_0\rangle, |\phi_2\rangle\}$), bases $|\phi_1\rangle$ and $|\phi_3\rangle$ have no contribution. The effective measurement problem reduces to finding optimal values for the six parameters $\{\lambda_{10}, \lambda_{12}, \lambda_{20}, \lambda_{22}, \lambda_{30}, \lambda_{32}\}$ that maximize Eve's probability of correctly identifying the transmitted state.

Eve measures Alice's three states $(|\psi_0\rangle, |\psi_+\rangle, |\psi_-\rangle)$ using a POVM with elements $\{|\alpha_i\rangle\langle\alpha_i|\}_{i=0}^4$.
The probability that Eve obtains measurement outcome $j$ when measuring state $|\psi_i\rangle$ is given by the Born rule:
\begin{equation}
P(j|i) = |\langle\alpha_j|\psi_i\rangle|^2 = |\lambda_{j0}\langle\phi_0|\psi_i\rangle + \lambda_{j2}\langle\phi_2|\psi_i\rangle|^2.
\end{equation}

Substituting the expressions for Alice's states in the orthonormal basis, we obtain: 
\begin{equation}
    \begin{aligned}
        \langle\phi_0|\psi_0\rangle &= 1, \quad \langle\phi_2|\psi_0\rangle = 0, \\
\langle\phi_0|\psi_+\rangle &= -\frac{1}{2}, \quad \langle\phi_2|\psi_+\rangle = \frac{\sqrt{3}}{2}, \\
\langle\phi_0|\psi_-\rangle &= -\frac{1}{2}, \quad \langle\phi_2|\psi_-\rangle = -\frac{\sqrt{3}}{2}.
    \end{aligned}
\end{equation}

The measurement outcomes are:
\begin{equation}
    \begin{aligned}
        P(j|0) &= |\lambda_{j0}|^2, \\
P(j|+) &= \left|-\frac{1}{2}\lambda_{j0} + \frac{\sqrt{3}}{2}\lambda_{j2}\right|^2, \\
P(j|-) &= \left|-\frac{1}{2}\lambda_{j0} - \frac{\sqrt{3}}{2}\lambda_{j2}\right|^2.
    \end{aligned}
\end{equation}

Explicitly, the measurement outcomes are shown in Table \ref{tab:meas_outcome}.
\begin{table}[h]
\centering
\caption{Probability squared amplitudes for measurement outcomes}
\resizebox{\columnwidth}{!}{%
\begin{tabular}{|c|c|c|c|}
\hline
& $|\alpha_1\rangle$ & $|\alpha_2\rangle$ & $|\alpha_3\rangle$ \\
\hline
$|\psi_0\rangle$ & $|\lambda_{10}|^2$ & $|\lambda_{20}|^2$ & $|\lambda_{30}|^2$ \\[8pt]
\hline
$|\psi_+\rangle$ & $\left|-\frac{1}{2}\lambda_{10} + \frac{\sqrt{3}}{2}\lambda_{12}\right|^2$ & $\left|-\frac{1}{2}\lambda_{20} + \frac{\sqrt{3}}{2}\lambda_{22}\right|^2$ & $\left|-\frac{1}{2}\lambda_{30} + \frac{\sqrt{3}}{2}\lambda_{32}\right|^2$ \\[8pt]
\hline
$|\psi_-\rangle$ & $\left|-\frac{1}{2}\lambda_{10} - \frac{\sqrt{3}}{2}\lambda_{12}\right|^2$ & $\left|-\frac{1}{2}\lambda_{20} - \frac{\sqrt{3}}{2}\lambda_{22}\right|^2$ & $\left|-\frac{1}{2}\lambda_{30} - \frac{\sqrt{3}}{2}\lambda_{32}\right|^2$ \\[8pt]
\hline
\end{tabular}%
}
\label{tab:meas_outcome}
\end{table}

These expressions show how the measurement probabilities depend on the coefficients of Eve's measurement operators in the relevant two-dimensional subspace. For a single measurement, the maximum probability that Eve correctly identifies state $|\psi_i\rangle$ when all three states are equally likely is:
\begin{equation}
\begin{aligned}
    P_{\text{max}} &= \frac{1}{3}\sum_{i} \max_j P(j|i)
    = \frac{1}{3}\sum_{i=1,2,3} \left[|\lambda_{i0}|^2\right. \\
    &+ \left.\left|-\frac{1}{2}\lambda_{i0} + \frac{\sqrt{3}}{2}\lambda_{i2}\right|^2 + \left|-\frac{1}{2}\lambda_{i0} - \frac{\sqrt{3}}{2}\lambda_{i2}\right|^2\right].
\end{aligned}
\end{equation}

This achieves its maximum value of $2/3$ when $\lambda_{i2} = 0$ for all $i = 1,2,3$, meaning Eve's optimal single-photon measurement has no component along $|\phi_2\rangle$. The same bound follows directly from the Helstrom/square-root measurement for symmetric trine ensembles, as shown in Appendix~\ref{app:ternary_povm}.

\subsection{Analysis of Three-Photon Groups}

A central element of Method~II is the secrecy of Alice's permutation $\sigma \in S_3$, her random ordering of three photons within each group. To derive Eve's maximum success probability in the three-photon protocol, we must analyze her ability to correctly identify all three states and their ordering when Alice sends a randomly ordered group. Eve's task is significantly more challenging than single-state discrimination because she must both identify each state correctly and determine $\sigma$. When Eve intercepts a group of three photons where each photon is in one of the states $\{|\psi_0\rangle, |\psi_+\rangle, |\psi_-\rangle\}$ with Alice's random ordering, she performs her optimal measurement on each photon independently. The configuration is shown in Figure~\ref{fig:ternary_polarization}. The measurement outcomes create different detection patterns that we must analyze systematically.

The possible detection patterns when Eve measures three photons can be categorized based on which detectors click. We denote a detection pattern as a sequence like $(i,j,k)$ where $i$, $j$, and $k$ indicate which of Eve's detectors (1, 2, or 3) registered each photon. The patterns fall into three distinct categories.

The first category consists of patterns where only one detector clicks for all three photons, $(1,1,1)$, $(2,2,2)$, or $(3,3,3)$, with probability $Q_1$. The second category includes patterns where one detector clicks twice and another clicks once, such as $(1,1,2)$, $(1,2,2)$, or $(2,2,3)$ in some order. This category is further subdivided based on which states cause the double click: $Q_{21}$ for when $|\psi_+\rangle$ and $|\psi_-\rangle$ hit the same detector, $Q_{22}$ for when $|\psi_0\rangle$ and $|\psi_+\rangle$ hit the same detector, and $Q_{23}$ for when $|\psi_0\rangle$ and $|\psi_-\rangle$ hit the same detector. The third category comprises patterns where all three detectors click once each, such as $(1,2,3)$ in some order, with probability denoted as $Q_3$.

To understand these probabilities intuitively, consider what each pattern reveals to Eve. When all three detectors click differently (probability $Q_3$), Eve potentially obtains maximum information since each detector uniquely identifies one state. However, she still faces the challenge of determining the temporal ordering. When one detector clicks multiple times (probabilities $Q_1$ and $Q_2$ components), Eve cannot distinguish which photon caused which click within the same detector, creating fundamental ambiguity about the ordering.

The total probability must satisfy the normalization condition: $Q_T = Q_1 + Q_{21} + Q_{22} + Q_{23} + Q_3$. This normalization $Q_T$ serves as a consistency check for our calculations and ensures that all possible detection outcomes are accounted for.

Let us now derive these probabilities systematically. Due to the three-fold symmetry of the protocol states-- as discussed in equation (\ref{eq:symmetry}), we expect Eve's optimal measurement to respect this symmetry. The measurement operators, expressed in the defined basis, have coefficients related by symmetry transformations. We parameterize Eve's measurement operators using the symmetry constraint by introducing transformed parameters:

\begin{equation}
\begin{aligned}
\eta_{20} &= -\tfrac{1}{2}\lambda_{20} + \tfrac{\sqrt{3}}{2}\lambda_{22}, &\quad
\lambda_{20} &= -\tfrac{1}{2}\eta_{20} - \tfrac{\sqrt{3}}{2}\eta_{22}, \\
\eta_{22} &= -\tfrac{\sqrt{3}}{2}\lambda_{20} - \tfrac{1}{2}\lambda_{22}, &\quad
\lambda_{22} &= \tfrac{\sqrt{3}}{2}\eta_{20} - \tfrac{1}{2}\eta_{22}.
\label{eq:eta_transform_1}
\end{aligned}
\end{equation}

Also:
\begin{equation}
\begin{aligned}
\eta_{30} &= -\tfrac{1}{2}\lambda_{30} - \tfrac{\sqrt{3}}{2}\lambda_{32}, &\quad
\lambda_{30} &= -\tfrac{1}{2}\eta_{30} + \tfrac{\sqrt{3}}{2}\eta_{32}, \\
\eta_{32} &= \tfrac{\sqrt{3}}{2}\lambda_{30} - \tfrac{1}{2}\lambda_{32}, &\quad
\lambda_{32} &= -\tfrac{\sqrt{3}}{2}\eta_{30} - \tfrac{1}{2}\eta_{32}.
\label{eq:eta_transform_2}
\end{aligned}
\end{equation}

These relations encode a $120^\circ$ rotation between the measurement operators, mirroring the symmetry of Alice's states. Under this symmetry assumption, we associate detector 1 with $|\psi_0\rangle$, detector 2 with $|\psi_+\rangle$, and detector 3 with $|\psi_-\rangle$ as the most likely identifications.

The measurement outcomes for the three states under the symmetry approximation are shown in Table~\ref{tab:sym_meas_outcomes}.
\begin{table}[h]
\centering
\caption{Probability squared amplitudes for measurement outcomes under symmetry approximations}
\resizebox{\columnwidth}{!}{%
\begin{tabular}{|c|c|c|c|}
\hline
 & $|\alpha_1\rangle$ & $|\alpha_2\rangle$ & $|\alpha_3\rangle$ \\
\hline
$|\psi_0\rangle$ & $|\lambda_{10}|^2$ & $\left|-\frac{1}{2}\eta_{20} - \frac{\sqrt{3}}{2}\eta_{22}\right|^2$ & $\left|-\frac{1}{2}\eta_{30} + \frac{\sqrt{3}}{2}\eta_{32}\right|^2$ \\[8pt]
\hline
$|\psi_+\rangle$ & $\left|-\frac{1}{2}\lambda_{10} + \frac{\sqrt{3}}{2}\lambda_{12}\right|^2$ & $|\eta_{20}|^2$ & $\left|-\frac{1}{2}\eta_{30} - \frac{\sqrt{3}}{2}\eta_{32}\right|^2$ \\[8pt]
\hline
$|\psi_-\rangle$ & $\left|-\frac{1}{2}\lambda_{10} - \frac{\sqrt{3}}{2}\lambda_{12}\right|^2$ & $\left|-\frac{1}{2}\eta_{20} + \frac{\sqrt{3}}{2}\eta_{22}\right|^2$ & $|\eta_{30}|^2$ \\[8pt]
\hline
\end{tabular}%
}
\label{tab:sym_meas_outcomes}
\end{table}

Due to the symmetry, we assume detector 1 clicking means Eve finds a $|\psi_0\rangle$ photon, detector 2 clicking means Eve finds a $|\psi_+\rangle$ photon, and detector 3 clicking means Eve finds a $|\psi_-\rangle$ photon.

For a single measurement, we analyze the maximum probability that Eve correctly identifies a state when any of her detectors clicks. The conditional probability that detector $i$ correctly identifies its target state is:
\begin{equation}
\begin{aligned}
    &P(\text{correct}|\text{detector } i \text{ clicks}) \\
    &= \frac{\frac{1}{3}\lambda_{i0}^2}{\frac{1}{3}\lambda_{i0}^2 + \frac{1}{3}\left|-\frac{1}{2}\lambda_{i0} + \frac{\sqrt{3}}{2}\lambda_{i2}\right|^2 + \frac{1}{3}\left|-\frac{1}{2}\lambda_{i0} - \frac{\sqrt{3}}{2}\lambda_{i2}\right|^2}.
\end{aligned}
\end{equation}

Taking the maximum over all of the three detectors, $\max_{i=1,2,3} P(\text{correct}|\text{detector } i) = \max_{i=1,2,3}\bigl\{\tfrac{2\lambda_{i0}^2}{3\lambda_{i0}^2 + 3\lambda_{i2}^2}\bigr\} = \max_{i=1,2,3}\bigl\{\tfrac{2}{3 + 3|\lambda_{i2}/\lambda_{i0}|^2}\bigr\} = \tfrac{2}{3}.$ This maximum is achieved when $\lambda_{i2} = 0$ for all $i = 1,2,3$. This means that if one of Eve's detectors clicks, her maximum correct identification rate is $2/3$ when her measurement operators have no component along $|\phi_2\rangle$. This result provides crucial insight into Eve's optimal attack strategy for single-photon measurements. However, the security of Method II essentially relies on Alice sending groups of three photons with different states in random order. Eve's 
measurements on these three photons are necessarily correlated. The maximum probability of correctly identifying all three states and their ordering is significantly more complex than the naive expectation of $(2/3)^3 = 4/9$ for independent measurements. The optimization problem is to find the values of $\lambda_0$ and $\lambda_2$ (subject to normalization in the relevant subspace) that maximize Eve's average success probability. This optimization, combined with the analysis of the three-photon protocol with random ordering, leads to the fundamental security limit derived below.

\paragraph{Case 1: One detector clicks three times ($Q_1$: $(j,j,j)$)}

When all three photons trigger the same detector $j$, Eve faces maximum ambiguity. While she knows all three photons caused the same detector to click, she cannot determine which click corresponds to which temporal position.

The probability $Q_1$ that only one detector clicks for all three photons is calculated by summing over all possible single-detector patterns:
\begin{equation}
\begin{aligned}
    Q_1 &= 3 \times P(\text{all three states} \rightarrow \text{same detector})\\
    &= P(\text{all three} \rightarrow \alpha_1) + P(\text{all three} \rightarrow \alpha_2) \\
    &+ P(\text{all three} \rightarrow \alpha_3), \\
&= 3 \times P(\alpha_1|\psi_0)  P(\alpha_1|\psi_+)  P(\alpha_1|\psi_-),
\end{aligned}
\end{equation}
where the factor of 3 accounts for the three equivalent detectors. In short:
\begin{equation}
Q_1 = \sum_{\text{perms}} P(\alpha_i|\psi_{i_1})  P(\alpha_i|\psi_{i_2})  P(\alpha_i|\psi_{i_3}),
\end{equation}
where the sum is over all permutations $(i_1, i_2, i_3)$ of the three states $(0, +, -)$ and $i$ represents the single clicking detector. The probability is given by:
\begin{equation}
    \begin{aligned}
        Q_1 &= \left[\lambda_{10}\left(-\frac{1}{2}\lambda_{10} + \frac{\sqrt{3}}{2}\lambda_{12}\right)\left(-\frac{1}{2}\lambda_{10} - \frac{\sqrt{3}}{2}\lambda_{12}\right)\right]^2 \\
&+ \left[\eta_{20}\left(-\frac{1}{2}\eta_{20} - \frac{\sqrt{3}}{2}\eta_{22}\right)\left(-\frac{1}{2}\eta_{20} + \frac{\sqrt{3}}{2}\eta_{22}\right)\right]^2 \\
&+ \left[\eta_{30}\left(-\frac{1}{2}\eta_{30} + \frac{\sqrt{3}}{2}\eta_{32}\right)\left(-\frac{1}{2}\eta_{30} - \frac{\sqrt{3}}{2}\eta_{32}\right)\right]^2.
    \end{aligned}
\end{equation}

This expression represents the sum over all permutations where each detector clicks once. To find the optimal measurement strategy for Eve, we exploit the symmetry approximations:
\begin{equation}
    \begin{aligned}
        \lambda_{10} &= \eta_{20} = \eta_{30},\\
        \lambda_{12} &= \eta_{22} = \eta_{32},
    \end{aligned}
    \label{eq:approx}
\end{equation}
where the $\eta$ terms represent the coefficients after applying the symmetry transformations. This symmetry ansatz is justified by the Helstrom theory for symmetric state ensembles: when the input states form a representation of a cyclic group, the optimal minimum-error measurement inherits the same symmetry~\cite{Helstrom_book}. Since Alice's three trine states are cyclically related by $120^\circ$ rotations, the optimal POVM must respect this three-fold symmetry, validating Eq.~(\ref{eq:approx}). The probability that pattern $Q_1$ occurs: 
\begin{equation}
    \frac{Q_1}{Q_T} = \frac{3(1-3r^2)^2}{4\,(27r^4 + 36r^2 + 13)}, \qquad r \equiv \frac{\lambda_{12}}{\lambda_{10}}.
\end{equation}

\begin{figure}[ht!]
\hspace*{-0.3cm}
\includegraphics[width=0.95\columnwidth]{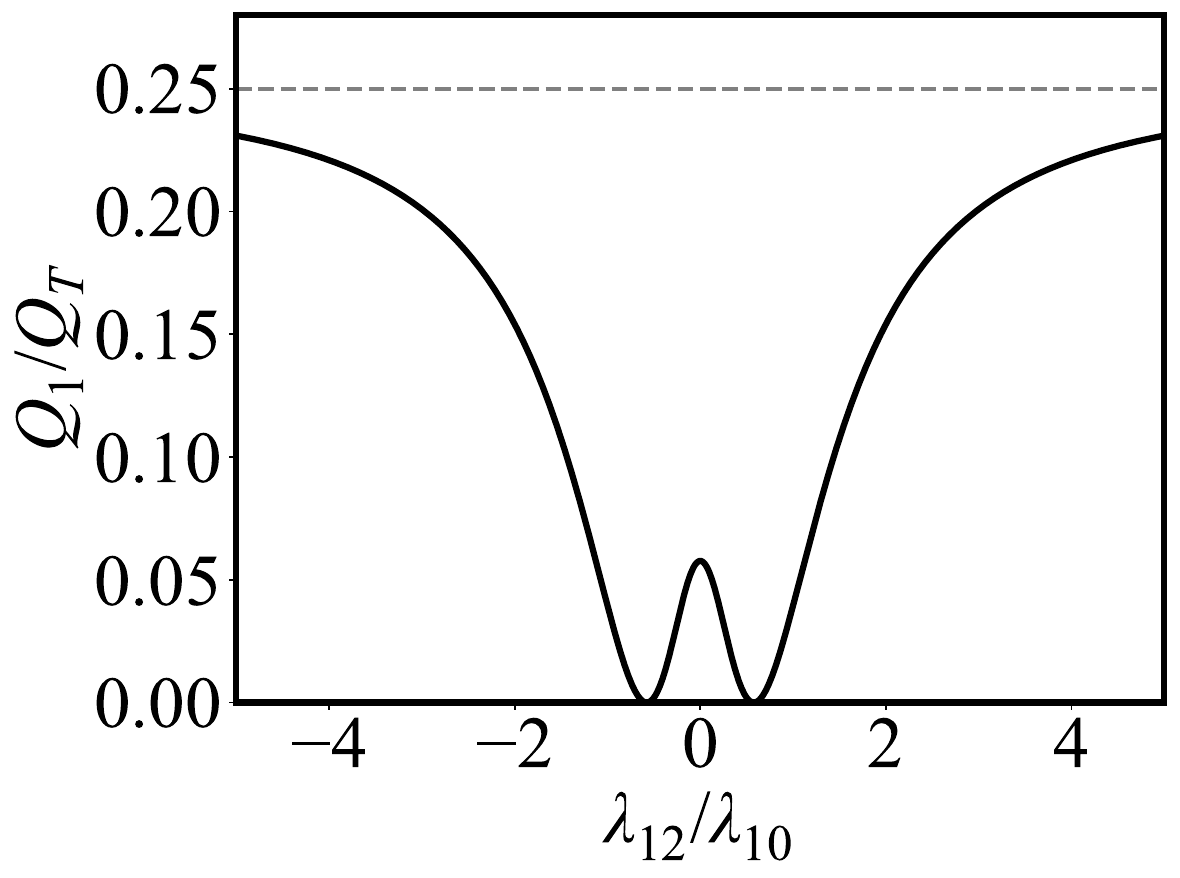}
\caption{Probability of single detector clicking $Q_1/Q_T$ as a function of $\lambda_{12}/\lambda_{10}$. The value at $\lambda_{12} = 0$ is 0.06.}
\label{fig:Q1_probability}
\end{figure}

Finally, we optimize the function with respect to the ratio parameter $|\lambda_{12}|/|\lambda_{10}|$. The optimization is shown in Figure~\ref{fig:Q1_probability}.

In this scenario, since Eve has no information about $\sigma$, she must guess uniformly at random, achieving success probability $1/|S_3| = 1/6$.\\

\paragraph{Case 2: One detector clicks twice, another once ($Q_2$: $(j,j,k), (j,k,j), (k,j,j)$)}

When one detector clicks twice and another clicks once, Eve has partial information about the photon ordering. She knows which state caused the unique click but cannot distinguish between the two positions of the repeated click.
The total probability $Q_2 = Q_{21} + Q_{22} + Q_{23}$ decomposes into three sub-patterns, where $Q_{21}$ corresponds to states $|\psi_+\rangle$ and $|\psi_-\rangle$ triggering the same detector, $Q_{22}$ to $|\psi_0\rangle$ and $|\psi_+\rangle$, and $Q_{23}$ to $|\psi_0\rangle$ and $|\psi_-\rangle$. Each term involves summing over the relevant permutations and detector combinations.

For this pattern, Eve obtains partial information about $\sigma$ but cannot fully determine it. Specifically, she knows which state triggered the unique detector but faces a two-fold ambiguity in assigning the remaining two states to their temporal positions. For example, if detectors 1 and 2 click with pattern $(1,1,2)$, Eve knows one state triggered detector 2, but cannot determine which of the two detector 1 clicks corresponds to which of the remaining two states.

Each sub-pattern has probability:
\begin{equation}
\begin{aligned}
    Q_{2i} = 2 &\times P(\text{one state} \rightarrow \text{unique detector}) \\
    &\times P(\text{two states} \rightarrow \text{same detector}).
\end{aligned}
\end{equation}

For compactness, we define:
\begin{equation}
\begin{aligned}
c_j &\equiv -\tfrac{1}{2}\eta_{j0} + \tfrac{\sqrt{3}}{2}\eta_{j2}, \\
s_j &\equiv -\tfrac{1}{2}\eta_{j0} - \tfrac{\sqrt{3}}{2}\eta_{j2},
\end{aligned}
\end{equation}
for $j \in \{2,3\}$, and similarly,
\begin{equation}
\begin{aligned}
c_1 &\equiv -\tfrac{1}{2}\lambda_{10} + \tfrac{\sqrt{3}}{2}\lambda_{12}, \\
s_1 &\equiv -\tfrac{1}{2}\lambda_{10} - \tfrac{\sqrt{3}}{2}\lambda_{12}.
\end{aligned}
\end{equation}

The probability $Q_{21}$ that states $|\psi_+\rangle$ and $|\psi_-\rangle$ trigger the same detector is:
\begin{equation}
\begin{aligned}
Q_{21} &= |\lambda_{10}\eta_{20}c_2|^2 + |\lambda_{10}\eta_{30}s_3|^2 \\
       & + |s_2  c_1  s_1|^2 + |s_2  s_3  \eta_{30}|^2 \\
       &+ |c_3  c_1  s_1|^2 + |c_3  \eta_{20}  c_2|^2.
\end{aligned}
\end{equation}

Each term represents a different assignment of the three states to the detectors, with two states going to detector 1 and one to either detector 2 or 3.

Similarly, the probability $Q_{22}$ that states $|\psi_0\rangle$ and $|\psi_+\rangle$ trigger the same detector is:
\begin{equation}
\begin{aligned}
Q_{22} &= |c_1  s_2  c_2|^2 + |c_1  c_3  \eta_{30}|^2 \\
       &+ |\eta_{20}  \lambda_{10}  s_1|^2 + |\eta_{20}  c_3  \eta_{30}|^2 \\
       & + |s_3  \lambda_{10}  s_1|^2 + |s_3  s_2  c_2|^2.
\end{aligned}
\end{equation}

The probability $Q_{23}$ that states $|\psi_0\rangle$ and $|\psi_-\rangle$ trigger the same detector is:
\begin{equation}
\begin{aligned}
Q_{23} &= |s_1  s_2  \eta_{20}|^2 + |s_1  c_3  s_3|^2 \\
       &+ |c_2  \lambda_{10}  c_1|^2 + |c_2  c_3  s_3|^2 \\
       &+ |\eta_{30}  \lambda_{10}  c_1|^2 + |\eta_{30}  s_2  \eta_{20}|^2.
\end{aligned}
\end{equation}

The conditional probability of correct identification given pattern $Q_2$ is:
\begin{equation}
P(\text{correct} \mid Q_2) = \frac{1}{2}  \frac{N_{Q_2}}{Q_{21} + Q_{22} + Q_{23}},
\label{eq:q2_info_rate}
\end{equation}
where the factor $\frac{1}{2}$ accounts for the ordering ambiguity when one detector clicks twice. The numerator contains terms where Eve correctly identifies all three states:
\begin{equation}
\begin{aligned}
N_{Q_2} &= |\lambda_{10}\eta_{20}c_2|^2 + |\lambda_{10}s_3\eta_{30}|^2 \\
        & + |\eta_{20}\lambda_{10}s_1|^2 + |\eta_{20}c_3\eta_{30}|^2 \\
        & + |\eta_{30}\lambda_{10}c_1|^2 + |\eta_{30}s_2\eta_{20}|^2.
\end{aligned}
\end{equation}

Under the symmetry approximations in Eq.\ref{eq:approx}, with $r \equiv \lambda_{12}/\lambda_{10}$ as before:
\begin{equation}
P(\text{correct} \mid Q_2) = \frac{1}{2}  \frac{(-\tfrac{1}{2} + \tfrac{\sqrt{3}}{2}r)^2 + (-\tfrac{1}{2} - \tfrac{\sqrt{3}}{2}r)^2}{(Q_{21} + Q_{22} + Q_{23})/\lambda_{10}^6}.
\end{equation}

With these symmetry approximations, all three $Q_{2i}$ terms are equal. Defining $u_\pm \equiv 1 \pm \sqrt{3}r$:
\begin{equation}
\begin{aligned}
\frac{Q_{21}}{\lambda_{10}^6} &= \frac{Q_{22}}{\lambda_{10}^6} = \frac{Q_{23}}{\lambda_{10}^6} \\
&= \tfrac{1}{4}u_+^2 + \tfrac{1}{4}u_-^2 + \tfrac{1}{16}u_-^4 + \tfrac{1}{16}u_+^4 \\
&+ \tfrac{1}{64}u_-^4 u_+^2 + \tfrac{1}{64}u_+^4 u_-^2.
\end{aligned}
\end{equation}

Neglecting higher-order terms: $\frac{Q_{2i}}{\lambda_{10}^6} \approx \tfrac{1}{4}(u_+^2 + u_-^2) + \tfrac{1}{16}(u_-^4 + u_+^4).$ Hence, Eq.~(\ref{eq:q2_info_rate}) simplifies to:
\begin{equation}
P(\text{correct} \mid Q_2) = \frac{1}{2}  \frac{u_+^2 + u_-^2}{u_+^2 + u_-^2 + \tfrac{1}{4}u_+^4 + \tfrac{1}{4}u_-^4}.
\label{eq:q2_simplified}
\end{equation}

\begin{figure}[ht]
\hspace*{-0.3cm}
\includegraphics[width=1\columnwidth]{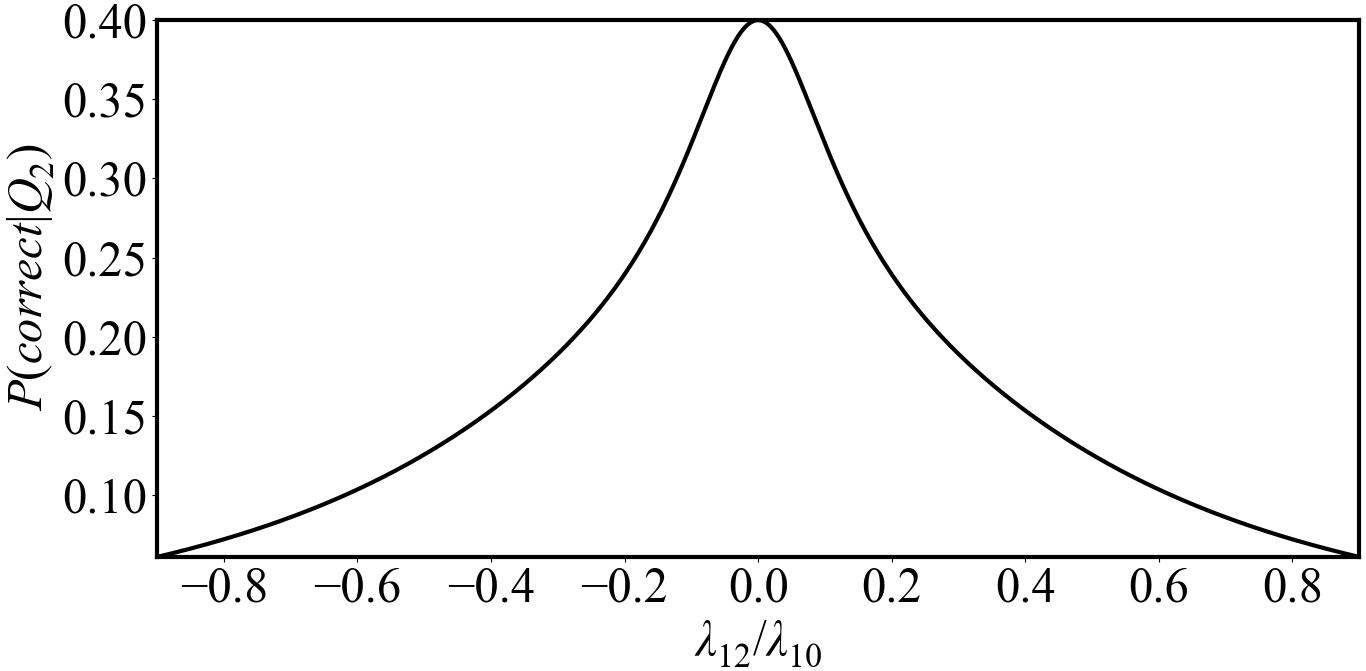}
\caption{Conditional probability of correct identification $P(\text{correct}|Q_2)$ as a function of the POVM parameter ratio $\lambda_{12}/\lambda_{10}$ for detection pattern $Q_2$ (one detector clicks twice). The factor of $\frac{1}{2}$ accounts for ordering ambiguity. The maximum value of 0.4 occurs at $\lambda_{12}/\lambda_{10} = 0$.}
\label{fig:pattern1_optimization}
\end{figure}

The analysis shows that the maximum occurs at $\lambda_{12} = 0$. This optimization is illustrated in Figure~\ref{fig:pattern1_optimization}, which shows the probability of correct identification rate as a function of $\lambda_{12}/\lambda_{10}$ when two similar detectors click. The maximum value of 0.4 occurs at $\lambda_{12}= 0$.

\paragraph{Case 3: Each detector clicks once ($Q_3$: permutations of $(i,j,k)$)}

The most informative scenario for Eve occurs when each photon triggers a different detector. The probability $Q_3$ that all three detectors click once each is obtained by summing over all permutations assigning states to detectors:
\begin{equation}
Q_3 = \sum_{\sigma \in S_3} \prod_{j=1}^{3} P(\alpha_j|\psi_{\sigma(j)}),
\end{equation}
where we identify $(\psi_1, \psi_2, \psi_3) \equiv (\psi_0, \psi_+, \psi_-)$. Expanding over all six permutations of the three states:
{\relsize{-1}
\begin{equation}
\begin{aligned}
Q_3 = 
P(\alpha_1|\psi_0) P(\alpha_2|\psi_+) P(\alpha_3|\psi_-) &+  P(\alpha_1|\psi_0) P(\alpha_2|\psi_-) P(\alpha_3|\psi_+) \\
+ P(\alpha_1|\psi_+) P(\alpha_2|\psi_0) P(\alpha_3|\psi_-) &+ P(\alpha_1|\psi_+) P(\alpha_2|\psi_-) P(\alpha_3|\psi_0) \\
+ P(\alpha_1|\psi_-) P(\alpha_2|\psi_0) P(\alpha_3|\psi_+) &+  P(\alpha_1|\psi_-) P(\alpha_2|\psi_+) P(\alpha_3|\psi_0).
\end{aligned}
\end{equation}
}

The probability $Q_3$ that all three detectors click once each involves summing over all six permutations of the three states:
\begin{equation}
\begin{aligned}
Q_3 &= |\lambda_{10}\eta_{20}\eta_{30}|^2 + |\lambda_{10}  s_3  c_2|^2 \\
    & + |c_1  s_2  \eta_{30}|^2 + |\eta_{20}  s_1  c_3|^2 \\
    & + |s_1  s_2  s_3|^2 + |c_1  c_2  c_3|^2.
\end{aligned}
\end{equation}

Eve correctly identifies $\sigma$
only when the state-to-detector assignment matches her labeling convention: $|\psi_0\rangle \to \alpha_1
,\; |\psi_+\rangle \to \alpha_2
, \; |\psi_-\rangle \to \alpha_3$
. The probability that Eve correctly identifies the photon ordering is:
\begin{equation}
P(\text{correct}|Q_3) = \frac{|\lambda_{10}|^2 |\eta_{20}|^2 |\eta_{30}|^2}{Q_3}.
\end{equation}

To evaluate this expression, we compute the normalized detection probability:

\begin{equation}
\begin{aligned}
\frac{Q_3}{(\lambda_{10}\eta_{20}\eta_{30})^2} &= 1 + \tfrac{1}{16}u_-^2 u_+^2 + \tfrac{1}{16}u_+^2 u_-^2 \\
    & + \tfrac{1}{16}u_-^2 u_+^2 + \tfrac{1}{64}u_-^6 + \tfrac{1}{64}u_+^6.
\end{aligned}
\end{equation}

Under the symmetry approximations (Eq.~\ref{eq:approx}), this simplifies to:
\begin{equation}
\frac{Q_3}{\lambda_{10}^6} = 1 + \tfrac{3}{16}u_+^2 u_-^2 + \tfrac{1}{64}u_+^6 + \tfrac{1}{64}u_-^6.
\end{equation}

In this case, if Eve's measurement is perfectly aligned with the state symmetry (meaning detector 1 is optimized for $|\psi_0\rangle$, detector 2 for $|\psi_+\rangle$, and detector 3 for $|\psi_-\rangle$), she can potentially determine the photon ordering with higher probability. However, even in this optimal scenario, the success rate is limited to approximately 0.82 due to the fundamental indistinguishability of the non-orthogonal states.

\begin{figure}[H]
\hspace*{-0.3cm}
\includegraphics[width=1\columnwidth]{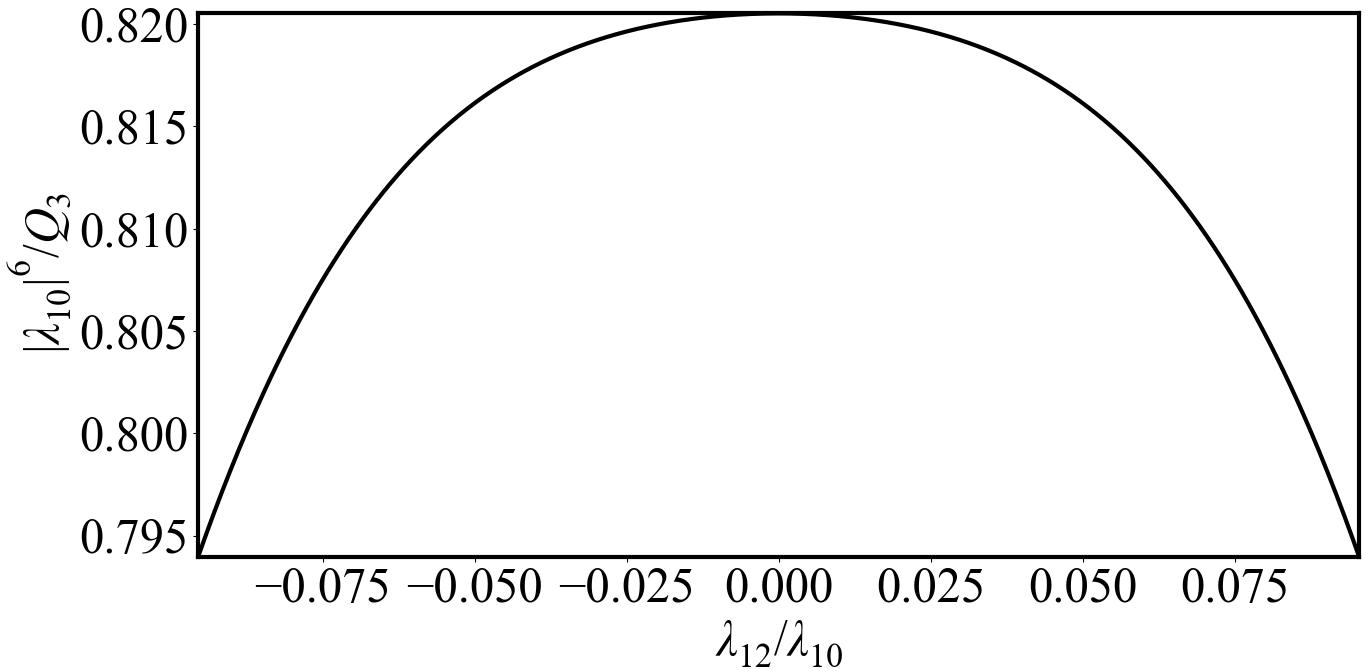}
\caption{Conditional probability of correct identification $P(\text{correct}|Q_3) = |\lambda_{10}|^6/Q_3$ as a function of the POVM parameter ratio $\lambda_{12}/\lambda_{10}$ for detection pattern $Q_3$ (all three detectors click). The maximum value of 0.82 occurs at $\lambda_{12}/\lambda_{10} = 0$.}
\label{fig:info_rate_Q3}
\end{figure}

Figure~\ref{fig:info_rate_Q3} shows the probability of correct identification as a function of $\lambda_{12}/\lambda_{10}$. Evaluating at $\lambda_{12} = 0$:

\begin{equation}
\left.\frac{|\lambda_{10}|^6}{Q_3}\right|_{\lambda_{12}=0} \approx 0.8205.
\end{equation}

This represents the highest conditional success probability among the three detection patterns. When all three detectors click distinctly, Eve gains maximal information about the individual states. Nevertheless, the success rate remains bounded at 0.82 due to the residual indistinguishability of the non-orthogonal states. Eve may hope $Q_3/Q_T$ reach its maximum value. Using $u_\pm = 1 \pm \sqrt{3}r$, the fraction of events where all three detectors click is
\begin{equation}
\frac{Q_3}{Q_T} = \frac{1 + \tfrac{3}{16}u_+^2 u_-^2}{1 + \tfrac{6}{16}u_+^2 u_-^2 + \tfrac{3}{4}(u_+^2 + u_-^2) + \tfrac{3}{16}(u_-^4 + u_+^4)}.
\end{equation}

\begin{figure}[H]
\hspace*{-0.1cm}\includegraphics[width=1\columnwidth]{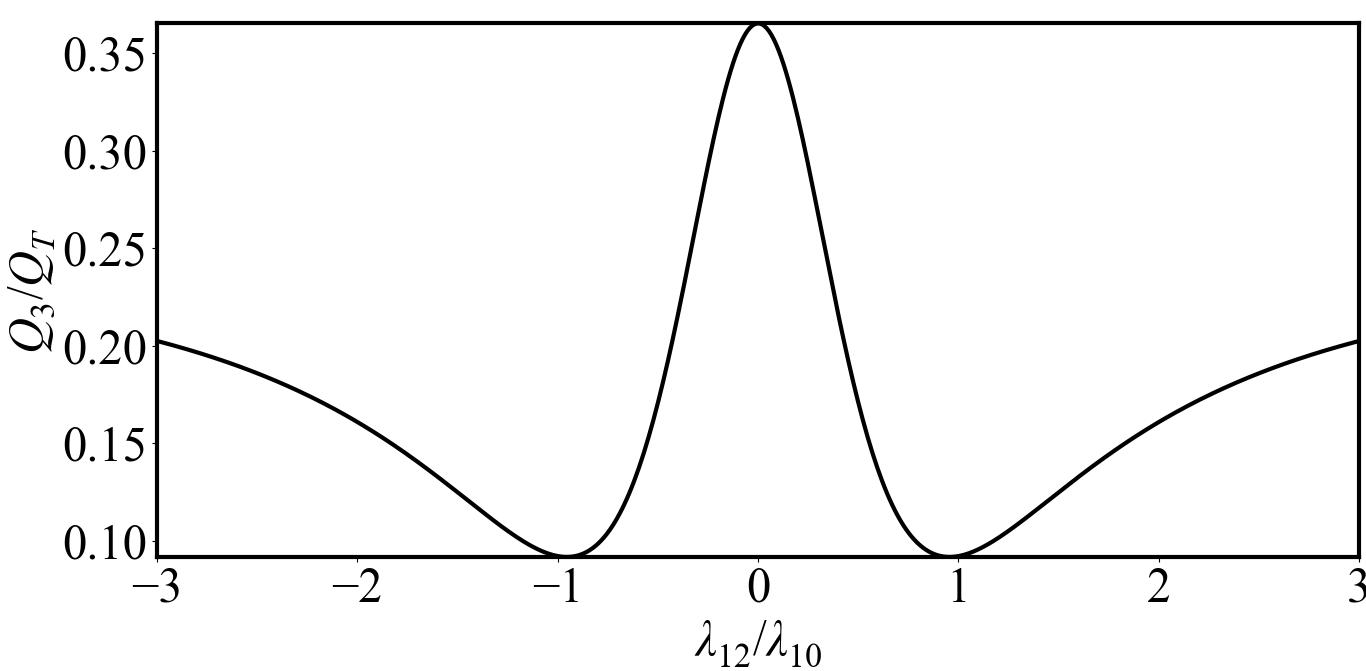}
\caption{The probability $Q_3/Q_T$ that three detectors all click as a function of $\lambda_{12}/\lambda_{10}$. The maximum value is 0.37 at $\lambda_{12}/\lambda_{10} = 0$.}
\label{fig:q3_probability}
\end{figure}

Figure~\ref{fig:q3_probability} shows the probability $Q_3/Q_T$ as a function of $\lambda_{12}/\lambda_{10}$, reaching its maximum of 0.37 at $\lambda_{12}/\lambda_{10} = 0$. Hence, the numerical evaluation yields:
\begin{equation}
    \begin{aligned}
        \frac{Q_1}{Q_T} &\approx 0.06 \quad \text{(one detector clicks three times)}, \\
\frac{Q_2}{Q_T} &\approx 0.57 \quad \text{(one detector clicks twice)}, \\
\frac{Q_3}{Q_T} &\approx 0.37 \quad \text{(all three detectors click differently)}.
    \end{aligned}
\end{equation}

\subsection{Total success probability}

Having derived the conditional success probabilities for each detection pattern class under Eve's optimal POVM (Figures~\ref{fig:Q1_probability}--\ref{fig:q3_probability}), we now combine them with the pattern occurrence  probabilities to obtain Eve's overall success rate.

\begin{table}[ht]
\centering
\caption{Summary of detection probabilities and information extraction rates under optimal measurement strategy} \scalebox{0.86}{
\begin{tabular}{|c|c|c|c|}
\hline
Pattern Type & Symbol & Probability & Correct Information  \\
 & & & Rate \\
\hline
Single detector & $Q_1$ & 0.06 & 1/6 \\
One detector twice & $Q_{21} + Q_{22} + Q_{23}$ & 0.57 & $0.4$ \\
All detectors once & $Q_3$ & 0.37 & 0.82 \\
\hline
\end{tabular}
\label{tab:detection_summary}
}
\end{table}

The overall maximum success probability for Eve is therefore:

\begin{equation}
    \begin{aligned}
        P_{\text{max}} &= Q_1 \times P_{\text{success|pattern 1}} + Q_2 \times P_{\text{success|pattern 2}} \\
        &+ Q_3 \times P_{\text{success|pattern 3}}\\
&= 0.06 \times \frac{1}{6} + 0.57 \times 0.4 + 0.37 \times 0.82 \approx 0.54.
    \end{aligned}
\end{equation}

\subsubsection{Role of the Fourth Detector and Completeness}

A complete measurement in the four-dimensional Hilbert space requires four orthogonal measurement operators $\{|\alpha_1\rangle, |\alpha_2\rangle, |\alpha_3\rangle, |\alpha_4\rangle\}$. Since Alice's three states lie in a two-dimensional subspace spanned by $\{|\phi_0\rangle, |\phi_2\rangle\}$, two important questions arise: can Eve gain additional information by using multiple detectors to measure a single state, and does the fourth detector $|\alpha_4\rangle$ contribute to Eve's ability to distinguish Alice's states? We address both questions through geometric and algebraic arguments.

\paragraph{Projection onto measurement subspaces.}

Consider first whether Eve could benefit from using two detectors, say $|\alpha_3\rangle$ and $|\alpha_4\rangle$, to collectively identify a single state such as $|\psi_-\rangle$. To analyze this, we express $|\psi_-\rangle$ in Eve's measurement basis:
\begin{equation}
|\psi_-\rangle = x_1|\alpha_1\rangle + x_2|\alpha_2\rangle + x_3|\alpha_3\rangle + x_4|\alpha_4\rangle,
\end{equation}
where $\sum_i |x_i|^2 = 1$. The total probability that either detector $|\alpha_3\rangle$ or $|\alpha_4\rangle$ registers the state is:
\begin{equation}
P^2 = |\langle\alpha_3|\psi_-\rangle|^2 + |\langle\alpha_4|\psi_-\rangle|^2 = |x_3|^2 + |x_4|^2.
\label{eq:projection_sum}
\end{equation}

This quantity represents the squared magnitude of $|\psi_-\rangle$'s projection onto the two-dimensional subspace spanned by $\{|\alpha_3\rangle, |\alpha_4\rangle\}$. A fundamental property of such projections is their independence from the specific basis choice within the subspace. To demonstrate this, consider an alternative orthonormal basis obtained by rotation through angle $y$:

\begin{equation}
\begin{aligned}
|Y_3\rangle &= \cos y\, |\alpha_3\rangle + \sin y\, |\alpha_4\rangle ,\\
|Y_4\rangle &= \cos y\, |\alpha_4\rangle - \sin y\, |\alpha_3\rangle.
\end{aligned}
\end{equation}

Computing the projection in this rotated basis:
\begin{equation}
\begin{aligned}
&|\langle Y_3|\psi_-\rangle|^2 + |\langle Y_4|\psi_-\rangle|^2 \\
&= |x_3\cos y + x_4\sin y|^2 + |x_4\cos y - x_3\sin y|^2 \\
&= |x_3|^2\cos^2 y + |x_4|^2\sin^2 y + |x_4|^2\cos^2 y + |x_3|^2\sin^2 y \\
&= |x_3|^2 + |x_4|^2.
\end{aligned}
\end{equation}

\begin{figure}[ht]
\centering
\includegraphics[width=0.6\columnwidth]{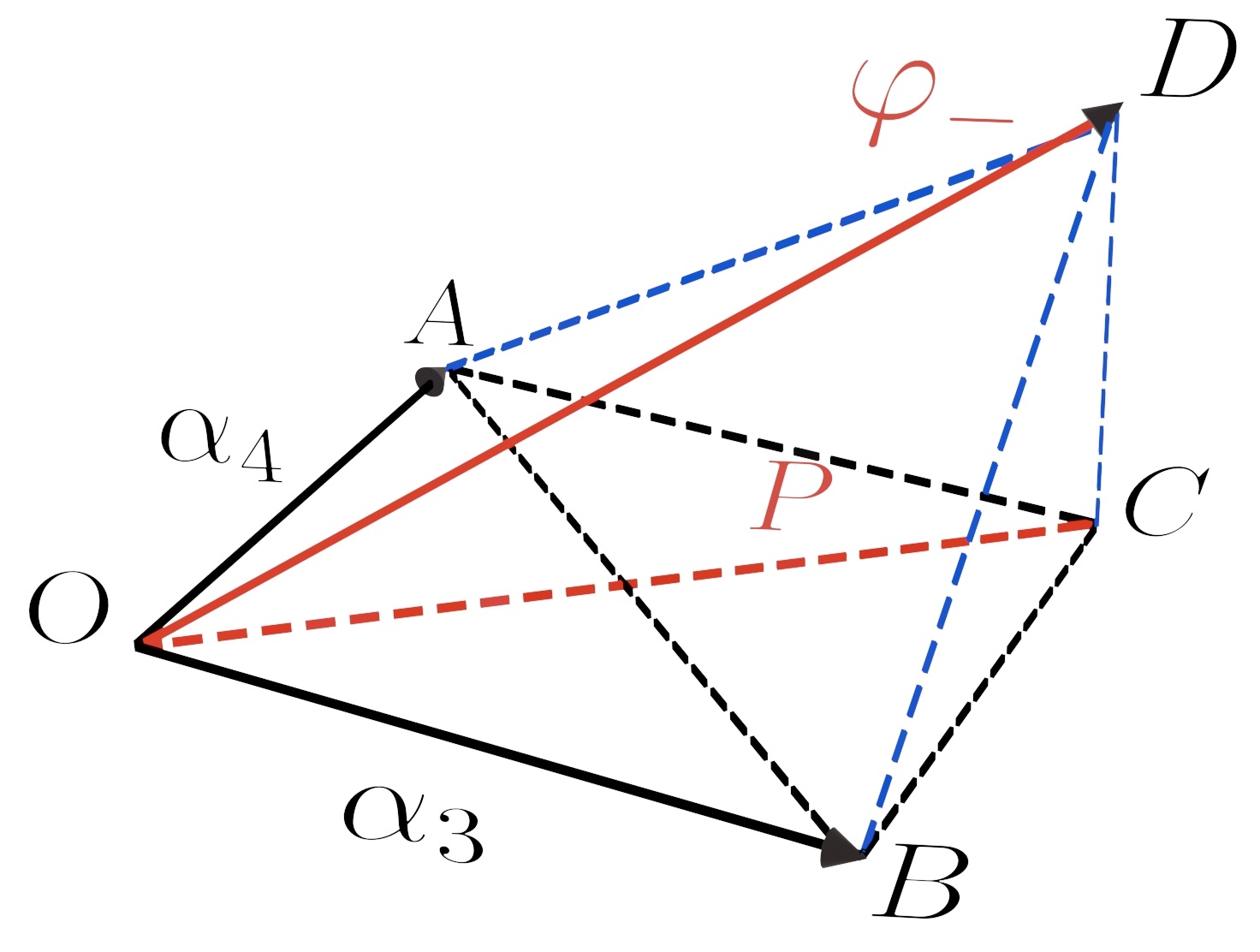}
\caption{Geometric interpretation of subspace projection. The state $|\psi_-\rangle$ (red vector OD) projects onto the subspace spanned by $\{|\alpha_3\rangle, |\alpha_4\rangle\}$ (the plane containing axes OA and OB). The projection (dashed red line OC) has magnitude $P = \sqrt{|x_3|^2 + |x_4|^2}$, computed as $\overline{OC} = \sqrt{\overline{OA}^2 + \overline{OB}^2}$ since $\overline{OA} \perp \overline{OB}$. Rotating the basis vectors within the plane changes OA and OB individually but preserves their sum of squares, leaving $\overline{OC}$ invariant.}
\label{fig:projection_geometry}
\end{figure}

The projection magnitude is invariant under rotations within the subspace. Figure~\ref{fig:projection_geometry} illustrates this geometrically: the state $|\psi_-\rangle$ (vector OD) projects onto the plane spanned by $|\alpha_3\rangle$ and $|\alpha_4\rangle$ with a fixed length $\overline{OC} = \sqrt{|x_3|^2 + |x_4|^2}$, regardless of how coordinate axes are oriented within that plane.

The physical implication is that the projection of $|\psi_-\rangle$ onto any subspace depends only on the subspace itself, not on how Eve partitions it between detectors. Consequently, Eve gains nothing by splitting detection of a single state across multiple detectors, which justifies analyzing her strategy in terms of three primary detectors aligned with Alice's three states.

\paragraph{Rotational symmetry and orthogonality of the fourth detector.}

To prove that the fourth detector does not contribute to state discrimination, we analyze the complete measurement structure under the symmetry approximation. Recall that Eve's measurement operators in the basis $\{|\phi_0\rangle, |\phi_1\rangle, |\phi_2\rangle, |\phi_3\rangle\}$ have the general form:
\begin{equation}
|\alpha_j\rangle = \sum_{i=0}^{3} \lambda_{ji}|\phi_i\rangle.
\end{equation}

Using the parameter transformations from Eqs.~(\ref{eq:eta_transform_1})--(\ref{eq:eta_transform_2}) and the symmetry approximation $\lambda_{10} = \eta_{20} = \eta_{30}$ and $\lambda_{12} = \eta_{22} = \eta_{32}$, we can express the four measurement operators as a matrix where rows correspond to detectors and columns to basis states:
{\relsize{-1}
\begin{equation}
\begin{pmatrix}
|\alpha_1\rangle \\
|\alpha_2\rangle \\
|\alpha_3\rangle \\
|\alpha_4\rangle
\end{pmatrix}
=
\begin{pmatrix}
\lambda_{10} & \lambda_{11} & \lambda_{12} & \lambda_{13} \\
-\frac{1}{2}\eta_{20} - \frac{\sqrt{3}}{2}\eta_{22} & \lambda_{21} & -\frac{1}{2}\eta_{22} + \frac{\sqrt{3}}{2}\eta_{20} & \lambda_{23} \\
-\frac{1}{2}\eta_{30} + \frac{\sqrt{3}}{2}\eta_{32} & \lambda_{31} & -\frac{1}{2}\eta_{32} - \frac{\sqrt{3}}{2}\eta_{30} & \lambda_{33} \\
\lambda_{40} & \lambda_{41} & \lambda_{42} & \lambda_{43}
\end{pmatrix}
\begin{pmatrix}
|\phi_0\rangle \\
|\phi_1\rangle \\
|\phi_2\rangle \\
|\phi_3\rangle
\end{pmatrix}.
\label{eq:measurement_matrix}
\end{equation}
}

The key observation is that Alice's states lie entirely in the subspace spanned by $\{|\phi_0\rangle, |\phi_2\rangle\}$, corresponding to the first and third columns. The relevant submatrix governing state discrimination is:

\begin{equation}
M_{\text{eff}} = 
\begin{pmatrix}
\lambda_{10} & \lambda_{12} \\
-\frac{1}{2}\eta_{20} - \frac{\sqrt{3}}{2}\eta_{22} & -\frac{1}{2}\eta_{22} + \frac{\sqrt{3}}{2}\eta_{20} \\
-\frac{1}{2}\eta_{30} + \frac{\sqrt{3}}{2}\eta_{32} & -\frac{1}{2}\eta_{32} - \frac{\sqrt{3}}{2}\eta_{30}
\end{pmatrix}.
\label{eq:effective_submatrix}
\end{equation}

\paragraph{Rotational invariance and Measurement-matrix structure.}

The three-fold symmetry of Alice's states suggests examining how the measurement structure transforms under $120^\circ$ rotations in the $\{|\phi_0\rangle, |\phi_2\rangle\}$ subspace. The relevant rotation operator is:
\begin{equation}
\label{eq:rotation_matrix}
R_{120^\circ} = 
\begin{pmatrix}
\cos\frac{2\pi}{3} & 0 & -\sin\frac{2\pi}{3} & 0 \\[3pt]
0 & 1 & 0 & 0 \\[3pt]
\sin\frac{2\pi}{3} & 0 & \cos\frac{2\pi}{3} & 0 \\[3pt]
0 & 0 & 0 & 1
\end{pmatrix}
=
\begin{pmatrix}
-\frac{1}{2} & 0 & -\frac{\sqrt{3}}{2} & 0 \\[3pt]
0 & 1 & 0 & 0 \\[3pt]
\frac{\sqrt{3}}{2} & 0 & -\frac{1}{2} & 0 \\[3pt]
0 & 0 & 0 & 1
\end{pmatrix}.
\end{equation}

This rotation acts only on the $|\phi_0\rangle$--$|\phi_2\rangle$ subspace while leaving $|\phi_1\rangle$ and $|\phi_3\rangle$ unchanged. Under this rotation, the effective submatrix $M_{\text{eff}}$ transforms as:
\begin{equation}
M_{\text{eff}} \rightarrow R_{120^\circ}^{(2\times2)} \; M_{\text{eff}} ,
\end{equation}
where $R_{120^\circ}^{(2\times2)}$ is the $2\times2$ rotation matrix acting on the subspace.

Direct calculation confirms that under the symmetry approximation, the effective submatrix is invariant under this rotation: the rows of $M_{\text{eff}}$ are cyclically permuted, which corresponds to relabeling the detectors rather than changing the measurement structure. This invariance has a profound consequence for the fourth detector.

Since the parameters $\{\lambda_{40}, \lambda_{42}\}$ defining $|\alpha_4\rangle$'s components in the relevant subspace are determined by orthogonality with $|\alpha_1\rangle$, $|\alpha_2\rangle$, and $|\alpha_3\rangle$, and since the structure of these three detectors is invariant under $120^\circ$ rotations, the fourth detector's subspace components must also be rotationally invariant. The only vector in a two-dimensional space that is invariant under $120^\circ$ rotations is the zero vector.

Therefore, under the symmetry approximation, $\lambda_{40} = \lambda_{42} = 0$. $|\alpha_4\rangle = \lambda_{41}|\phi_1\rangle + \lambda_{43}|\phi_3\rangle$ lies entirely in the orthogonal complement spanned by $\{|\phi_1\rangle, |\phi_3\rangle\}$.

Since Alice's states have no components along $|\phi_1\rangle$ or $|\phi_3\rangle$, the fourth detector never clicks when measuring any of Alice's states:
\begin{equation}
\langle\alpha_4|\psi_0\rangle = \langle\alpha_4|\psi_+\rangle = \langle\alpha_4|\psi_-\rangle = 0.
\end{equation}

The orthogonality of $|\alpha_4\rangle$ to Alice's state space has two important implications:

First, it validates our security analysis, which focused on only three measurement operators. The fourth detector, required for mathematical completeness of the POVM, contributes nothing to Eve's information gain about the transmitted states.

Second, and perhaps more practically, this result implies that Eve's optimal measurement apparatus effectively operates as a three-outcome measurement when intercepting the ternary protocol. The fourth detector serves only to account for states outside Alice's encoding subspace, states that never occur in the legitimate protocol operation. This dimensional reduction from four to three measurement outcomes mirrors the dimensional reduction of Alice's encoding from the full four-dimensional Hilbert space to a two-dimensional subspace, reflecting a deep structural property of the protocol's security.

The maximum probability that Eve correctly identifies Alice's photon ordering therefore remains bounded, as derived in the preceding analysis, with no possibility of improvement through exploitation of the fourth measurement degree of freedom.

\subsubsection{Final Security Bound and Comparison}

Eve's maximum success probability of 54\%, the ternary discrimination bound, cannot be exceeded regardless of computational resources or measurement technology. This represents a substantial reduction from the binary discrimination bound. Appendix~\ref{app:ternary_povm} corroborates this bound through an independent POVM/permutation-uncertainty argument.

The mathematical analysis can be understood through a simple physical picture. In the ternary protocol, the three photons encoding each logical symbol carry no intrinsic labels identifying their position within the ordered triple $(A_1, A_2, A_3)$. Alice's encoding comprises three non-orthogonal polarization states transmitted in a secret temporal order specified by a permutation $\sigma \in S_3$, where $S_3$ denotes the symmetric group containing the six possible orderings of three elements. Bob learns the correct permutation only during the classical sifting stage. Eve, however, must perform her measurements before this announcement. Her optimal POVM on each photon collapses the individual trine state but destroys any coherence that could preserve ordering information. She obtains an unordered multiset of outcomes and must guess which measurement result originated from which temporal slot.

The security bound reflects both quantum and combinatorial contributions. The quantum mechanical limit on discriminating three symmetric trine states bounds Eve's single-photon success probability to 2/3, already below the binary
discrimination bound. The combinatorial uncertainty then further degrades her performance. When all three of Eve's detectors click differently (probability $Q_3 \approx 0.37$), she gains maximal information about the states but still faces residual ordering ambiguity, achieving 82\% success. When one detector clicks twice (probability $Q_2 \approx 0.57$), she cannot determine which click corresponds to which state, reducing her success to 40\%. When a single detector clicks three times (probability $Q_1 \approx 0.06$), she can only guess randomly among six orderings. The weighted combination yields the ternary
discrimination bound.

A notable feature of this security mechanism is that no coherence across the three photons is required. Only the path-polarization coherence within each individual photon, essential for the quantum eraser interference, must be maintained. The combinatorial protection arises entirely from classical uncertainty about temporal ordering, making the protocol robust against decoherence between photons in the same group.

The security analysis also reveals that the optimal eavesdropping strategy requires Eve to measure in a basis aligned with the natural symmetry of the quantum states. This confirms that the three-fold symmetric configuration is not merely convenient but fundamentally optimal for security. Any deviation from this symmetry, either in Alice's state preparation or Eve's measurement strategy, only serves to reduce information transfer without improving security, validating the protocol's design principles. 

Our results highlight an intermediate security layer (protocol-level physical indistinguishability) bridging abstract QKD security proofs and practical implementations, whereby engineered state geometry and quantum-eraser constraints limit Eve's optimal measurements before classical post-processing.

\section{Conclusion}
\label{sec:conclusion}

This work has established both the fundamental security limitations of binary quantum eraser cryptography and a practical pathway to enhanced security through ternary encoding.
The binary protocol permits eavesdropping success of $85\%$ — the binary discrimination bound — in the two-state, four-state, and randomized-polarization variants. On the trade-off curve of Section~\ref{sec:efficiency-binary}, the standard binary eraser occupies the point $(B_{\text{pole}}, E_{ff}) = (0.85, 0.25)$.

The ternary protocol overcomes this limitation through symmetric three-state encoding with randomized photon ordering. Two mechanisms combine: quantum indistinguishability limits single-photon discrimination to 2/3, while combinatorial uncertainty from unknown ordering further constrains multi-photon attacks. The resulting 54\% eavesdropping bound against individual attacks represents substantial improvement while maintaining 0.30 bits per photon efficiency and preserving automatic sifting without basis reconciliation.

Several directions merit future investigation. Extension to higher-dimensional encoding may yield further security improvements, though implementation complexity requires careful analysis. Hybrid protocols combining quantum eraser mechanisms with other degrees of freedom (orbital angular momentum, time-bin encoding) present opportunities for enhanced capabilities. Experimental implementation would validate theoretical bounds under realistic conditions, including detector inefficiencies and channel losses; Appendix~\ref{app:imperfections} analyzes how beam-splitter imbalance, decoherence, and rotator calibration errors modify the detection statistics.

A central priority for future work is integrating the physical-layer bounds established here into a full composable security framework, incorporating privacy amplification, finite-key analysis, and resistance to general collective and coherent attacks. A detailed information-theoretic analysis, quantifying Eve's mutual information with the key and the achievable secret-key rate under the ternary protocol, will be presented elsewhere~\cite{Halawani_inprep}. Despite these open questions, the ternary quantum eraser protocol demonstrates that expanding the encoding alphabet beyond binary states can enhance security while preserving the operational elegance that distinguishes quantum eraser cryptography.

\section{Acknowledgment}

This work is supported by the King Abdulaziz city for Science and Technology (KACST). Z.-H. Li is supported by the Natural Science Foundation of Shanghai (Grant No. 25ZR1401140).

\appendix
\raggedbottom

\section{Operators $T_A$ and $T_B$ cancel each other}
\label{app:operators}

\subsection{Correct Derivation for $T_A$ and $T_B$ With Non-Ideal Angles}

\subsubsection{Cancellation of the which-path tags in the ideal case}

In the ideal version of the protocol, Alice and Bob use polarization rotators $T_A$ and $T_B$ that apply opposite rotations of equal magnitude,
\begin{equation}
\begin{aligned}
T_A = R(\theta), \quad T_B = R(-\theta),
\end{aligned}
\end{equation}
hence that the combined action on the two interferometer arms is:
\begin{equation}
\begin{aligned}
(T_B \otimes T_A)|\psi\rangle = (R(-\theta) \otimes R(\theta))|\psi\rangle.
\end{aligned}
\end{equation}

Because the photon's polarization state is ultimately recombined at BS2, the net effect of the two rotations is proportional to the identity on the polarization subspace:
\begin{equation}
\begin{aligned}
R(-\theta)R(\theta) = I,
\end{aligned}
\end{equation}
hence the path--polarization tagging introduced by $T_A$ is exactly undone by $T_B$. As a result, the two paths remain indistinguishable in polarization, and full interference is recovered at the final beam splitter.

\subsubsection{Non-ideal rotators: derivation of residual distinguishability}

In practice, neither rotator is perfect. Let:
\begin{equation}
\begin{aligned}
T_A = R(\theta + \delta_A), \quad T_B = R(-\theta + \delta_B),
\end{aligned}
\end{equation}
where $\delta_A$ and $\delta_B$ denote (possibly small) calibration errors. Then the combined transformation is:
\begin{equation}
\begin{aligned}
R(-\theta + \delta_B)R(\theta + \delta_A) = R(\delta_A + \delta_B).
\label{eq:combined_rotation}
\end{aligned}
\end{equation}

Thus any deviation from perfect cancellation results in a net rotation:
\begin{equation}
\begin{aligned}
R_{\text{net}} = R(\Delta\theta), \quad \Delta\theta = \delta_A + \delta_B.
\label{eq:net_rotation}
\end{aligned}
\end{equation}

This residual rotation couples the two polarization amplitudes according to:
\begin{equation}
\begin{aligned}
R(\Delta\theta)\begin{pmatrix} H \\ V \end{pmatrix} = \begin{pmatrix} \cos\Delta\theta & -\sin\Delta\theta \\ \sin\Delta\theta & \cos\Delta\theta \end{pmatrix}\begin{pmatrix} H \\ V \end{pmatrix},
\label{eq:residual_rotation}
\end{aligned}
\end{equation}
as such, the upper and lower interferometer arms emerge from Bob's station with polarization states separated by a nonzero angle $\Delta\theta$.

\subsubsection{Effect on interference visibility}

Because the visibility of the interference at BS2 is proportional to the inner product of the two polarization states, as detailed in \ref{sec:security-binary},
\begin{equation}
\begin{aligned}
V = |\langle\phi_u|\phi_l\rangle|,
\label{eq:visibility}
\end{aligned}
\end{equation}
and the overlap of two pure qubit states separated by angle $\Delta\theta$ is:
\begin{equation}
\begin{aligned}
|\langle\phi_u|\phi_l\rangle| = |\cos\Delta\theta|,
\label{eq:overlap}
\end{aligned}
\end{equation}
we obtain the visibility reduction,
\begin{equation}
\begin{aligned}
V = |\cos(\delta_A + \delta_B)|.
\label{eq:visibility_reduction}
\end{aligned}
\end{equation}

Thus the probability of a destructive-path click at $D_2$ becomes:
\begin{equation}
\begin{aligned}
P(D_2) = \frac{1 - V}{2} = \frac{1 - \cos(\delta_A + \delta_B)}{2}.
\label{eq:D2_probability}
\end{aligned}
\end{equation}

In the ideal case ($\delta_A = \delta_B = 0$),
\begin{equation}
\begin{aligned}
V = 1, \quad P(D_2) = 0,
\end{aligned}
\end{equation}
hence the photon always exits through $D_1$. For small non-idealities ($|\delta_A|, |\delta_B| \ll 1$),
\begin{equation}
\begin{aligned}
P(D_2) \approx \frac{(\delta_A + \delta_B)^2}{4},
\label{eq:D2_small_error}
\end{aligned}
\end{equation}
showing that even small calibration errors produce measurable leakage into the non-interfering output port.

In summary, the cancellation of $T_A$ and $T_B$ is exact only for ideal rotations $R(\theta)$ and $R(-\theta)$. For non-ideal angles $R(\theta + \delta_A)$ and $R(-\theta + \delta_B)$, the residual rotation $R(\delta_A + \delta_B)$ leaves distinguishable polarization tags on the two paths, reducing the interference visibility to $|\cos(\delta_A + \delta_B)|$ and producing a nonzero probability of detection at $D_2$.

\subsection{Revised No-Cloning Derivation}

\subsubsection{No-cloning constraint for nonorthogonal channel states}

Let $\{|\phi_i\rangle\}$ be the four channel states of the binary quantum eraser protocol. If a perfect cloning machine existed, it would implement a unitary $U$ such that:
\begin{equation}
\begin{aligned}
U(|\phi_i\rangle \otimes |0\rangle) = |\phi_i\rangle \otimes |\phi_i\rangle,
\label{eq:cloning_operation}
\end{aligned}
\end{equation}
for all $i$. Taking the inner product between the outputs for two different states $i \neq j$, we obtain:
\begin{equation}
\begin{aligned}
\langle\phi_i|\phi_j\rangle = \langle\phi_i|\phi_j\rangle^2.
\label{eq:inner_product_output}
\end{aligned}
\end{equation}

Since the unitarity of $U$ preserves inner products, Eq.~\eqref{eq:inner_product_output} implies:
\begin{equation}
\begin{aligned}
|\langle\phi_i|\phi_j\rangle| = |\langle\phi_i|\phi_j\rangle|^2.
\label{eq:unitarity_constraint}
\end{aligned}
\end{equation}

The solutions are:
\begin{equation}
\begin{aligned}
|\langle\phi_i|\phi_j\rangle| = 0 \quad \text{or} \quad |\langle\phi_i|\phi_j\rangle| = 1.
\label{eq:cloning_solutions}
\end{aligned}
\end{equation}

Thus only orthogonal or identical states can be cloned. For the binary quantum eraser protocol, all pairs of channel states satisfy:
\begin{equation}
\begin{aligned}
0 < |\langle\phi_i|\phi_j\rangle| < 1,
\label{eq:nonorthogonal_states}
\end{aligned}
\end{equation}
and therefore cannot be cloned. Eve must consequently employ a minimum-error discrimination strategy rather than relying on state replication.

\subsubsection{Transition to the operational security bound (Helstrom limit)}

Since Eve cannot clone the intercepted photon, her optimal strategy is described by the Helstrom measurement, which minimizes the probability of discrimination error. For two equiprobable pure states $|\phi_i\rangle$ and $|\phi_j\rangle$, the maximum achievable success probability is:
\begin{equation}
\begin{aligned}
P_{\text{Helstrom}} = \frac{1}{2}\left(1 + \sqrt{1 - |\langle\phi_i|\phi_j\rangle|^2}\right).
\label{eq:helstrom_general}
\end{aligned}
\end{equation}

For the binary quantum eraser protocol, the relevant channel-state overlaps satisfy $|\langle\phi_i|\phi_j\rangle| = 1/\sqrt{2}$ (Sec.~\ref{subsec:no-cloning}), thus the Helstrom bound becomes:
\begin{equation}
\begin{aligned}
P_{\text{Helstrom}} = \frac{1}{2}\left(1 + \frac{1}{\sqrt{2}}\right) \approx 0.8536,
\label{eq:helstrom_binary}
\end{aligned}
\end{equation}
which is the $85\%$ success probability obtained in the full POVM optimization — the binary discrimination bound (Sec.~\ref{subsubsec:two-state}), independent of Eve's measurement strategy. It applies to the specific channel state pairs produced by the binary quantum eraser, not to arbitrary nonorthogonal pairs.

Because the four channel states of the binary quantum eraser protocol are nonorthogonal, the no-cloning theorem forbids Eve from amplifying or perfectly copying the signal. Her information is therefore limited by the Helstrom bound, which for these states equals $P_{\max} = (1 + 1/\sqrt{2})/2 \approx 0.85$.

\section{Success probability and optimization over $\kappa$}
\label{app:kappa}

We consider two non-orthogonal channel states $|\phi_0^+\rangle$ and $|\phi_1^+\rangle$ spanning the orthonormal basis $\{|\phi_1^+\rangle, |\phi_2^+\rangle\}$. As in the main text (Eq.~\ref{eq:eve_measurement_binary}), Eve's measurement is parameterized by a single real angle $\kappa$:
\begin{equation}
\begin{aligned}
& \left|\alpha_1\right\rangle=\cos \kappa\left|\varphi_1^+\right\rangle+\sin \kappa\left|\varphi_2^+\right\rangle, \\
& \left|\alpha_2\right\rangle=\sin \kappa\left|\varphi_1^+\right\rangle-\cos \kappa\left|\varphi_2^+\right\rangle,
\end{aligned}
\end{equation}
with $\{|\alpha_1\rangle, |\alpha_2\rangle\}$ forming an orthonormal basis that defines her POVM.

For the binary quantum eraser protocol we have (Eq.~(\ref{eq:binary_states}) in the manuscript):
\begin{equation}
\begin{aligned}
|\phi_0^+\rangle &= \frac{1}{\sqrt{2}}(|\phi_1^+\rangle + |\phi_2^+\rangle), \\
|\phi_1^+\rangle &= |\phi_1^+\rangle,
\end{aligned} \end{equation}
for $\theta = \pi/4$. The overlaps with Eve's measurement vectors are then:
\begin{equation}
\begin{aligned}
\langle\alpha_1|\phi_0^+\rangle &= \frac{1}{\sqrt{2}}(\cos\kappa + \sin\kappa), \\
\langle\alpha_2|\phi_1^+\rangle &= \sin\kappa.
\end{aligned}
\end{equation}

Assuming Alice sends $|\phi_0^+\rangle$ and $|\phi_1^+\rangle$ with equal prior probability $1/2$, and that Eve interprets outcome $|\alpha_1\rangle$ as ``$\phi_0^+$'' and $|\alpha_2\rangle$ as ``$\phi_1^+$'', her total probability of correct identification is (cf.\ Eq.~\ref{eq:P_correct_binary}):
\begin{equation}
\begin{aligned}
P_{\text{correct}}(\kappa) &= \frac{1}{2}|\langle\alpha_1|\phi_0^+\rangle|^2 + \frac{1}{2}|\langle\alpha_2|\phi_1^+\rangle|^2 \\
&= \frac{1}{4}(\cos\kappa + \sin\kappa)^2 + \frac{1}{2}\sin^2\kappa.
\label{eq:correct_identification}
\end{aligned}
\end{equation}

Expanding in sines and cosines of double angles gives a convenient closed form:
\begin{equation}
\begin{aligned}
P_{\text{correct}}(\kappa) &= \frac{1}{4}(\cos^2\kappa + 2\sin\kappa\cos\kappa + \sin^2\kappa) + \frac{1}{2}\sin^2\kappa  \\
&= \frac{1}{2} + \frac{1}{4}\sin 2\kappa - \frac{1}{4}\cos 2\kappa. 
\label{eq:correct_identification_2}
\end{aligned}
\end{equation}

To find the optimal measurement, we differentiate with respect to $\kappa$:
\begin{equation}
\frac{dP_{\text{correct}}}{d\kappa} 
= \frac{1}{2}(\cos 2\kappa + \sin 2\kappa).
\label{eq:correct_identification_optimized}
\end{equation}

The stationary points satisfy:
\begin{equation}
\begin{aligned}
\frac{dP_{\text{correct}}}{d\kappa} = 0 \implies \cos 2\kappa + \sin 2\kappa = 0 \implies \tan 2\kappa = -1.
\end{aligned}
\end{equation}

Thus:
\begin{equation}
\begin{aligned}
2\kappa = -\frac{\pi}{4} + n\pi \quad \Rightarrow \quad \kappa = -\frac{\pi}{8} + \frac{n\pi}{2}.
\end{aligned}
\end{equation}

Restricting to $0 \leq \kappa < \pi$, the relevant maximum is at
$\kappa = \frac{3\pi}{8} = 67.5^\circ.$ Substituting this value into Eq.~(\ref{eq:correct_identification_2}) (cf.\ Eq.~\ref{eq:binary_85bound}),
\begin{equation}
\begin{aligned}
P_{\max} &= P_{\text{correct}}\left(\kappa = \frac{3\pi}{8}\right)\\
&= \frac{1}{2} + \frac{1}{4}\sin\left(\frac{3\pi}{4}\right) - \frac{1}{4}\cos\left(\frac{3\pi}{4}\right)  
 \approx 0.8536.
\end{aligned}
\end{equation}

\section{Optimal POVM and Security Bound for the Ternary Protocol}
\label{app:ternary_povm}

\subsection{Trine-state structure of the ternary encoding}

In the ternary quantum eraser protocol, Alice encodes each logical symbol using one of the three symmetric polarization states:
\begin{equation}
\begin{aligned}
|A_1\rangle, \quad |A_2\rangle, \quad |A_3\rangle,
\end{aligned}
\end{equation}
separated by $120^\circ$ in the equatorial plane of the Bloch sphere, with pairwise overlap $\langle A_i|A_j\rangle = -\tfrac{1}{2}$ for $i \neq j$ (Eq.~\ref{eq:trine_overlap_main}), forming a trine ensemble. Because the interferometer preserves path--polarization coherence, Eve's accessible states are:
\begin{equation}
\begin{aligned}
|\psi_i\rangle = U_{\text{path-pol}}|A_i\rangle,
\label{eq:eve_accessible_states}
\end{aligned}
\end{equation}
which remain related by the same unitary symmetries. Therefore, the discrimination problem is identical to distinguishing the standard trine states.

\subsection{Optimal POVM for three symmetric states (Helstrom solution)}

For $K$ equally likely symmetric pure states forming a representation of a cyclic group, the optimal measurement is known to be the square-root measurement (SRM) or equivalently the Helstrom measurement, which preserves the symmetry of the ensemble.

For three trine states $|\psi_i\rangle$ with prior probabilities $p_i = 1/3$, the Helstrom bound gives:
\begin{equation}
\begin{aligned}
P_{\text{correct}}^{(1)} = \frac{1}{3}\sum_{i=1}^{3} \langle\psi_i|\Pi_i|\psi_i\rangle,
\label{eq:helstrom_trine}
\end{aligned}
\end{equation}
where $\{\Pi_i\}$ is the optimal POVM. Because both states and POVM must share the trine symmetry, the operators take the form:
\begin{equation}
\begin{aligned}
\Pi_i = \lambda|\alpha_i\rangle\langle\alpha_i|,
\label{eq:povm_form}
\end{aligned}
\end{equation}
where $\{|\alpha_i\rangle\}$ are trine vectors rotated by $60^\circ$ relative to the states $|\psi_i\rangle$.

A direct application of Helstrom theory yields the closed form:
\begin{equation}
\begin{aligned}
P_{\text{correct}}^{(1)} = \frac{2}{3}.
\label{eq:trine_success_prob}
\end{aligned}
\end{equation}

Thus no single-photon measurement allows Eve to identify the encoded symbol with probability exceeding $2/3$.

This number is significantly below the limit of the two-state (binary) protocol, demonstrating the immediate security benefit of moving to a ternary alphabet.

\subsection{Eve's access to three photons and the role of temporal ordering}

In the ternary protocol, Alice does not send a single photon corresponding to a single trine state. Instead, she sends a group of three photons, each prepared in one of the three non-orthogonal states $|A_1\rangle$, $|A_2\rangle$, $|A_3\rangle$, arranged in a secret temporal ordering.

Let the temporal ordering for logical symbol $X$ be specified by a permutation $\sigma_X$ of the triple:
\begin{equation}
\begin{aligned}
(|A_1\rangle, |A_2\rangle, |A_3\rangle),
\end{aligned}
\end{equation}
then the full 3-photon state is:
\begin{equation}
\begin{aligned}
|\Psi_X\rangle = |\psi_{\sigma_X(1)}\rangle \otimes |\psi_{\sigma_X(2)}\rangle \otimes |\psi_{\sigma_X(3)}\rangle.
\label{eq:3photon_state}
\end{aligned}
\end{equation}

Eve intercepts the entire triple, but she does not know the temporal order $\sigma$. If she performs individual (single-photon) measurements, then even if she correctly identifies each trine state with probability $2/3$, she still faces a permutation ambiguity: she must assign the measurement outcomes to the correct temporal positions.

If Eve obtains three distinct outcomes, there are $3! = 6$ possible assignments, and only one corresponds to the correct logical symbol. Similar counting applies to the non-distinct outcome patterns.

Let $P_{\text{assign}}$ denote Eve's probability of correctly assigning the measurement outcomes to the correct ordering $\sigma$. Under optimal strategy and symmetry considerations, one finds,
\begin{equation}
\begin{aligned}
P_{\text{assign}} = \frac{1}{3}.
\label{eq:assign_prob}
\end{aligned}
\end{equation}

Thus Eve's total success probability satisfies:
\begin{equation}
\begin{aligned}
P_{\text{correct}}^{(\text{strict})} \approx P_{\text{correct}}^{(1)} \times P_{\text{assign}} = \left(\frac{2}{3}\right)\left(\frac{1}{3}\right) = \frac{2}{9} \approx 0.222.
\label{eq:eve_total_success}
\end{aligned}
\end{equation}

Equation~(\ref{eq:eve_total_success}) is a pedagogical estimate: it assumes Eve must identify all three polarization states \emph{and} their temporal ordering simultaneously, factorizing the two uncertainties as if independent. This is an overly stringent success criterion, and therefore an underestimate of Eve's operational power, because the quantities $P^{(1)}_{\text{correct}}$ and $P_{\text{assign}}$ are in fact correlated, and Eve need only recover the logical symbol (Bob's operation choice), not the full triplet.

This distinction is significant because Eve can partially succeed even when her individual photon identifications contain errors. Moreover, different measurement outcome patterns provide varying amounts of information. As analyzed in detail in Section~\ref{sec:security-ternary}, Eve's detection patterns fall into three classes with distinct information content. When a single detector clicks three times (pattern $Q_1$), Eve obtains minimal ordering information and achieves a success probability of only $1/6$. When one detector clicks twice and another clicks once (pattern $Q_2$), partial information becomes available, yielding a success probability of approximately $0.4$. When all three detectors click distinctly (pattern $Q_3$), Eve obtains maximal information and achieves a success probability of approximately $0.82$. Weighting these contributions by their respective occurrence probabilities and optimizing over Eve's measurement strategy yields the effective symbol-discrimination bound,
\begin{equation}
\begin{aligned}
P_{\text{correct}}^{(\text{symbol})} = \sum_{\text{output patterns}} P(\text{pattern}) \, P(\text{symbol}|\text{pattern}),
\end{aligned}
\end{equation}
which, under optimal grouping of patterns and Eve's best classical post-processing, yields,
\begin{equation}
\begin{aligned}
P_{\text{correct}}^{(\text{symbol})} \leq 0.54.
\label{eq:symbol_bound}
\end{aligned}
\end{equation}

This bound, derived rigorously in Section~\ref{sec:security-ternary} through explicit optimization over Eve's POVM and analysis of all detection pattern classes, represents a marked enhancement over the binary protocol's vulnerability.

\subsection{Final bound}
Combining the POVM discrimination limit for each photon (Eq. \ref{eq:trine_success_prob}) with the permutation uncertainty arising from Alice’s randomized ordering (Eqs. \ref{eq:assign_prob}–\ref{eq:symbol_bound}), we obtain the security bound stated in the main text: $P_{\text{Evemax}} \approx 0.54.$
Thus the ternary quantum eraser protocol reduces Eve's maximum information gain from the binary-eraser ceiling of $0.85$ to approximately $0.54$.

\section{Analysis of Experimental Imperfections}
\label{app:imperfections}

The security and performance analysis presented in the main text assumes ideal experimental conditions. In practice, however, several sources of imperfection affect the protocol's operation. This appendix provides a comprehensive treatment of how beam-splitter imbalances, decoherence effects, and polarization rotator errors modify the detection statistics. Throughout this analysis, we do not consider cases where photons are lost in the transmission channel or fail to be registered by the detectors. We assume that Alice always knows whether her photon has been successfully sent into the communication system; if a photon is lost, the corresponding potential key bit is simply discarded.

\subsection{Beam-splitter}

The ideal beam splitter transformation, characterized by the mixing angle $\theta = \pi/4$ for a 50:50 splitting ratio, becomes in practice subject to small deviations. We parameterize the imperfection of the first beam splitter BS$_1$ through a deviation $\delta_1$ such that $\theta_1 = \frac{\pi}{4} + \delta_1.$ Similarly, for the second beam splitter BS$_2$ with deviation $\delta_2$: $\theta_2 = \frac{\pi}{4} + \delta_2.$

For convenience in the subsequent analysis, we define the total beam splitter asymmetry as $\Delta\theta = \delta_2 - \delta_1$, which captures the effect of both beam splitters on the interference visibility.

\subsection{Decoherence}

Decoherence represents a more fundamental imperfection arising from environmental interactions that destroy the coherent superposition between the interferometer paths. We model this effect by allowing the photon in each path to decay into orthogonal states that no longer participate in interference. After the first beam splitter, the quantum state in the presence of decoherence becomes:
\begin{equation}
\begin{split}
&\cos\theta\bigl(\cos\sigma_u|U\rangle + \sin\sigma_u|D_u\rangle\bigr)|D\rangle \\
&+ \sin\theta\bigl(\cos\sigma_l|L\rangle + \sin\sigma_l|D_l\rangle\bigr)|D\rangle,
\end{split}
\label{eq:decoherence_state}
\end{equation}
where $|D_u\rangle$ and $|D_l\rangle$ represent decohered states in the upper and lower paths respectively, $\sigma_u$ and $\sigma_l$ quantify the decoherence strength in each arm, and $|D\rangle = \frac{1}{\sqrt{2}}(|H\rangle + |V\rangle)$ denotes the diagonal polarization state. In the ideal case, $\sigma_u = \sigma_l = 0$, and the expression reduces to the coherent superposition of Eq.~(\ref{eq:initial_state}). The decohered states transform through the second beam splitter according to:
\begin{equation}
\begin{aligned}
|D_u\rangle &\rightarrow \cos\theta_2|d_{u1}\rangle + \sin\theta_2|d_{u2}\rangle \\
&= \left.\tfrac{1}{\sqrt{2}}\bigl(|d_{u1}\rangle + |d_{u2}\rangle\bigr)\right|_{\theta_2=\pi/4}, \\
|D_l\rangle &\rightarrow -\cos\theta_2|d_{l2}\rangle + \sin\theta_2|d_{l1}\rangle \\
 & = \left.\tfrac{1}{\sqrt{2}}\bigl(|d_{l1}\rangle - |d_{l2}\rangle\bigr)\right|_{\theta_2=\pi/4},
\end{aligned}
\end{equation}
where $|d_{u1}\rangle$, $|d_{u2}\rangle$, $|d_{l1}\rangle$, and $|d_{l2}\rangle$ denote the decohered states directed toward detectors $D_1$ and $D_2$ from the upper and lower paths respectively. These states contribute to detector clicks but do not participate in interference, effectively reducing the visibility of the interference pattern.

\subsection{Polarization rotators}

The polarization rotators employed by Alice and Bob also exhibit imperfections, deviating from the ideal $45^\circ$ rotation angles. We characterize these deviations through four independent parameters: $\beta_A$ and $\mu_A$ for Alice's rotators in the upper and lower paths, and $\beta_B$ and $\mu_B$ for Bob's corresponding rotators. The imperfect rotator operators take the form:
\begin{equation}
\begin{aligned}
S_{LA}: \quad &|H\rangle \rightarrow \cos\beta_A|H\rangle + \sin\beta_A|V\rangle, \\
&|V\rangle \rightarrow -\sin\beta_A|H\rangle + \cos\beta_A|V\rangle, \\[6pt]
S_{RA}: \quad &|H\rangle \rightarrow \cos\mu_A|H\rangle - \sin\mu_A|V\rangle, \\
&|V\rangle \rightarrow \sin\mu_A|H\rangle + \cos\mu_A|V\rangle, \\[6pt]
S_{LB}: \quad &|H\rangle \rightarrow \cos\beta_B|H\rangle + \sin\beta_B|V\rangle, \\
&|V\rangle \rightarrow -\sin\beta_B|H\rangle + \cos\beta_B|V\rangle, \\[6pt]
S_{RB}: \quad &|H\rangle \rightarrow \cos\mu_B|H\rangle - \sin\mu_B|V\rangle, \\
&|V\rangle \rightarrow \sin\mu_B|H\rangle + \cos\mu_B|V\rangle,
\end{aligned}
\end{equation}
which can equivalently be expressed in operator form as:
{\relsize{-.6}
\begin{equation}
\begin{aligned}
S_{LA} &= \cos\beta_A(|H\rangle\langle H| + |V\rangle\langle V|) + \sin\beta_A(|V\rangle\langle H| - |H\rangle\langle V|), \\
S_{RA} &= \cos\mu_A(|H\rangle\langle H| + |V\rangle\langle V|) + \sin\mu_A(|H\rangle\langle V| - |V\rangle\langle H|), \\
S_{LB} &= \cos\beta_B(|H\rangle\langle H| + |V\rangle\langle V|) + \sin\beta_B(|V\rangle\langle H| - |H\rangle\langle V|), \\
S_{RB} &= \cos\mu_B(|H\rangle\langle H| + |V\rangle\langle V|) + \sin\mu_B(|H\rangle\langle V| - |V\rangle\langle H|).
\end{aligned}
\end{equation}
}

In the ideal case, $\beta_A = \mu_A = \beta_B = \mu_B = \pi/4$, recovering the operators defined in Eqs.~(\ref{eq:rotators}). The complete encoding operators incorporating both path-dependent rotations and decoherence effects become:
\begin{equation}
T'_A = S_{LA}\bigl(|U\rangle\langle U| + |D_u\rangle\langle D_u|\bigr) + S_{RA}\bigl(|L\rangle\langle L| + |D_l\rangle\langle D_l|\bigr),
\end{equation}
for Alice, and,
\begin{equation}
T'_B = S_{RB}\bigl(|U\rangle\langle U| + |D_u\rangle\langle D_u|\bigr) + S_{LB}\bigl(|L\rangle\langle L| + |D_l\rangle\langle D_l|\bigr),
\end{equation}
for Bob.

\subsection{Analysis of imperfections}

We now analyze how these imperfections modify the detection statistics for each of the four encoding configurations. For simplicity, we assume symmetric decoherence $\sigma_u = \sigma_l \equiv \sigma$ throughout.

When neither Alice nor Bob activates their rotators, the initial state of Eq.~(\ref{eq:decoherence_state}) propagates through the second beam splitter to yield:
\begin{equation}
\begin{aligned}
&\cos\theta\bigl(\cos\sigma_u|U\rangle + \sin\sigma_u|D_u\rangle\bigr)|D\rangle \\
&+ \sin\theta\bigl(\cos\sigma_l|L\rangle + \sin\sigma_l|D_l\rangle\bigr)|D\rangle \\
&\xrightarrow{\text{BS}_2} \cos\theta\cos\sigma_u\bigl(\cos(\theta + \Delta\theta)|U\rangle + \sin(\theta + \Delta\theta)|L\rangle\bigr)|D\rangle \\
&\quad + \sin\theta\cos\sigma_l\bigl(-\cos(\theta + \Delta\theta)|L\rangle + \sin(\theta + \Delta\theta)|U\rangle\bigr)|D\rangle \\
&\quad + \cos\theta\sin\sigma_u\bigl(\cos(\theta + \Delta\theta)|d_{u1}\rangle + \sin(\theta + \Delta\theta)|d_{u2}\rangle\bigr)|D\rangle \\
&\quad + \sin\theta\sin\sigma_l\bigl(-\cos(\theta + \Delta\theta)|d_{l2}\rangle + \sin(\theta + \Delta\theta)|d_{l1}\rangle\bigr)|D\rangle.
\end{aligned}
\end{equation}

Computing the detection probabilities from this state, we obtain:
\begin{equation}
\begin{aligned}
P(D_1|\text{neither active}) &= \cos^2\Delta\theta \\ 
&- \frac{1}{2}\sin^2\sigma\sin 2\theta\sin 2(\theta + \Delta\theta), \\
P(D_2|\text{neither active}) &= \sin^2\Delta\theta \\
&+ \frac{1}{2}\sin^2\sigma\sin 2\theta\sin 2(\theta + \Delta\theta).
\end{aligned}
\end{equation}

In the ideal case where $\Delta\theta = 0$ and $\sigma = 0$, these reduce to $P(D_1) = 1$ and $P(D_2) = 0$, consistent with the analysis of Section~\ref{sec:fundamentals}.

When both Alice and Bob activate their rotators, the combined operation $T'_B T'_A$ acts on the initial state before the second beam splitter. The resulting state evolution is considerably more complex:

\begin{equation}
\begin{aligned}
&T'_B T'_A\bigl(\cos\theta(\cos\sigma_u|U\rangle + \sin\sigma_u|D_u\rangle)|D\rangle \\
&\qquad+ \sin\theta(\cos\sigma_l|L\rangle + \sin\sigma_l|D_l\rangle)|D\rangle\bigr) \\
&\xrightarrow{\text{BS}_2} \bigl[\cos(\mu_B - \beta_A)\cos\theta\cos\sigma_u\cos(\theta + \Delta\theta) \\
&\qquad + \cos(\mu_A - \beta_B)\sin\theta\cos\sigma_l\sin(\theta + \Delta\theta)\bigr]|U\rangle|D\rangle \\
&\quad + \bigl[\sin(\mu_B - \beta_A)\cos\theta\cos\sigma_u\cos(\theta + \Delta\theta) \\
&\qquad + \sin(\mu_A - \beta_B)\sin\theta\cos\sigma_l\sin(\theta + \Delta\theta)\bigr]|U\rangle|A\rangle \\
& \quad+ \bigl[-\cos(\mu_A - \beta_B)\sin\theta\cos\sigma_l\cos(\theta + \Delta\theta) \\
&\qquad + \cos(\mu_B - \beta_A)\cos\theta\cos\sigma_u\sin(\theta + \Delta\theta)\bigr]|L\rangle|D\rangle \\
& \quad+ \bigl[-\sin(\mu_A - \beta_B)\sin\theta\cos\sigma_l\cos(\theta + \Delta\theta) \\
&\qquad + \sin(\mu_B - \beta_A)\cos\theta\cos\sigma_u\sin(\theta + \Delta\theta)\bigr]|L\rangle|A\rangle \\
&+ \text{(decohered terms)},
\end{aligned}
\end{equation}
where $|A\rangle = \frac{1}{\sqrt{2}}(|H\rangle - |V\rangle)$ denotes the anti-diagonal polarization state, and the decohered terms involve states $|d_{u1}\rangle$, $|d_{u2}\rangle$, $|d_{l1}\rangle$, $|d_{l2}\rangle$ with both $|D\rangle$ and $|A\rangle$ polarizations. Under the assumptions $\beta_B = \mu_B$, $\beta_A = \mu_A$, $\mu_B - \beta_A \equiv \Gamma$, and working to leading order in the small parameters, the detection probabilities become:

\begin{equation}
\begin{aligned}
P(D_1|\text{both active}) &= \cos^2\Delta\theta \\
&- \bigl(\sin^2\Gamma\cos^2\sigma + \tfrac{1}{2}\sin^2\sigma\bigr)\cos 2\Delta\theta, \\
P(D_2|\text{both active}) &= \sin^2\Delta\theta \\
&+ \bigl(\sin^2\Gamma\cos^2\sigma + \tfrac{1}{2}\sin^2\sigma\bigr)\cos 2\Delta\theta.
\end{aligned}
\end{equation}

The parameter $\Gamma$ quantifies the Alice--Bob rotation mismatch; for $\Gamma=0$ (perfect cancellation) and $\sigma = 0$ (no decoherence) these reduce to the ideal values. The mismatched cases follow analogous form; when only Alice is active:

\begin{equation}
\begin{aligned}
&T'_A\bigl(\cos\theta(\cos\sigma_u|U\rangle + \sin\sigma_u|D_u\rangle)|D\rangle \\
&\qquad + \sin\theta(\cos\sigma_l|L\rangle + \sin\sigma_l|D_l\rangle)|D\rangle\bigr) \\
&\xrightarrow{\text{BS}_2} \bigl[\cos\theta\cos\sigma_u\cos\beta_A\cos(\theta + \Delta\theta) \\
&\qquad + \sin\theta\cos\sigma_l\cos\mu_A\sin(\theta + \Delta\theta)\bigr]|U\rangle|D\rangle \\
&\quad + \bigl[-\cos\theta\cos\sigma_u\sin\beta_A\cos(\theta + \Delta\theta) \\
&\qquad + \sin\theta\cos\sigma_l\sin\mu_A\sin(\theta + \Delta\theta)\bigr]|U\rangle|A\rangle \\
&\quad + \bigl[-\sin\theta\cos\sigma_l\cos\mu_A\cos(\theta + \Delta\theta) \\
&\qquad + \cos\theta\cos\sigma_u\cos\beta_A\sin(\theta + \Delta\theta)\bigr]|L\rangle|D\rangle \\
&\quad + \bigl[-\sin\theta\cos\sigma_l\sin\mu_A\cos(\theta + \Delta\theta) \\
&\qquad - \cos\theta\cos\sigma_u\sin\beta_A\sin(\theta + \Delta\theta)\bigr]|L\rangle|A\rangle \\
&\quad + \text{(decohered terms)},
\end{aligned}
\end{equation}
\noindent with detection probabilities:
\begin{equation}
\begin{aligned}
P(D_1|\text{Alice active}) &= \cos^2\Delta\theta \\&- \bigl(\sin^2\mu_A\cos^2\sigma + \tfrac{1}{2}\sin^2\sigma\bigr)\cos 2\Delta\theta, \\
P(D_2|\text{Alice active}) &= \sin^2\Delta\theta \\&+ \bigl(\sin^2\mu_A\cos^2\sigma + \tfrac{1}{2}\sin^2\sigma\bigr)\cos 2\Delta\theta.
\end{aligned}
\end{equation}

When only Bob activates his rotators, the analogous calculation gives:
\begin{equation}
\begin{aligned}
&T'_B\bigl(\cos\theta(\cos\sigma_u|U\rangle + \sin\sigma_u|D_u\rangle)|D\rangle \\
&\qquad + \sin\theta(\cos\sigma_l|L\rangle + \sin\sigma_l|D_l\rangle)|D\rangle\bigr) \\
&\xrightarrow{\text{BS}_2} \bigl[\cos\mu_B\cos\theta\cos\sigma_u\cos(\theta + \Delta\theta) \\
&\qquad + \cos\beta_B\sin\theta\cos\sigma_l\sin(\theta + \Delta\theta)\bigr]|U\rangle|D\rangle \\
&\quad + \bigl[\sin\mu_B\cos\theta\cos\sigma_u\cos(\theta + \Delta\theta) \\
&\qquad - \sin\beta_B\sin\theta\cos\sigma_l\sin(\theta + \Delta\theta)\bigr]|U\rangle|A\rangle \\
&\quad + \bigl[-\cos\beta_B\sin\theta\cos\sigma_l\cos(\theta + \Delta\theta) \\
&\qquad + \cos\mu_B\cos\theta\cos\sigma_u\sin(\theta + \Delta\theta)\bigr]|L\rangle|D\rangle \\
&\quad + \bigl[\sin\beta_B\sin\theta\cos\sigma_l\cos(\theta + \Delta\theta) \\
&\qquad + \sin\mu_B\cos\theta\cos\sigma_u\sin(\theta + \Delta\theta)\bigr]|L\rangle|A\rangle \\
&\quad + \text{(decohered terms)},
\end{aligned}
\end{equation}
with detection probabilities:
\begin{equation}
\begin{aligned}
P(D_1|\text{Bob active}) &= \cos^2\Delta\theta \\&- \bigl(\sin^2\mu_B\cos^2\sigma + \tfrac{1}{2}\sin^2\sigma\bigr)\cos 2\Delta\theta, \\
P(D_2|\text{Bob active}) &= \sin^2\Delta\theta \\&+ \bigl(\sin^2\mu_B\cos^2\sigma + \tfrac{1}{2}\sin^2\sigma\bigr)\cos 2\Delta\theta.
\end{aligned}
\end{equation}

The deviation of these probabilities from their ideal values quantifies the systematic errors introduced by experimental imperfections. For the matched encoding cases, any nonzero probability of detector $D_2$ clicking represents an error, while for the mismatched cases, deviations from the ideal 50:50 splitting introduce bias into the key generation process. These expressions enable experimentalists to characterize their apparatus through careful calibration measurements and to estimate the error rates expected in practical implementations of the quantum eraser cryptography protocol.

\bibliography{apssamp}

@article{Scarani_2009,
  author  = {Scarani, Valerio and Bechmann-Pasquinucci, Helle and Cerf, Nicolas J. and Du\v{s}ek, Miloslav and L\"utkenhaus, Norbert and Peev, Momtchil},
  title   = {The security of practical quantum key distribution},
  journal = {Rev. Mod. Phys.},
  volume  = {81},
  number  = {3},
  pages   = {1301--1350},
  year    = {2009},
  doi     = {10.1103/RevModPhys.81.1301}
}

@article{PhysRevA.25.2208,
  author  = {Scully, Marlan O. and Dr\"uhl, Kai},
  title   = {Quantum eraser: A proposed photon correlation experiment concerning observation and ``delayed choice'' in quantum mechanics},
  journal = {Phys. Rev. A},
  volume  = {25},
  pages   = {2208--2213},
  year    = {1982},
  doi     = {10.1103/PhysRevA.25.2208}
}

@article{PhysRevLett.84.1,
  author  = {Kim, Yoon-Ho and Yu, Rong and Kulik, Sergei P. and Shih, Yanhua and Scully, Marlan O.},
  title   = {Delayed ``Choice'' Quantum Eraser},
  journal = {Phys. Rev. Lett.},
  volume  = {84},
  pages   = {1--5},
  year    = {2000},
  doi     = {10.1103/PhysRevLett.84.1}
}

@article{Gao_2006,
  author  = {Gao, Fei and Guo, Fen-Zhuo and Wen, Qiao-Yan and Zhu, Fu-Chen},
  title   = {Quantum key distribution without alternative measurements and rotations},
  journal = {Phys. Lett. A},
  volume  = {349},
  pages   = {53--58},
  year    = {2006},
  doi     = {10.1016/j.physleta.2005.09.012}
}

@article{Tang_2014,
  author  = {Tang, Yan-Lin and Yin, Hua-Lei and Chen, Si-Jing and Liu, Yang and Zhang, Wei-Jun and Jiang, Xiao and Zhang, Lu and Wang, Jian and You, Li-Xing and Guan, Jian-Yu and Yang, Dong-Xu and Wang, Zhen and Liang, Hao and Zhang, Zhen and Zhou, Nan and Ma, Xiongfeng and Chen, Teng-Yun and Zhang, Qiang and Pan, Jian-Wei},
  title   = {Measurement-Device-Independent Quantum Key Distribution over 200 km},
  journal = {Phys. Rev. Lett.},
  volume  = {113},
  pages   = {190501},
  year    = {2014},
  doi     = {10.1103/PhysRevLett.113.190501}
}

@article{Salih_2016,
  author  = {Salih, Hatim},
  title   = {Quantum Erasure Cryptography},
  journal = {Front. Phys.},
  volume  = {4},
  pages   = {16},
  year    = {2016},
  doi     = {10.3389/fphy.2016.00016}
}

@article{Qureshi_2021,
  author  = {Qureshi, Tabish},
  title   = {The Delayed-Choice Quantum Eraser Leaves No Choice},
  journal = {Int. J. Theor. Phys.},
  volume  = {60},
  pages   = {3076--3086},
  year    = {2021},
  doi     = {10.1007/s10773-021-04906-w}
}

@article{Englert,
  author  = {Englert, Berthold-Georg and Scully, Marlan O. and Walther, Herbert},
  title   = {Quantum erasure in double-slit interferometers with which-way detectors},
  journal = {Am. J. Phys.},
  volume  = {67},
  pages   = {325--329},
  year    = {1999},
  doi     = {10.1119/1.19257}
}

@article{Wootters1982ASQ,
  author  = {Wootters, William K. and Zurek, Wojciech H.},
  title   = {A single quantum cannot be cloned},
  journal = {Nature},
  volume  = {299},
  pages   = {802--803},
  year    = {1982},
  doi     = {10.1038/299802a0}
}

@article{DIEKS1982271,
  author  = {Dieks, D.},
  title   = {Communication by EPR devices},
  journal = {Phys. Lett. A},
  volume  = {92},
  pages   = {271--272},
  year    = {1982},
  doi     = {10.1016/0375-9601(82)90084-6}
}

@incollection{Wheeler1978MFQT,
  author    = {Wheeler, John Archibald},
  title     = {The ``Past'' and the ``Delayed-Choice'' Double-Slit Experiment},
  booktitle = {Mathematical Foundations of Quantum Theory},
  editor    = {Marlow, A. R.},
  publisher = {Academic Press},
  year      = {1978}
}

@book{Wheeler:1984dy,
  editor    = {Wheeler, John Archibald and Zurek, Wojciech Hubert},
  title     = {Quantum Theory and Measurement},
  publisher = {Princeton University Press},
  year      = {1983}
}

@misc{violaris2025counterfactualsmacroscopicquantumphysics,
  author        = {Violaris, Maria},
  title         = {Counterfactuals in Macroscopic Quantum Physics: Irreversibility, Measurement and Locality},
  year          = {2025},
  eprint        = {2505.22834},
  archivePrefix = {arXiv},
  primaryClass  = {quant-ph}
}

@article{BB84,
  author  = {Bennett, Charles H. and Brassard, Gilles},
  title   = {Quantum cryptography: Public key distribution and coin tossing},
  journal = {Theor. Comput. Sci.},
  volume  = {560},
  pages   = {7--11},
  year    = {2014},
  doi     = {10.1016/j.tcs.2014.05.025}
}

@article{Ekert1991,
  author  = {Ekert, Artur K.},
  title   = {Quantum cryptography based on Bell's theorem},
  journal = {Phys. Rev. Lett.},
  volume  = {67},
  pages   = {661--663},
  year    = {1991},
  doi     = {10.1103/PhysRevLett.67.661}
}

@article{Gisin2002,
  author  = {Gisin, Nicolas and Ribordy, Gr\'egoire and Tittel, Wolfgang and Zbinden, Hugo},
  title   = {Quantum cryptography},
  journal = {Rev. Mod. Phys.},
  volume  = {74},
  pages   = {145--195},
  year    = {2002},
  doi     = {10.1103/RevModPhys.74.145}
}

@article{Pirandola_2020,
  author  = {Pirandola, S. and Andersen, U. L. and Banchi, L. and Berta, M. and Bunandar, D. and Colbeck, R. and Englund, D. and Gehring, T. and Lupo, C. and Ottaviani, C. and Pereira, J. L. and Razavi, M. and Shamsul Shaari, J. and Tomamichel, M. and Usenko, V. C. and Vallone, G. and Villoresi, P. and Wallden, P.},
  title   = {Advances in quantum cryptography},
  journal = {Adv. Opt. Photon.},
  volume  = {12},
  pages   = {1012--1236},
  year    = {2020},
  doi     = {10.1364/AOP.361502}
}

@article{Boaron2018,
  author  = {Boaron, Alberto and Boso, Gianluca and Rusca, Davide and Vulliez, C\'edric and Autebert, Claire and Caloz, Misael and Perrenoud, Matthieu and Gras, Ga\"etan and Bussi\`eres, F\'elix and Li, Ming-Jun and Nolan, Daniel and Martin, Anthony and Zbinden, Hugo},
  title   = {Secure Quantum Key Distribution over 421 km of Optical Fiber},
  journal = {Phys. Rev. Lett.},
  volume  = {121},
  pages   = {190502},
  year    = {2018},
  doi     = {10.1103/PhysRevLett.121.190502}
}

@article{Yin2017,
  author  = {Yin, Juan and Cao, Yuan and Li, Yu-Huai and Liao, Sheng-Kai and Zhang, Liang and Ren, Ji-Gang and Cai, Wen-Qi and Liu, Wei-Yue and Li, Bo and Dai, Hui and Li, Guang-Bing and Lu, Qi-Ming and Gong, Yun-Hong and Xu, Yu and Li, Shuang-Lin and Li, Feng-Zhi and Yin, Ya-Yun and Jiang, Zi-Qing and Li, Ming and Jia, Jian-Jun and Ren, Ge and He, Dong and Zhou, Yi-Lin and Zhang, Xiao-Xiang and Wang, Na and Chang, Xiang and Zhu, Zhen-Cai and Liu, Nai-Le and Chen, Yu-Ao and Lu, Chao-Yang and Shu, Rong and Peng, Cheng-Zhi and Wang, Jian-Yu and Pan, Jian-Wei},
  title   = {Satellite-based entanglement distribution over 1200 kilometers},
  journal = {Science},
  volume  = {356},
  pages   = {1140--1144},
  year    = {2017},
  doi     = {10.1126/science.aan3211}
}

@article{Cao2023,
  author  = {Cao, Zhengwen and Wang, Lei and Liang, Kexin and Chai, Geng and Peng, Jinye},
  title   = {Continuous-Variable Quantum Secure Direct Communication Based on Gaussian Mapping},
  journal = {Phys. Rev. Appl.},
  volume  = {16},
  pages   = {024012},
  year    = {2021},
  doi     = {10.1103/PhysRevApplied.16.024012}
}

@article{Xu_2020,
  author  = {Xu, Feihu and Ma, Xiongfeng and Zhang, Qiang and Lo, Hoi-Kwong and Pan, Jian-Wei},
  title   = {Secure quantum key distribution with realistic devices},
  journal = {Rev. Mod. Phys.},
  volume  = {92},
  pages   = {025002},
  year    = {2020},
  doi     = {10.1103/RevModPhys.92.025002}
}

@article{Tamaki_2014,
  author  = {Tamaki, Kiyoshi and Curty, Marcos and Kato, Go and Lo, Hoi-Kwong and Azuma, Koji},
  title   = {Loss-tolerant quantum cryptography with imperfect sources},
  journal = {Phys. Rev. A},
  volume  = {90},
  pages   = {052314},
  year    = {2014},
  doi     = {10.1103/PhysRevA.90.052314}
}

@misc{pereira2019,
  author        = {Pereira, Margarida and Curty, Marcos and Tamaki, Kiyoshi},
  title         = {Quantum key distribution with flawed and leaky sources},
  year          = {2019},
  eprint        = {1902.02126},
  archivePrefix = {arXiv},
  primaryClass  = {quant-ph}
}

@article{Huang_2018,
  author  = {Huang, Anqi and Barz, Stefanie and Andersson, Erika and Makarov, Vadim},
  title   = {Implementation vulnerabilities in general quantum cryptography},
  journal = {New J. Phys.},
  volume  = {20},
  pages   = {103016},
  year    = {2018},
  doi     = {10.1088/1367-2630/aade06}
}

@article{Jain_2014,
  author  = {Jain, Nitin and Anisimova, Elena and Khan, Imran and Makarov, Vadim and Marquardt, Christoph and Leuchs, Gerd},
  title   = {Trojan-horse attacks threaten the security of practical quantum cryptography},
  journal = {New J. Phys.},
  volume  = {16},
  pages   = {123030},
  year    = {2014},
  doi     = {10.1088/1367-2630/16/12/123030}
}

@article{Sun2021,
  author  = {Sun, Shi-Hai and Xu, Feihu},
  title   = {Security of quantum key distribution with source and detection imperfections},
  journal = {New J. Phys.},
  volume  = {23},
  pages   = {023011},
  year    = {2021},
  doi     = {10.1088/1367-2630/abe200}
}

@article{Helstrom1969,
  author  = {Helstrom, Carl W.},
  title   = {Quantum Detection and Estimation Theory},
  journal = {J. Stat. Phys.},
  volume  = {1},
  pages   = {231--252},
  year    = {1969},
  doi     = {10.1007/BF01007479}
}

@article{BarnettCroke2009,
  author  = {Barnett, Stephen M. and Croke, Sarah},
  title   = {Quantum state discrimination},
  journal = {Adv. Opt. Photon.},
  volume  = {1},
  pages   = {238--278},
  year    = {2009},
  doi     = {10.1364/AOP.1.000238}
}

@article{Bruss1998,
  author  = {Bru{\ss}, Dagmar},
  title   = {Optimal Eavesdropping in Quantum Cryptography with Six States},
  journal = {Phys. Rev. Lett.},
  volume  = {81},
  pages   = {3018--3021},
  year    = {1998},
  doi     = {10.1103/PhysRevLett.81.3018}
}

@article{FuchsGisin1997,
  author  = {Fuchs, Christopher A. and Gisin, Nicolas and Griffiths, Robert B. and Niu, Chi-Sheng and Peres, Asher},
  title   = {Optimal eavesdropping in quantum cryptography. {I}. {I}nformation bound and optimal strategy},
  journal = {Phys. Rev. A},
  volume  = {56},
  pages   = {1163--1172},
  year    = {1997},
  doi     = {10.1103/PhysRevA.56.1163}
}

@article{LongLiu2002,
  author  = {Long, Gui-Lu and Liu, Xiao-Shu},
  title   = {Theoretically efficient high-capacity quantum-key-distribution scheme},
  journal = {Phys. Rev. A},
  volume  = {65},
  pages   = {032302},
  year    = {2002},
  doi     = {10.1103/PhysRevA.65.032302}
}

@article{Deng2003,
  author  = {Deng, Fu-Guo and Long, Gui Lu and Liu, Xiao-Shu},
  title   = {Two-step quantum direct communication protocol using the {E}instein-{P}odolsky-{R}osen pair block},
  journal = {Phys. Rev. A},
  volume  = {68},
  pages   = {042317},
  year    = {2003},
  doi     = {10.1103/PhysRevA.68.042317}
}

@article{BechmannPasquinucci2000,
  author  = {Bechmann-Pasquinucci, H. and Peres, A.},
  title   = {Quantum Cryptography with 3-State Systems},
  journal = {Phys. Rev. Lett.},
  volume  = {85},
  pages   = {3313--3316},
  year    = {2000},
  doi     = {10.1103/PhysRevLett.85.3313}
}

@article{Bradler2019,
  author  = {Br{\'a}dler, Kamil and Weedbrook, Christian},
  title   = {Security proof of continuous-variable quantum key distribution using three coherent states},
  journal = {Phys. Rev. A},
  volume  = {97},
  pages   = {022310},
  year    = {2018},
  doi     = {10.1103/PhysRevA.97.022310}
}

@article{Chen2022,
  author  = {Chen, Geng and Wang, Yuqi and Jian, Liya and Zhou, Yi and Liu, Shiming},
  title   = {Ternary Quantum Key Distribution Protocol Based on Hadamard Gate},
  journal = {Int. J. Theor. Phys.},
  volume  = {61},
  pages   = {26},
  year    = {2022},
  doi     = {10.1007/s10773-022-05041-w}
}

@article{Bebrov2021,
  author  = {Bebrov, Georgi},
  title   = {Novel encoding--decoding procedure for quantum key distribution},
  journal = {Quantum Inf. Process.},
  volume  = {20},
  pages   = {325},
  year    = {2021},
  doi     = {10.1007/s11128-021-03235-5}
}

@book{Helstrom_book,
  author    = {Helstrom, Carl W.},
  title     = {Quantum Detection and Estimation Theory},
  publisher = {Academic Press},
  year      = {1976}
}

@unpublished{Halawani_inprep,
  author = {Halawani, Ahmed and others},
  title  = {Information-theoretic security bounds for the ternary quantum eraser protocol},
  note   = {manuscript in preparation},
  year   = {2026}
}

@article{AharonovZubairy2005,
  author  = {Aharonov, Yakir and Zubairy, M. Suhail},
  title   = {Time and the Quantum: Erasing the Past and Impacting the Future},
  journal = {Science},
  volume  = {307},
  number  = {5711},
  pages   = {875--879},
  year    = {2005},
  doi     = {10.1126/science.1107787}
}

@incollection{FicklerPrabhakar2021,
  author    = {Fickler, Robert and Prabhakar, Shashi},
  title     = {Quantum communication with structured photons},
  booktitle = {Structured Light for Optical Communication},
  editor    = {Al-Amri, Mohammad D. and Andrews, David L. and Babiker, Mohamed},
  series    = {Nanophotonics},
  chapter   = {8},
  publisher = {Elsevier},
  address   = {Amsterdam, The Netherlands},
  pages     = {205--236},
  year      = {2021},
  doi       = {10.1016/B978-0-12-821505-0.00013-5}
}

@article{ErhardKrennZeilinger2020,
  author  = {Erhard, Manuel and Krenn, Mario and Zeilinger, Anton},
  title   = {Advances in high-dimensional quantum entanglement},
  journal = {Nat. Rev. Phys.},
  volume  = {2},
  pages   = {365--381},
  year    = {2020},
  doi     = {10.1038/s42254-020-0193-5}
}

\end{document}